\documentclass{emulateapj}
\usepackage{apjfonts}
\usepackage{graphicx}
\pdfoutput=1 

\shortauthors{Sullivan et al.}
\shorttitle{{\it TESS}:~Simulated Detections}

\begin{document}

%
\def\ltsima{$\; \buildrel < \over \sim \;$}
\def\lsim{\lower.5ex\hbox{\ltsima}}
\def\gtsima{$\; \buildrel > \over \sim \;$}
\def\gsim{\lower.5ex\hbox{\gtsima}}
\def\tess{{\it TESS}}
\def\kepler{{\it Kepler}}
\def \jwst {{\it JWST}}
\def \teff {T_{\rm eff}}
\def \phir {\Phi_{\rm R}}
\def \fov {24$^{\circ}$}
\def \pixsz {$21\farcs 1$}
\def \aeff {69~cm$^2$}    
\def \epd {105~mm}      
\def \cool {4000}       
\def \snrthresh {7.1}  
\def \snrcr {7.3}  
\def \multm {55$\%$} 
\def \multfgk {53$\%$} 
\def \tmultfgk {26.2$\%$}
\def \tmultm {33.6$\%$}
\def \tmultkep {46$\%$}
\def \pointdur {27.4}
\def \binpct{48$\%$}
      
\def \nearths {70}   
\def \nspeths {486}   
\def \nsmplas {556}
\def \nsmneps {1100}  
\def \ngiantps {67}
\def \nhz {48}
\def \nhzbright {between 2 and 7}
\def \neartherr {70$\pm$9}   
\def \nspetherr {486$\pm$22}   
\def \nsmplaerr {556$\pm$24}
\def \nsmneperr {1111$\pm$122}  
\def \ngiantpserr {67$\pm$8}  
\def \nspethfgk {137}  
\def \nspethm {419}      
\def \nsmbright {130}
\def \nsingles {110}
\def \nhzerr {48$\pm$7}
\def \njwsterr {18$\pm$5}
\def \nkopphz{14$\pm$4}
\def \nebin {97461}
\def \nhebin {21441}
\def \nebs {250 $\pm$16}
\def \nbebs {443 $\pm$20}
\def \nhebs {410 $\pm$20} 
\def \nplatot {$734^{+784}{-510}$}
\def \nebtot {1103$\pm$33}
\def \hbebcenpct{69$\%$}
\def \nfpimg{150}
\def \bebimgpct{78$\%$}
\def \fppdpct{83$\%$}
\def \plancenpct{6$\%$}
\def \nebsec {58$\%$}
\def \nplasec {0.01$\%$}
\def \nebgress {53$\%$}
\def \nebellip {34$\%$}
\def \satmag {6.8}
\def \satpct {$3\%$}
\def \spethicone {4.2}
\def \spethicten {6.3}
\def \smnepicone{5.7}
\def \smnepicten {7.3}
\def \hzicone {9.5}
\def \hzicten {11.6}
\def \nbrightcat {1.58$\times 10^7$}

\def \underadten {12$\%$}
\def \underadtwy {6$\%$}
                                                                                          
%

\bibliographystyle{apj}

\title{The Transiting Exoplanet Survey Satellite:\\
Simulations of planet detections and astrophysical false positives}

\author{
Peter W.\ Sullivan\altaffilmark{1,2},
Joshua N.\ Winn\altaffilmark{1,2},
Zachory K.\ Berta-Thompson\altaffilmark{2}, 
David Charbonneau\altaffilmark{3}, 
Drake Deming\altaffilmark{4}, 
Courtney D.\ Dressing\altaffilmark{3},
David W.\ Latham\altaffilmark{3},
Alan M.\ Levine\altaffilmark{2},  
Peter R.\ McCullough\altaffilmark{5,6},
Timothy Morton\altaffilmark{7}, \\
George R.\ Ricker\altaffilmark{2},
Roland Vanderspek\altaffilmark{2},  
Deborah Woods\altaffilmark{8}
}

\slugcomment{{\it Astroph.~J.}, 809, 1 (2015); corrected as described
  in {\it Errata}, {\it Astroph.~J.}, 837, 1 (2017)}

\altaffiltext{1}{Department of Physics, 77 Massachusetts Ave.,
  Massachusetts Institute of Technology, Cambridge, MA 02139}

\altaffiltext{2}{MIT Kavli Institute for Astrophysics and Space
  Research, 70 Vassar St., Cambridge, MA 02139}

\altaffiltext{3}{Harvard-Smithsonian Center for Astrophysics, 60
  Garden St., Cambridge, MA 02138}

\altaffiltext{4}{Department of Astronomy, University of Maryland,
  College Park, MD 20742}

\altaffiltext{5}{Space Telescope Science Institute, Baltimore, MD
  21218}

\altaffiltext{6}{Department of Physics and Astronomy, Johns Hopkins
  University, 3400 North Charles Street, Baltimore, MD 21218}

\altaffiltext{7}{Department of Astrophysical Sciences, 4 Ivy Lane,
  Peyton Hall, Princeton University, Princeton, NJ 08544}

\altaffiltext{8}{MIT Lincoln Laboratory, 244 Wood St., Lexington, MA
  02420}

\begin{abstract}

  The {\it Transiting Exoplanet Survey Satellite} ({\it TESS}) is a
  NASA-sponsored Explorer mission that will perform a wide-field
  survey for planets that transit bright host stars. Here, we predict
  the properties of the transiting planets that {\it TESS} will detect
  along with the eclipsing binary stars that produce false-positive
  photometric signals.  The predictions are based on Monte Carlo
  simulations of the nearby population of stars, occurrence rates of
  planets derived from {\it Kepler}, and models for the photometric
  performance and sky coverage of the {\it TESS} cameras.  We expect
  that {\it TESS} will find approximately 1700 transiting planets from
  2$\times 10^5$ pre-selected target stars.  This includes \nsmplas{}
  planets smaller than twice the size of Earth, of which \nspethm{}
  are hosted by M dwarf stars and \nspethfgk{} are hosted by FGK
  dwarfs. Approximately \nsmbright{} of the $R<2R_{\oplus}$ planets
  will have host stars brighter than $K_s = 9$. Approximately \nhz{}
  of the planets with $R<2R_{\oplus}$ lie within or near the habitable
  zone ($0.2 < S/S_\oplus < 2$); \nhzbright{} such planets have host
  stars brighter than $K_s = 9$.  We also expect approximately
  \nsmneps{} detections of planets with radii 2-4~$R_{\oplus}$, and
  \ngiantps{} planets larger than 4~$R_{\oplus}$. Additional planets
  larger than 2 $R_{\oplus}$ can be detected around stars that are not
  among the pre-selected target stars, because {\it TESS} will also
  deliver full-frame images at a 30~min cadence. The planet detections
  are accompanied by over one thousand astrophysical false
  positives. We discuss how {\it TESS} data and ground-based
  observations can be used to distinguish the false positives from
  genuine planets. We also discuss the prospects for follow-up
  observations to measure the masses and atmospheres of the \tess{}
  planets.
  
\end{abstract}


\section{INTRODUCTION}

Transiting exoplanets offer opportunities to explore the compositions,
atmospheres, and orbital dynamics of planets beyond the solar
system. The Transiting Exoplanet Survey Satellite (\tess{}) is a
NASA-sponsored Explorer mission that will monitor several hundred
thousand Sun-like and smaller stars for transiting planets
\citep{ricker2015}. The brightest dwarf stars in the sky are the
highest priority for \tess{} because they facilitate follow-up
measurements of the planet masses and atmospheres. After launch
(currently scheduled for late 2017), \tess{} will spend two years
observing nearly the entire sky using four wide-field cameras.

Previous wide-field transit surveys, such as HAT \citep{hat}, TrES
\citep{tres}, XO \citep{xo}, WASP \citep{wasp}, and KELT \citep{kelt},
have been conducted with ground-based telescopes. These surveys have
been very successful in finding giant planets that orbit bright host
stars, but they have struggled to find planets smaller than Neptune
because of the obstacles to achieving fine photometric precision
beneath the Earth's atmosphere.  In contrast, the space missions {\it
  CoRoT} \citep{corot} and \kepler{} \citep{kepler} achieved
outstanding photometric precision, but targeted relatively faint stars
within restricted regions of the sky.  This has made it difficult to
measure the masses or study the atmospheres of the small planets
discovered by {\it CoRoT} and \kepler{}, except for the brightest
systems in each sample.
 
\tess{} aims to combine the merits of wide-field surveys with the fine
photometric precision and long intervals of uninterrupted observation
that are possible in a space mission.  Compared to \kepler{}, \tess{}
will examine stars that are generally brighter by 3~magnitudes 
over a solid angle that is larger by a factor of 400. However, in
order to complete the survey within the primary mission duration of two
years, \tess{} will not monitor stars for nearly as long as \kepler{}
did; it will mainly be sensitive to planets with periods
$\lesssim$20 days.

This paper presents simulations of the population of transiting
planets that \tess{} will detect and the population of eclipsing
binary stars that produce photometric signals resembling those of
transiting planets. These simulations were originally developed to
inform the design of the mission. They are also being used to plan the
campaign of ground-based observations required to distinguish planets
from eclipsing binaries as well as follow-up measurements of planetary
masses and atmospheres. In the future, these simulations could inform
proposals for an extended mission.

Pioneering work on calculating the yield of all-sky transit surveys
was carried out by \cite{pepper2003}. Subsequently, \cite{beattygaudi}
simulated in greater detail the planet yield for several ground-based
and space-based transit surveys, but not including \tess{} (which had
not yet been selected by NASA). \citet{deming2009} considered \tess{}
specifically, but those calculations were based on an earlier design
for the mission with different choices for the observing interval and
duty cycle, the number of cameras and collecting area, and other key
parameters. Furthermore, the occurrence rates of planets have since
been clarified by the {\it Kepler} mission. We have therefore built
our simulation from scratch rather than adapting this previous work.

We have organized this paper as follows:

Section~\ref{sec:pre} provides an overview of \tess{} and the types of
stars that will be searched for transiting planets.

Sections~\ref{sec:catalog}-\ref{sec:best} present our model for the
relevant stellar and planetary populations.  Section~\ref{sec:catalog}
describes the properties and luminosity function of the stars in our
simulation.  Section \ref{sec:eclip} describes the assignment of
transiting planets and eclipsing binary companions to these stars.
Section~\ref{sec:best} combines these results to forecast the
properties of the brightest transiting planet systems on the sky,
regardless of how they might be detected.  This information helps to
set expectations for the yield of any wide-field transit survey, and
for the properties of the most favorable transiting planets for
characterization.

Sections~\ref{sec:inst}-\ref{sec:fp} then describe the detection of
the simulated planets specifically with \tess{}.
Section~\ref{sec:inst} details our model for the photometric
performance of the \tess{} cameras.  Section~\ref{sec:yield} presents
the simulated detections of planets and their properties.
Section~\ref{sec:yield} also shows the detections of astrophysical
false-positives, and Section~\ref{sec:fp} investigates the
possibilities for distinguishing them from planets using \tess{} data
and supplementary data from ground-based telescopes.

Finally, Section~\ref{sec:followup} discusses the prospects for
following up the \tess{} planets to study their masses and
atmospheres.

\section{BRIEF OVERVIEW OF \tess}
\label{sec:pre}

\tess{} employs four refractive cameras, each with a field of view of
\fov{}~$\times$~\fov{} imaged by an array of four 2k$\times$2k
charge-coupled devices (CCD).  This gives a pixel scale of \pixsz{}.
The four camera fields are stacked vertically to create a combined
field that is 24$^{\circ}$ wide and 96$^{\circ}$ tall, captured by
64~Mpixels.  Each camera has an entrance pupil diameter of 105~mm and
an effective collecting area of \aeff{} after accounting for
transmissive losses in the lenses and their coatings. (The relative
spectral response functions of the camera and CCD will be considered
separately.)

Each camera will acquire a new image every 2 seconds. The readout
noise, for which the design goal has a root-mean-square (RMS) level of
10~$e^-$~pix$^{-1}$, is incurred with every 2~sec image.  This places
the read noise at or below the zodiacal photon-counting noise, which
ranges from 10-16~$e^-$~pix$^{-1}$ RMS for a 2~sec integration time
(see Section~\ref{sec:zodi}).

Due to limitations in data storage and telemetry, it will not be
possible to transmit all the 2~sec images back to Earth.  Instead,
\tess{} will stack these images to create two basic data products with
longer effective exposure times.  First, the subset of pixels that
surround several hundred thousand pre-selected ``target stars'' will
be stacked at a 2~min cadence. Second, the full-frame images
(``FFIs'') will be stacked at a 30~min cadence.  The selection of the
target stars will be based on the detectability of small planets; this
described further in Section~\ref{sec:target_selection}.  The FFIs
will allow a wider range of stars to be searched for transits, and
they will also enable many other scientific investigations that
require time-domain photometry of bright sources.

\begin{figure}[htb]
\epsscale{1.0}
\plotone{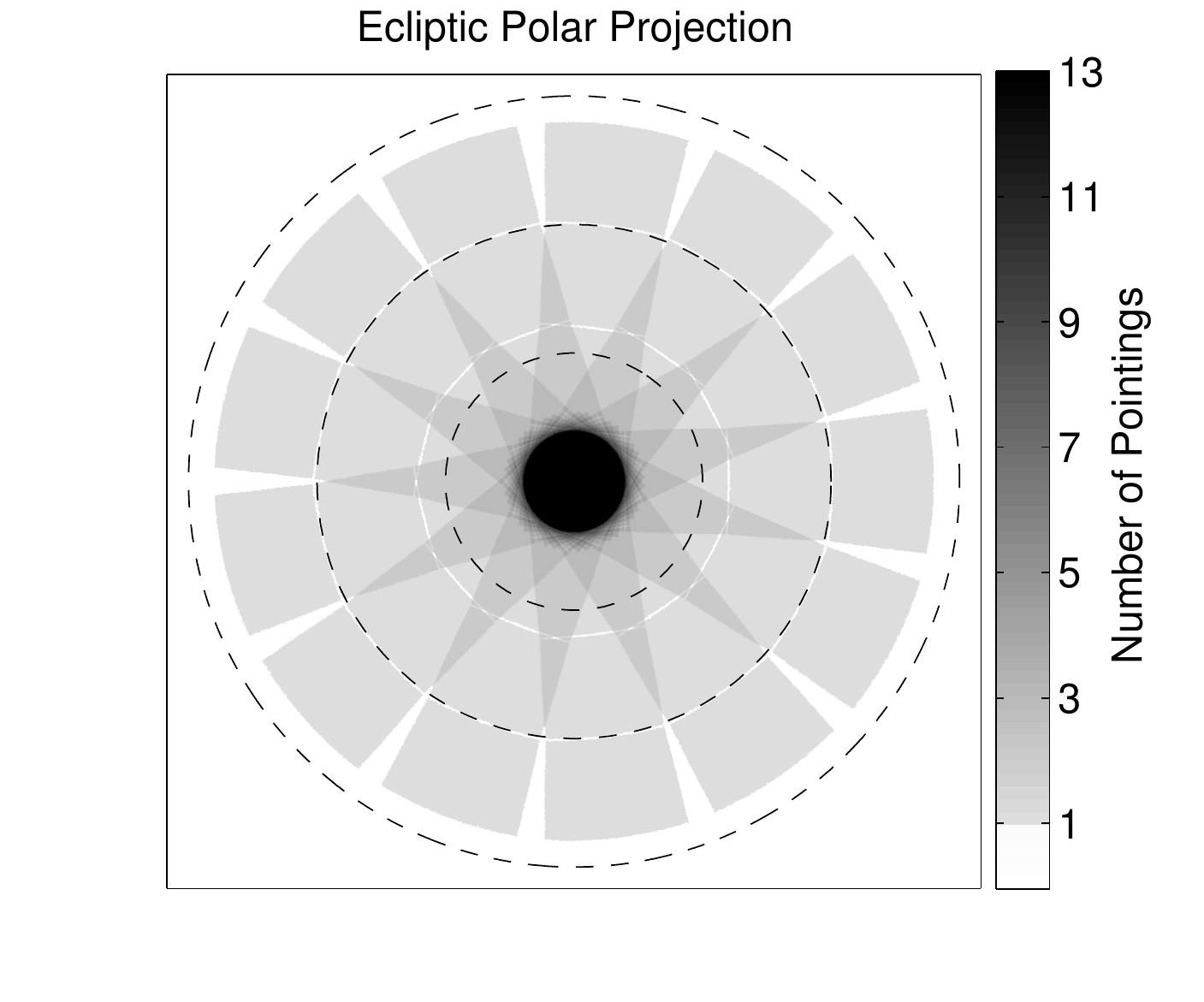}
\caption{Polar projection illustrating how each ecliptic hemisphere is
  divided into 13 pointings. At each pointing, \tess{} observes for a
  duration of \pointdur{} days, or two spacecraft orbits. The four
  \tess{} cameras have a combined field-of-view of
  24$^{\circ}\times$96$^{\circ}$.  The number of pointings that
  encompass a given star is primarily a function of the
  star's ecliptic latitude. The dashed lines show $0^{\circ}$,
  $30^{\circ}$, and $60^{\circ}$ of ecliptic latitude. Coverage near
  the ecliptic ($0^{\circ}$) is sacrificed in favor of coverage near
  the ecliptic poles, which receive nearly continuous coverage for 355
  days.}
\label{fig:npoint}
\end{figure}

\break
\subsection{Sky Coverage}
\tess{} will observe from a 13.7-day elliptical orbit around the
Earth.  Over two years, it will observe the sky using 26
pointings. Two spacecraft orbits (\pointdur{}~days) are devoted to
each pointing.  Because the cameras are fixed to the spacecraft, the
spacecraft must re-orient for every pointing.  The pointings are
spaced equally in ecliptic longitude, and they are positioned such
that the top camera is centered on the ecliptic pole and the bottom
camera reaches down to an ecliptic latitude of 6$^\circ$.  Figure
\ref{fig:npoint} shows the hemispherical coverage resulting from this
arrangement.

\begin{figure}[ht]
\epsscale{1.0}
\plotone{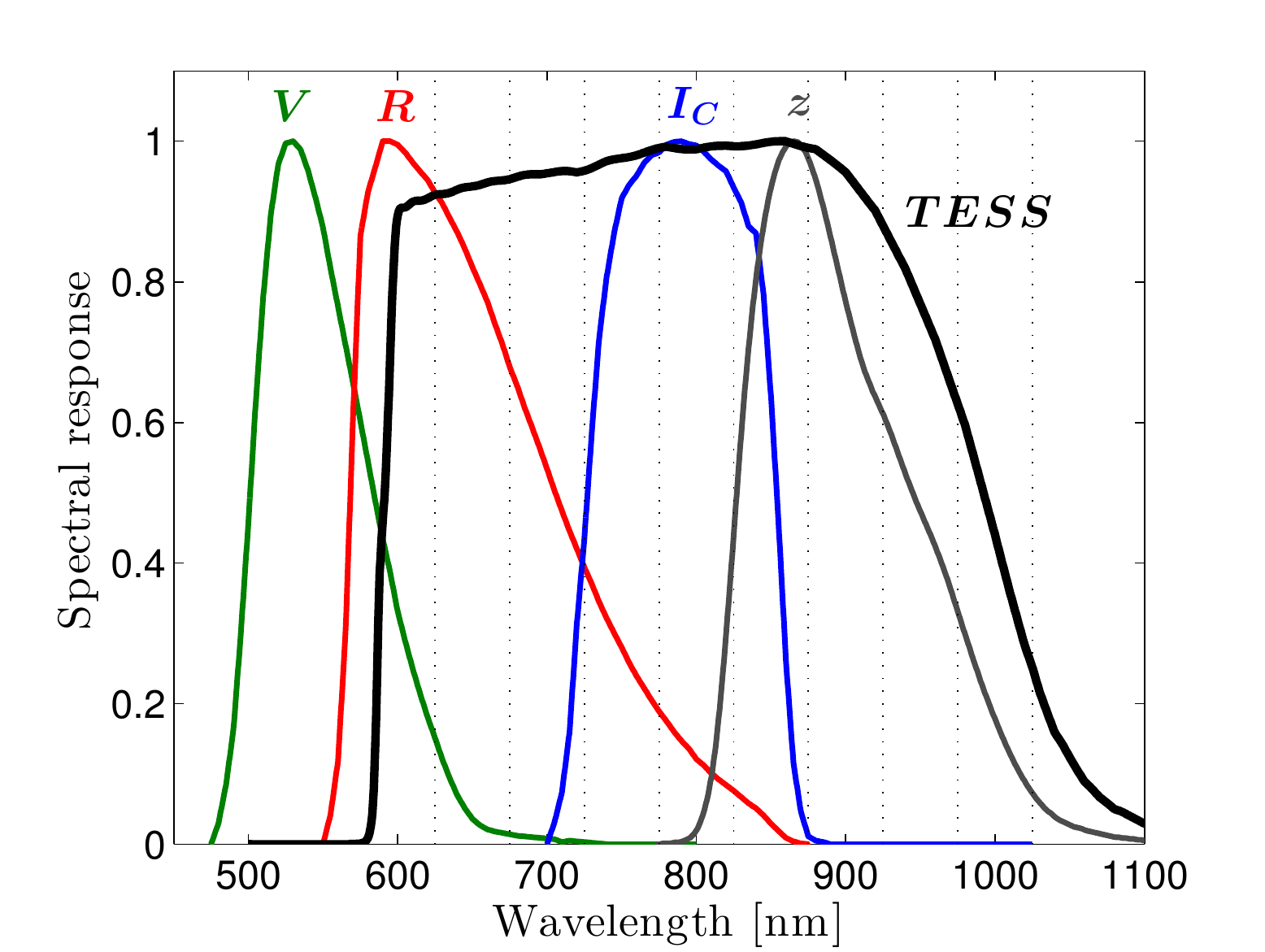}
\caption{The \tess{} spectral response, which is the product of the
  CCD quantum efficiency and the longpass filter curve. Shown for
  comparison are the filter curves for the familiar Johnson-Cousins
  $V$, $R$, and $I_C$ filters as well as the SDSS $z$ filter. Each 
  curve is normalized to have a maximum value of unity. The vertical dotted lines 
  indicate the wavelengths at which the point-spread function 
  is evaluated for our optical model (see Section~\ref{sec:image}).}
\label{fig:filt}
\end{figure}

\subsection{Spectral Response}
\label{sec:bandpass}

The spectral response of the \tess{} cameras is limited at its red end
by the quantum efficiency of the CCDs. \tess{} employs the MIT Lincoln
Laboratory CCID-80 detector, a back-illuminated CCD with a depletion
depth of 100~$\mu$m.  This relatively deep depletion allows for
sensitivity to wavelengths slightly longer than 1000~nm.

At its blue end, the spectral response is limited by a longpass filter
with a cut-on wavelength of 600~nm. Figure \ref{fig:filt} shows the
the complete spectral response, defined as the product of the quantum
efficiency and filter transmission curves.

\begin{deluxetable}{cccc}

\tabletypesize{\scriptsize}
\tablecolumns{4}
\tablewidth{0pt}
\tablecaption{Fluxes in the $\tess{}$ bandpass and $I_C-T$ colors.\label{tbl:phot}}

\tablehead{
\colhead{Spectral Type\tablenotemark{a}} &
\colhead{$\teff$} &
\colhead{$I_C=0$ photon flux\tablenotemark{b}} &
\colhead{$I_C-T$} \\
\colhead{ } &
\colhead{[K]} &
\colhead{[$10^6$~ph~s$^{-1}$~cm$^{-2}$]} &
\colhead{[mmag]}
}
 
\startdata
M9V & 2450 & 2.38 & $ 306$ \\ 
M5V & 3000 & 1.43 & $-191$ \\ 
M4V & 3200 & 1.40 & $-202$ \\ 
M3V & 3400 & 1.38 & $-201$ \\ 
M1V & 3700 & 1.39 & $-174$ \\ 
K5V & 4100 & 1.41 & $-132$ \\ 
K3V & 4500 & 1.43 & $-101$ \\ 
K1V & 5000 & 1.45 & $-80.0$ \\ 
G2V & 5777 & 1.45 & $-69.5$ \\ 
F5V & 6500 & 1.48 & $-40.0$ \\ 
F0V & 7200 & 1.48 & $-34.1$ \\ 
A0V & 9700 & 1.56 & $ 35.0$
\enddata

\tablenotetext{a}{The mapping between $\teff$ and spectral type is
  based on data compiled by E.~Mamajek.}

\tablenotetext{b}{The photon flux at $T=0$ is $1.514 \times
  10^6$~ph~s$^{-1}$~cm$^{-2}$.}

\end{deluxetable}
 
It is convenient to define a \tess{} magnitude $T$ normalized such
that Vega has $T=0$.  We calculate the $T=0$ photon flux by
multiplying the template A0V spectrum provided by \citet{pickles} by
the \tess{} spectral response curve and then integrating over
wavelength. We assume Vega has a flux density of $F_\lambda =
3.44\times 10^{-9}$~erg~s$^{-1}$~cm$^{-2}$~\AA$^{-1}$ at $\lambda =
5556$~\AA{} \citep{vega}. We find that $T=0$ corresponds to a flux of
$4.03 \times 10^{-6}$~erg~s$^{-1}$~cm$^{-2}$, and a photon flux of
$1.514 \times 10^6$~ph~s$^{-1}$~cm$^{-2}$.

By repeating the calculation for different template spectra from the
\citet{pickles} library, we obtain the photon fluxes for stars of
other spectral types.  These are shown in Table~\ref{tbl:phot}. To
facilitate comparisons with the standard Johnson-Cousins $I_C$ band
(which is nearly centered within the $T$-band), Table~\ref{tbl:phot}
also provides synthetic $I_C-T$ colors. We note that the $I_C-T$ color
for an A0V star is $+0.035$, which is equal to the apparent $I_C$
magnitude defined for Vega.

\subsection{Simplified model for the sensitivity of \tess{}}
\label{sec:maglim}

The most important stellar characteristics that affect planet
detectability are apparent magnitude and stellar radius. Here we
provide a simple calculation for the limiting apparent magnitude (as a
function of stellar radius) that permits \tess{} to detect planets
smaller than Neptune ($R_p<4~R_\oplus$). This gives an overview of
{\it TESS}'s planet detection capabilities and establishes the
necessary depth of our more detailed simulations of the population of
nearby stars.

\begin{figure}[htb]
\epsscale{1.0}
\plotone{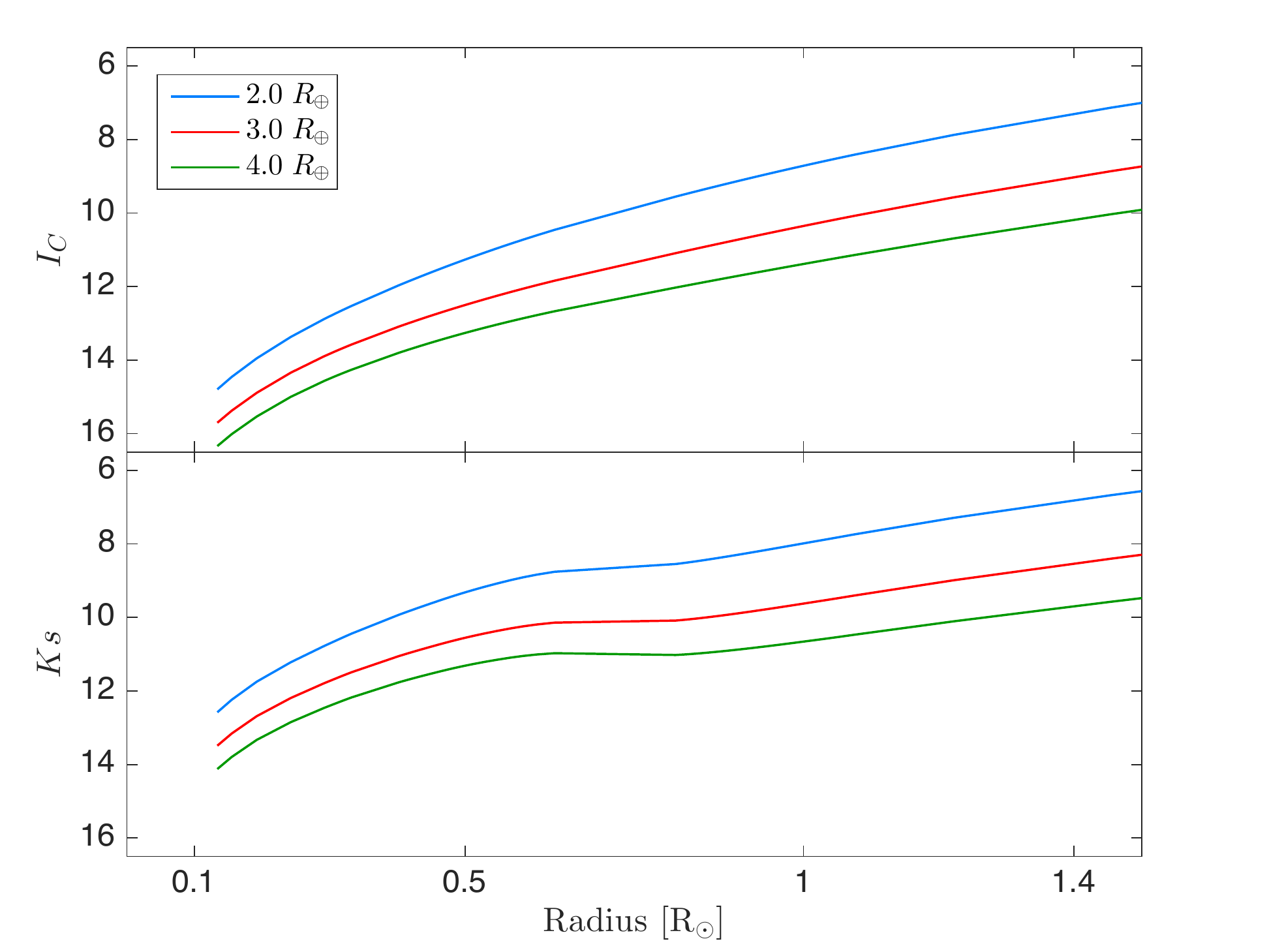}
\caption{The limiting magnitude for planet detection as a function of
  stellar radius for three planetary radii. Here,
  detection is defined as achieving a signal-to-noise ratio greater
  than $\snrcr{}$ from 6~hours of integration time during
  transits. The noise model includes read noise and photon-counting
  noise from the target star and a typical level of zodiacal light.
  While the \tess{} bandpass is similar to the $I_C$ band, the sensitivity
  curve is flatter in $K_s$ magnitudes.}
\label{fig:ikdet}
\end{figure}

We assume the noise in the photometric observations to be the
quadrature sum of read noise and the photon-counting noise from the
target star and the zodiacal background (see Section~\ref{sec:noise}
for the more comprehensive noise model). We require a signal-to-noise
ratio of $\snrcr{}$ for detection (see Section~\ref{sec:threshold} for
the rationale). We assume that the total integration time during
transits is 6~hours, which may represent two or more transits of
shorter duration. Using these assumptions, Figure~\ref{fig:ikdet}
shows the limiting apparent magnitude as a function of stellar radius
at which transiting planets of various sizes can be detected.

To gauge the necessary depth of the detailed simulations, we consider
the detection of small planets around two types of stars represented in 
Figure~\ref{fig:ikdet}, a Sun-like star and an M dwarf with $\teff=$3200 K. 
These two choices span the range of spectral types that \tess{}
will prioritize; stars just larger than the Sun give transit depths
that are too shallow, and dwarf stars just cooler than 3200 K
are too faint in the \tess{} bandpass.

For the Sun-like star, a 4~$R_\oplus$ planet produces a transit depth
of 0.13\%. The limiting magnitude for transits to be detectable is
about $I_C=11.4$.  This also corresponds to $K_s \approx 10.6$ and a
maximum distance of 290~pc, assuming no extinction. 

For the M dwarf with $T_{\rm eff}=$~3200 K, we assume $R_\star =
0.155~R_{\odot}$, based on the Dartmouth Stellar Evolution Database
(Dotter et al. 2008) for solar metallicity and an age of 1~Gyr. Since
Dressing \& Charbonneau (2015) found that M dwarfs very rarely have close-in
planets larger than 3~$R_{\oplus}$, we consider a planet of this size
rather than 4~$R_{\oplus}$. At 3~$R_{\oplus}$, the transit depth is
3.1\% and the limiting apparent magnitude for detection is
$I_C=15.2$. This corresponds to $K_s \approx 13$ and a maximum
distance of 120 pc, assuming no extinction.''

A similar calculation can be carried out for eclipsing binary
stars. Some \tess{} target stars will turn out to be eclipsing
binaries, and others will be blended with faint binaries in the
background.  The maximum eclipse depth for an eclipsing binary is
approximately 50\%, which occurs when two identical stars undergo a
total eclipse.  Assuming the period is 1~day, and that \tess{}
observes the system for \pointdur{}~days, the limiting apparent
magnitude for detection of the eclipse signals is $T<21$,
corresponding to many kiloparsecs.

To summarize, \tess{} is sensitive to small planets around Sun-like
stars within $\lsim$300~pc. For M dwarfs, the search distance is
$\lsim$100~pc.'' Eclipsing binaries can be detected across the Milky
Way. These considerations set the required depth of our simulations of
the stellar population, which must also take into account the
structure of the galaxy and extinction.

\section{STAR CATALOG}
\label{sec:catalog}

Due to the wide range of apparent magnitudes that we need to consider,
and the sensitivity of transit detections to stellar radii, we use a
synthetic stellar population rather than a real catalog. The basis for
our stellar population is TRILEGAL, an abbreviation for the
TRIdimensional modeL of thE GALaxy \citep{trilegal}. TRILEGAL is a
Monte Carlo population synthesis code that models the Milky Way with
four components: a thin disk, a thick disk, a halo, and a bulge. Each
of these components contains stars with the same initial mass function
but with a different spatial distribution, star formation rate, and
age-metallicity relation. For stars with masses 0.2-7~$M_{\odot}$,
TRILEGAL uses the Padova evolutionary tracks \citep{padova} to
determine the stellar radius, surface gravity, and luminosity as a
function of age. For stars less massive than 0.2~$M_{\odot}$, TRILEGAL
uses a brown dwarf model \citep{dusty}. Apparent magnitudes in various
photometric bands are computed using a spectral library drawing upon
several theoretical and empirical sources. A disk extinction model is
used to redden the apparent magnitudes depending on the location of
the star. TRILEGAL does not include the Magellanic Clouds, nor does it
model any star clusters.

The star counts predicted by the TRILEGAL model were originally
calibrated against the Deep Multicolor Survey (DMS) and ESO Imaging
Survey (EIS) of the South Galactic Pole. The model was also found to
be consistent with the EIS coverage of the Chandra Deep Field South
\citep{trilegalold}. More recently, TRILEGAL was updated and
re-calibrated against the shallower 2MASS and Hipparcos catalogs while
maintaining agreement with the DMS and EIS catalogs \citep{trilegal}.

Given a specified line of sight and solid angle, TRILEGAL returns a
magnitude-limited catalog of simulated stars, including properties
such as mass, age, metallicity, surface gravity, distance, and
extinction. Apparent magnitudes are reported in the Sloan $griz$,
2MASS $JHK_s$, and \kepler{} bandpasses; at our request, L. Girardi
kindly added the \tess{} bandpass to TRILEGAL.  When necessary, we
translate between the Sloan and Johnson-Cousins filters using the
transformations for Population I stars provided by \citet{jordi}.

We find it necessary to adjust the properties of the population of
low-mass stars ($M<0.78~M_\odot$) to bring them into satisfactory
agreement with more recent determinations of the absolute radii and
luminosity function of these stars. These modifications are described
in Sections~\ref{sec:dwarfs} and \ref{sec:lf}.  In addition, we employ
our own model for stellar multiplicity that is described in Section
\ref{sec:mult}.

\subsection{Model Queries}

The TRILEGAL simulation is accessed through a web-based
interface.\footnote{{\tt http://stev.oapd.inaf.it/cgi-bin/trilegal}}
We use the default input parameters for the simulation (Table
\ref{tbl:trilegal}); the {\it post facto} adjustments that we make to
dwarf properties, binarity, and the disk luminosity function are
discussed below.  The runtime of a TRILEGAL query is limited to 10
minutes, so we build an all-sky catalog by performing repeated queries
over regions with small solid angles.

\begin{deluxetable}{rl}
\tabletypesize{\scriptsize}
\tablecolumns{2}
\tablewidth{0pt}
\tablecaption{TRILEGAL input settings. \label{tbl:trilegal}}

\tablehead{
\colhead{Parameter} &
\colhead{Value}
}

\startdata

Galactic radius of Sun & 8.70 kpc \\
Galactic height of Sun & 24.2 pc \\

\cutinhead{IMF (log-normal, \citealt{chabrier})}
Characteristic mass & 0.1 $M_{\odot}$ \\
Dispersion & 0.627 $M_{\odot}$ \\

\cutinhead{Thin Disk}
Scale height (sech$^2$) & 94.69 pc \\
Scale radius (exponential) & 2.913 kpc \\
Surface density at Sun & 55.4 $M_{\odot}$ pc$^{-2}$ \\

\cutinhead{Thick Disk}
Scale height (sech$^2$) & 800 pc \\
Scale radius (exponential) & 2.394 kpc \\
Density at Sun & 10$^{-3} M_{\odot}$ pc$^{-3}$ \\

\cutinhead{Halo ($R^{1/4}$ Oblate Spheriod)}
Major axis & 2.699 kpc \\
Oblateness & 0.583 \\
Density at Sun & 10$^{-4} M_{\odot}$ pc$^{-3}$ \\

\cutinhead{Bulge (Triaxial, \citealt{bulge})}
Scale length & 2.5 kpc \\
truncation length &  95 pc \\
Bar: $y/x$ aspect ratio & 0.68 \\
Bar-Sun angle & 15$^{\circ}$ \\
$z/x$ ratio & 0.31 \\
Central Density & 406 $M_{\odot}$ pc$^{-3}$ \\

\cutinhead{Disk Extinction}
Scale height (exponential) & 110 pc \\
Scale radius (exponential) & 100 kpc \\
Extinction at Sun ($dA_V/dR$) & 0.15 mag kpc$^{-1}$ \\
$A_V(z=\infty)$ & 0.0378 mag \\
Randomization (RMS) & 10$\%$
\enddata
\end{deluxetable}

We divide the sky into 3072 equal-area tiles using the HEALPix scheme
\citep{healpix}. Each tile subtends a solid angle of 13.4 deg$^2$.
For the 164 tiles closest to the galactic disk and bulge, the stellar
surface density is too large for the necessary TRILEGAL computations
to complete within the runtime limit.  The high background level and
high incidence of eclipsing binaries will also make these areas
difficult to search for transiting planets, so we simply omit these
tiles from consideration. This leaves 2908 tiles covering 95$\%$ of
the sky.

For each of the 2908 sightlines through the centers of tiles,
we make three queries to TRILEGAL:
\begin{enumerate}

\item The ``bright catalog'' with $K_s<15$ and a solid angle of
  6.7~deg. This is intended to include any star that could be searched
  for transiting planets; the magnitude limit of $K_s<15$ is based on
  the considerations in Section~\ref{sec:maglim}. Using the $K_s$ band
  to set the limiting magnitude is a convenient way to allow the
  catalog to have a fainter $T$~magnitude limit for M stars than for
  FGK stars. The full solid angle of 13.4 deg$^2$ cannot be simulated
  due to the 10-minute maximum runtime of the simulation.  Instead, we
  simulate a 6.7~deg$^2$ field and simply duplicate each star in the
  catalog. Once duplicated, we assign coordinates to each star
  randomly from a probability distribution that is spatially uniform
  across the entire tile. Across all of the tiles, this catalog
  contains \nbrightcat{} stars.

\item The ``intermediate catalog'' with $T<21$ and a solid angle of
  0.134 deg$^2$. This is intended to include stars for which \tess{}
  would be able to detect a deep eclipse of a binary star. We use this
  catalog to assign blended background binaries to the target stars in
  the bright catalog and also to evaluate background fluxes. This
  deeper query is limited to a smaller solid angle (1/100th of the
  area of the tile) to limit computational time. The simulation then
  re-samples from these stars 100 times when assigning background
  stars to the target stars. We also restrict this catalog to $K_s>15$
  in the simulation to avoid double-counting stars from the bright
  catalog. Across all tiles, this catalog contains $1.81\times 10^9$
  stars.

\item The ``faint catalog'' with $21<T<27$ and a solid angle of
  0.0134~deg$^2$. This is used only to calculate background fluxes due
  to unresolved background stars. The limiting magnitude is not
  critical because the surface brightness due to unresolved stars is
  dominated by stars at the brighter end rather than the fainter end
  of the population of unresolved stars. Stars from this catalog are
  re-sampled 1000 times. Across all tiles, this catalog contains
  $6.18\times 10^9$ stars.

\end{enumerate}

\subsection{Properties of low-mass stars}
\label{sec:dwarfs}

Low-mass dwarf stars are of particular importance for \tess{} because
they are abundant in the solar neighborhood and their small sizes
facilitate the detection of small transiting planets. Although the
TRILEGAL model is designed to provide simulated stellar populations
with realistic distributions in spatial coordinates, mass, age, and
metallicity, we noticed that the radii of low-mass stars for a given
luminosity or $\teff$ in the TRILEGAL output were smaller than have
been measured in recent observations or calculated in recent
theoretical models.

\begin{figure}[ht]
\epsscale{1.0}
\plotone{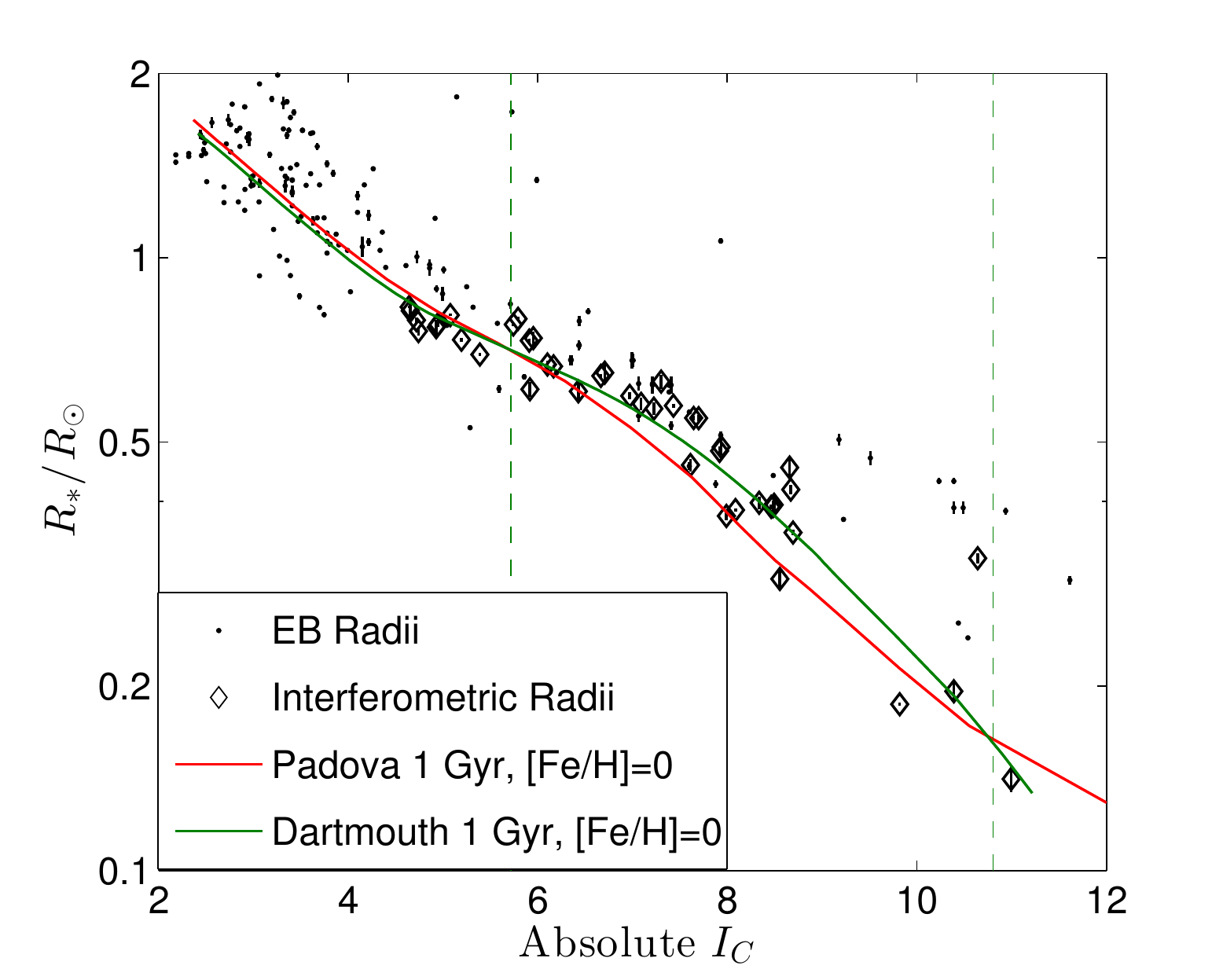}
\caption{The radius-magnitude relation for simulated stars compared to
  empirical observations. The Padova models (red curve) are employed
  by default within the TRILEGAL simulation. These models seem to
  underestimate the radii of low-mass stars; the Dartmouth models
  (green curve) give better agreement.  For stars of mass
  0.14-0.78~$M_{\odot}$ (dashed boundaries) we overwrite the
  TRILEGAL-supplied properties with Dartmouth-based properties for a
  star of the given mass, age, and metallicity. The interferometric
  measurements plotted here are from \citet{chara}, and the
  eclipsing-binary measurements come from a variety of sources (see text). The
  scatter in radius for $I_C\lesssim5$ arises from stellar evolution.}
\label{fig:rmi}
\end{figure}

Figure~\ref{fig:rmi} illustrates the discrepancy. It compares the
radius-magnitude relation employed by TRILEGAL with that of the more
recent Dartmouth models \citep{dartmouth} as well as empirical data
based on optical interferometry of field stars and analysis of
eclipsing binary stars. The interferometric radius measurements are
from \citet{chara}. The measurements based on eclipsing binaries are
from the compilation of \citet{ebs} that has since been maintained by
J.~Southworth\footnote{{\scriptsize \tt
    http://www.astro.keele.ac.uk/jkt/debcat/}\normalsize}.  We also
include the systems tabulated by \citet{winn16} in their study of
Kepler-16.  The published data specify $\teff$ rather than absolute
$I_C$ magnitude; in preparing Figure~\ref{fig:rmi}, we converted
$\teff$ into absolute $I_C$ using the temperature-magnitude data
compiled by E.~Mamajek\footnote{\label{mamajek}{\tiny \tt
    http://www.pas.rochester.edu/$\sim$emamajek/EEM\_dwarf\_UBVIJHK\_colors\_Teff.txt}
  \normalsize} and \citet{pecaut}.

Figure \ref{fig:rmi} shows that the Dartmouth stellar-evolutionary
models give better agreement with measured radii, especially those
from interferometry.  Therefore, to bring the key properties of the
simulated stars into better agreement with the data, we replaced the
TRILEGAL output for the apparent magnitudes and radii of low-mass
stars (0.15-0.78~$M_{\odot}$) with the properties calculated with the
Dartmouth models.  To make these replacements, we use a trilateral
interpolation in mass, age, and metallicity to determine the absolute
magnitudes, $\teff$, and radii from the grid of Dartmouth models. For
simplicity, we assume the helium abundance is solar for all stars.
Furthermore, motivated by \citet{fuhrmann}, we only select the grid
points that adhere to the following one-to-one relation between
[$\alpha$/Fe] and [Fe/H]:
\begin{equation}
\mathrm{[Fe/H]} \geq 0 \iff [\alpha/\mathrm{Fe}] = 0.0
\end{equation}
\begin{equation}
\mathrm{[Fe/H]} = -0.05 \iff [\alpha/\mathrm{Fe}] = +0.2
\end{equation}
\begin{equation}
\mathrm{[Fe/H]} \leq -0.1 \iff [\alpha/\mathrm{Fe}] =+0.4
\end{equation}

In calculating the apparent magnitudes of the stars with properties
overwritten from the Dartmouth models, we preserve the distance
modulus from TRILEGAL and apply reddening corrections using the same
extinction model that TRILEGAL uses. TRILEGAL reports the extinction
$A_V$ for each star, and for bands other than $V$, we use the
$A_{\lambda}/A_V$ ratios from \citet{cardelli}.

\subsection{Stellar Multiplicity}
\label{sec:mult}

Binary companions to the \tess{} target stars have three important
impacts on the detection of transiting planets. First, whenever a
``target star'' is really a binary, there are potentially two stars
that can be searched for transiting planets. The effective size of the
search sample is thereby increased. However, there is a second effect
that decreases the effective size of the search sample: if there is a
transit around one star, the constant light from the unresolved
companion diminishes the observed transit depth, making it more
difficult to detect the transit. Even if the transit is still
detectable, the radius of the planet may be underestimated due to the
diminished (or ``diluted'') depth.  The third effect is that a planet
around one member of a close binary has a limited range of periods
within which its orbit would be dynamically stable.

Furthermore, eclipsing binaries that are blended with target stars, or
that are bound to the target star in hierarchical triple or quadruple
systems, can produce eclipses that resemble planetary
transits. Because eclipsing binaries produce larger signals than
planetary transits, the population of eclipsing binaries needs to be
simulated down to fainter apparent magnitudes than the target stars.

To capture these effects in our simulations, we need a realistic
description of stellar multiplicity. We are guided by the review of
\citet{duchene}. The multiplicity fraction (MF) is defined as the
fraction of systems that have more than one star; it is the sum of the
binary fraction (BF), triple fraction (TF), quadruple fraction (QF),
and so on. Our simulations consider systems with up to 4 stars.

The MF has been observed to increase with the mass of the primary,
which is reflected in our simulation. In our TRILEGAL queries, {\it
  every} star is originally a binary, and we decide randomly whether
to keep the secondary based on the primary mass and the MF values in
Table~\ref{tbl:binary}.  Next, we turn a fraction of the remaining
binaries into triple and quadruple systems according to the desired TF
and QF. The MF, TF, and QF are adopted as follows:
\begin{enumerate}

\item For primary stars of mass 0.1-0.6~$M_{\odot}$, we adopt the MF of 26\%
  from \cite{delfosse2004}. For systems with $n=3$ or 4 components,
  the fraction of higher-order systems is taken to be $3.9^{2-n}$
  from \citet{duchene}.

\item For stars of mass 0.8-1.4~$M_{\odot}$, we draw on the results of
  \citet{raghavan2010}. Primary masses of 0.8-1.0~$M_{\odot}$ have a MF of 41$\%$, 
  while primary masses of 1.0-1.4~$M_{\odot}$ have a MF of 50$\%$. The fraction of
  higher-order systems is $3.8^{2-n}$ for both ranges \citep{duchene}.

\item For stars of mass 0.6-0.8~$M_\odot$, we adopt an intermediate MF
  of 34\%. The fraction of higher-order systems is $3.7^{2-n}$.
 
\item For primaries more massive than 1.4~$M_{\odot}$, we use the
  results for A stars from \citet{kouwenhoven}, giving a MF of 75\%.
  We assume that the fraction of higher-order systems is $3.7^{2-n}$.

\end{enumerate}

Next, we consider the properties of the binary systems.  TRILEGAL
originally creates binaries with a uniform distribution in the mass
ratio between the secondary and the primary, $q$, between 0.1 and 1.
However, a more realistic distribution in $q$ is
\begin{equation}
\frac{dN}{dq} \propto q^{\gamma},
\end{equation}
where the power-law index $\gamma$ is allowed to vary with the primary
mass, as specified in Table~\ref{tbl:binary}.  When we select the
binary systems to obtain the desired MF, we choose the systems to
re-create this distribution in $q$ over the range 0.1$<q<$1.0.
 
The period $P$ is not specified by TRILEGAL, so we assign it from a
log-normal distribution. \citet{duchene} parametrizes the distribution
in terms of the mean semimajor axis ($\bar{a}$) and the standard
deviation in $\log{P}$; both parameters vary with the primary mass as
shown in Table~\ref{tbl:binary}. We convert from $\bar{a}$ to
$\bar{P}$ with Kepler's third law.

The orbital inclination $i$ is drawn
randomly from a uniform distribution in $\cos i$. The orbital
eccentricity $e$ is drawn randomly from a uniform distribution,
between zero and a maximum value
\begin{equation}
e_{\rm max} = \frac{1}{\pi}\tan^{-1}\left(2\left[\log{P}-1.5\right]\right)+\frac{1}{2},
\end{equation}
where $P$ is specified in days, to provide a good fit to the range 
of eccentricities shown in Figure~14 of \citet{raghavan2010}.
The argument of pericenter $\omega$ is drawn randomly from
a uniform distribution between 0$^\circ$ and 360$^\circ$.

For the systems that are designated as triples, we assign the
properties using the approach originally suggested by
\citet{eggleton}. Although there is no physical reason why this method
should work well, it has been found to reproduce the multiplicity
properties of a sample of {\it Hipparcos} stars \citep{et2008}. First,
we create a binary according to the prescriptions described above with
a period $P_0$.  Then, we split the primary or secondary star (chosen
randomly) into a new pair of stars.  The new pair of stars orbit their
barycenter with a higher-order period $P_{\rm HOP}$ according to
\begin{equation}
\frac{P_{\rm HOP}}{P_0} = 0.2 \times 10^{-2u},
\end{equation}
where $u$ is uniformly distributed between 0 and 1.  This procedure
ensures that $P_{\rm HOP}$ is $<1/5$ the orbital period of the
original binary system, a rudimentary method for enforcing dynamical
stability.  The mass of a star is conserved when it is split, so the
barycenter of the original binary remains the same, and the orbital
period of the companion star about this barycenter is unchanged.

The original prescription given by \citet{eggleton} assigns $P_0$ from
a distribution peaking at $10^5$ days and allows the new period to
vary over 5 decades. Since our assumed distribution for $\log(P_0)$
peaks at a shorter period (for stars $\lesssim$1~$M_{\odot}$), we only
allow the higher-order orbital period to vary over 2 decades in our
implementation. In this way, we avoid generating unphysically short
periods.

The total mass of a new pair of stars is set equal to that of the
original star, and the mass ratio $q$ is assigned in the following
manner.  The parent distribution of $q$ is taken from the sample of
triples presented in Figure~16 of \citet{raghavan2010}.  We model this
distribution by setting $q=1.0$ for 23$\%$ of the pairs and drawing
$q$ from a normal distribution with $(\mu, \sigma^2)=(0.5, 0.04)$ for
the other 77\% of the pairs. Finally, for each star in a higher-order
pair, we calculate the absolute and apparent magnitudes, radius, and
$\teff$ from the new stellar mass in combination with the age and
metallicity inherited from the original star. We do so using the same
interpolation onto the Dartmouth model grid described in Section
\ref{sec:dwarfs}.

For the systems that are turned into quadruples, we create a binary
and then split {\it both} stars using the procedure described above.
This results in two higher-order pairs that orbit one another with the
original binary period $P_0$.

\begin{deluxetable}{ccccccc}
\tabletypesize{\scriptsize}
\tablecolumns{6}
\tablewidth{0pt}
\tablecaption{Binary properties as function of the mass of the primary. \label{tbl:binary}}

\tablehead{
\colhead{Mass [$M_\odot$]} &
\colhead{MF} &
\colhead{$\bar{a}$~[AU]} &
\colhead{$\sigma(\log P)$} & 
\colhead{$\gamma$} & 
\colhead{TF} & 
\colhead{QF}
}

\startdata
<0.1    & 0.22 & 4.5 & 0.5 & 4.0  & n/a & n/a \\
0.1-0.6 & 0.26 & 5.3 & 1.3 & 0.4  & 0.067 & 0.017 \\
0.6-0.8 & 0.34 & 20  & 2.0 & 0.35 & 0.089 & 0.023 \\
0.8-1.0 & 0.41 & 45  & 2.3 & 0.3  & 0.11  & 0.030 \\
1.0-1.4 & 0.50 & 45  & 2.3 & 0.3  & 0.14  & 0.037 \\
>1.4    & 0.75 & 350 & 3.0 & $-0.5$ & 0.20  & 0.055
\enddata
\end{deluxetable}

\subsection{Luminosity Function}
\label{sec:lf}

\begin{deluxetable}{lccc}

\tabletypesize{\scriptsize}
\tablecolumns{4}
\tablewidth{0pt}
\tablecaption{$J$-band luminosity function in $10^{-3}$~stars~pc$^{-3}$. \label{tbl:jlf}}

\tablehead{
\colhead{$M_J$} &
\colhead{Primaries and Singles} &
\colhead{Systems} &
\colhead{Individual Stars}
}

\startdata
3.25 & 0.85 & 0.94 & 1.08\\ 
3.75 & 1.44 & 1.74 & 1.72\\ 
4.25 & 2.74 & 2.87 & 3.10\\ 
4.75 & 3.85 & 3.38 & 4.55\\ 
5.25 & 1.55 & 1.54 & 2.19\\ 
5.75 & 1.79 & 1.91 & 2.27\\ 
6.25 & 3.01 & 3.12 & 3.57\\ 
6.75 & 3.37 & 4.04 & 4.15\\ 
7.25 & 7.74 & 7.90 & 8.82\\ 
7.75 & 7.15 & 7.10 & 8.57\\ 
8.25 & 7.62 & 7.03 & 9.29\\ 
8.75 & 4.84 & 4.89 & 6.64\\ 
9.25 & 5.25 & 4.75 & 6.50\\ 
9.75 & 3.56 & 3.49 & 4.72\\ 
10.25 & 1.95 & 2.11 & 2.68\\ 
10.75 & 2.16 & 2.10 & 2.67\\ 
11.25 & 1.75 & 1.56 & 2.21\\ 
11.75 & 1.11 & 1.07 & 1.52\\ 
12.25 & 0.73 & 0.76 & 1.08\\ 
12.75 & 0.55 & 0.52 & 0.84\\ 
13.25 & 0.45 & 0.36 & 0.69\\ 
13.75 & 0.02 & 0.02 & 0.06\\ 
14.25 & 0.00 & 0.00 & 0.02\\ 
14.75 & 0.00 & 0.00 & 0.00\\ 
15.25 & 0.00 & 0.00 & 0.02
\enddata

\end{deluxetable}

After modifying the TRILEGAL simulation to improve upon the properties
of low-mass stars and assign multiple-star systems, we ensure that the
luminosity function (LF) is in agreement with observations. For this
purpose, we rely on two independent $J$-band LFs reported in the
literature. The first LF is from \citet{cruz}. It is based on
volume-limited samples: a 20~pc sample for $M_J > 11$ and an 8~pc
sample for $M_J < 11$ \citep{reid2003}. Both samples use 2MASS
photometry and are limited to $J \lesssim 16$.  The second LF, from
\citet{bochanski}, is based on data from the Sloan Digital Sky Survey
for stars with $16 < r < 22$. The resulting LF is reported for the
range $5 < M_J < 10$. Where the \citet{cruz} and \citet{bochanski} LFs
overlap, we use the mean of the two LFs reported for single and
primary stars (the brightest member of a multiple system). This
results in the ``empirical LF'' to which the TRILEGAL LF is adjusted.

Next, we compute the LF of our TRILEGAL-based catalog by selecting all
of the single and primary disk stars with distances within
30~pc. Then, we bin the stars according to $M_J$ and compare the
result to the empirical LF. For each $M_J$ bin, we find the ratio of
the TRILEGAL LF to the empirical LF. This ratio ranges from 0.5 to 11
across all of the magnitude bins.

We then return to each HEALPix tile individually, and we bin the stars
by $M_J$.  Using the ratio computed above for each $M_J$ bin, we
select stars at random for duplication or deletion to bring the
simulated LF into agreement with the empirical LF. This process
results in a net reduction of $\approx$30\% in the total number of
stars in the catalog and a shift in the LF peak towards brighter
absolute magnitudes.
 
\begin{figure*}[htb]
\epsscale{1.0}
\plotone{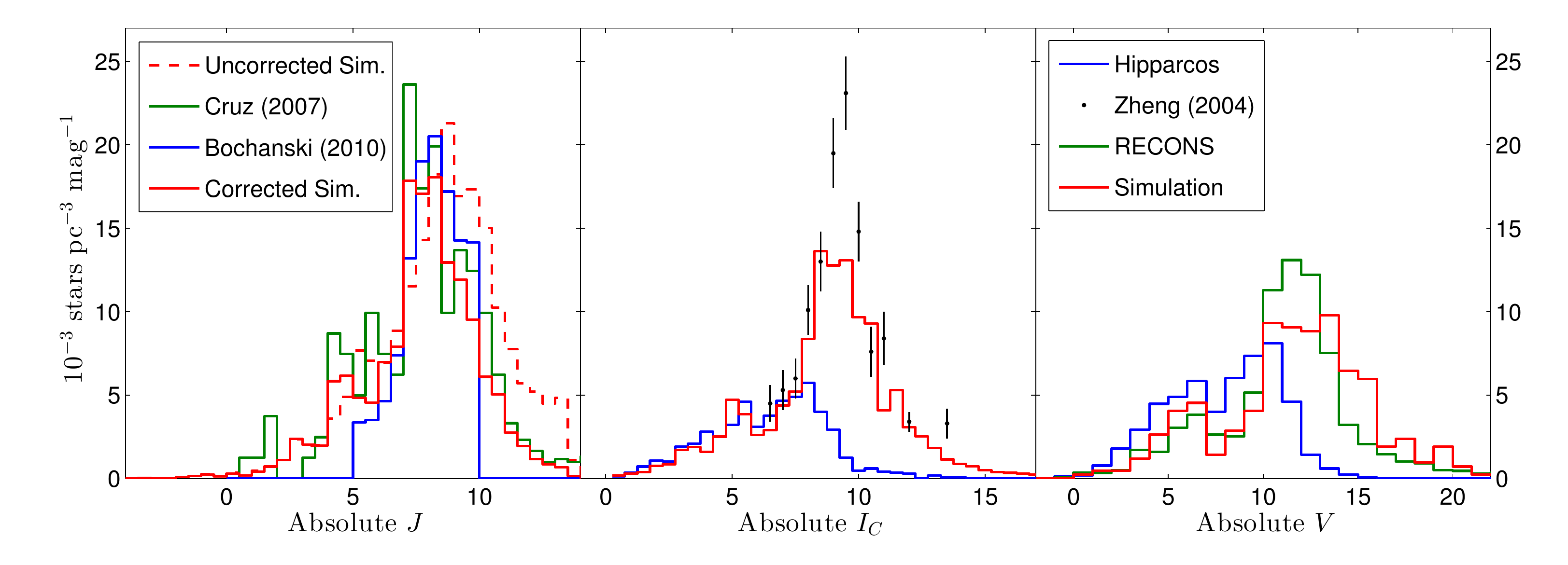}
\caption{The luminosity function of the simulated stellar population
  compared with various published determinations.
  \textit{Left.}---Comparison with the $J$-band LFs of \citet{cruz}
  and \citet{bochanski} before and after we correct the LF of the
  simulation. The stellar multiplicity and dwarf properties have
  already been adjusted in the ``Uncorrected'' LF.
  \textit{Center.}---Comparison with the $I_C$-band LF of
  \citet{zheng2004} and the Hipparcos sample (\citealt{hip1} and
  \citealt{hip2}).  \textit{Right.}---Comparison with Hipparcos and
  the 10~pc RECONS sample \citep{recons}.  For the $J$- and $V$-band
  LFs, we count the single, primary, and secondary stars separately,
  since binaries are generally resolved in the surveys with which we
  are comparing.  For the $I_C$ band, we count the system magnitude of
  binary systems since we assume they are unresolved in the
  \citet{zheng2004} survey. The range of absolute magnitudes from the
  Hipparcos catalog are dominated by single and primary stars, so this
  distinction is less important.}
\label{fig:lf}
\end{figure*}

The left panel of Figure \ref{fig:lf} shows the LF of the TRILEGAL
simulation before and after this adjustment. The final LF is also
quantified in Table \ref{tbl:jlf}.  Each column of the table considers
stellar multiplicity in a different fashion: ``Singles and Primaries''
counts single stars and the brightest member of a multiple system;
``Systems'' counts the combined flux of all stars in a system,
regardless of whether it is single or multiple; and ``Individual
Stars'' counts the primary and secondary members separately.

As a sanity check, we make some further comparisons between our
simulated LF and other published luminosity functions.  Figure
\ref{fig:lf} shows a comparison to the 10~pc RECONS sample
\citep{recons}, the Hipparcos catalog (\citealt{hip1} and
\citealt{hip2}), and the $I_C$-band LF of \citet{zheng2004}. The
agreement with the Hipparcos sample is good up until $V \approx 8$,
where the Hipparcos sample becomes incomplete. The RECONS LF has a
lower and blunter peak, and the \citet{zheng2004} LF has a sharper and
taller peak than the simulated LF, but are otherwise in reasonable
agreement.

\begin{figure*}[htb]
\epsscale{1.0}
\plotone{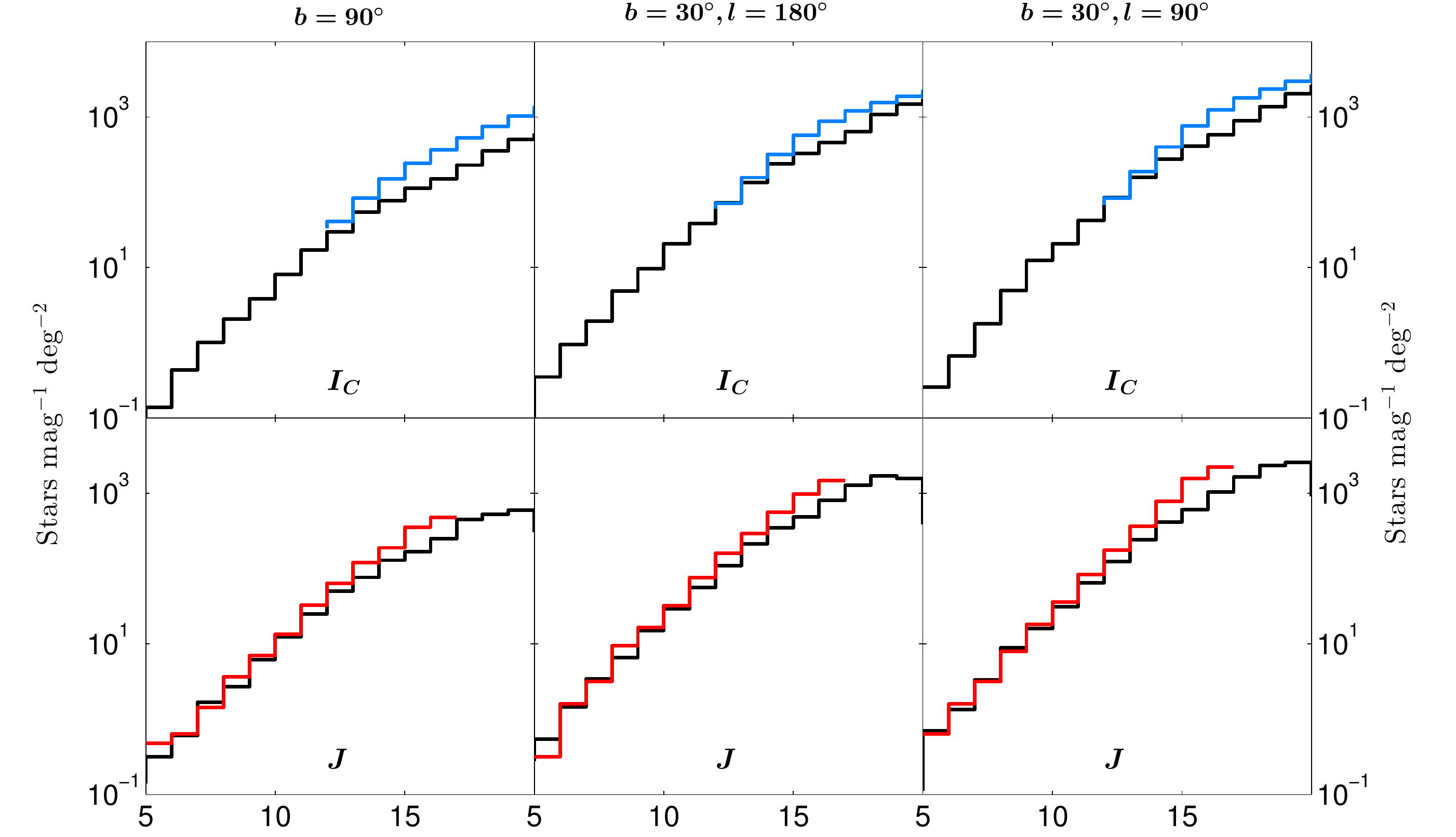}
\caption{Star counts as function of apparent magnitude and galactic
  coordinates. In the $I_C$ band ({\it top row}), we compare the star
  counts in our simulated catalog (black) to those from
  \citet{bahcall1981} (blue). In the $J$ band ({\it bottom row}), we
  compare our catalog (black) to the 2MASS point source catalog
  (red).}
\label{fig:dndij}
\end{figure*}

As another sanity check, we examine star counts as a function of
limiting apparent magnitude in Figure \ref{fig:dndij}.  We compare the
number of stars per unit magnitude per square degree in the simulated
stellar population against star counts from the classic
\citet{bahcall1981} star-count model in the $I_C$ band as well as
actual star counts from the 2MASS point source catalog \citep{2mass}
in the $J$ band.  In all cases, multiple systems are counted as a
single ``star'' with a magnitude equal to the total system
magnitude. The agreement seems satisfactory; we note that the
comparison with 2MASS becomes less reliable at faint magnitudes
because of photometric uncertainties as well as extra-galactic objects
in the 2MASS catalog.

\subsection{Stellar Variability}
\label{sec:starvar}

Intrinsic stellar variability is a potentially significant source of
photometric noise for the brightest stars that \tess{} observes.  To
each star in the simulation, we assign a level of intrinsic
photometric variability from a distribution corresponding to the
spectral type.  Our assignments are based on the variability of
\textit{Kepler} stars reported by \citet{basri}. For each star, they
calculated the median differential variability (MDV) on a 3-hour
timescale by binning the light curve into 3-hour segments and then
calculating the median of the absolute differences between adjacent
bins. Since each transit is a flux decrement between one segment of a
light curve relative to a much longer timeseries, rather than two
adjacent segments of equal length, the noise statistic relevant to
transit detection is approximately $\sqrt{2}$ smaller than the MDV.

G.\ Basri kindly provided the data from their Figures 7-10.  Their
sample is divided into four subsamples according to stellar
$\teff$. We select 100 stars in each subsample with $m_{\rm Kep}<11.5$
to minimize the contributions of instrumental noise from \kepler{}.
Since red giants exhibiting pulsations can contaminate the subsample
with $\teff<4500$ K, particularly at brighter apparent magnitudes, we
select stars with $12.5<m_{\rm Kep}<13.1$ for these temperatures.

\begin{figure}[ht]
\epsscale{1.0}
\plotone{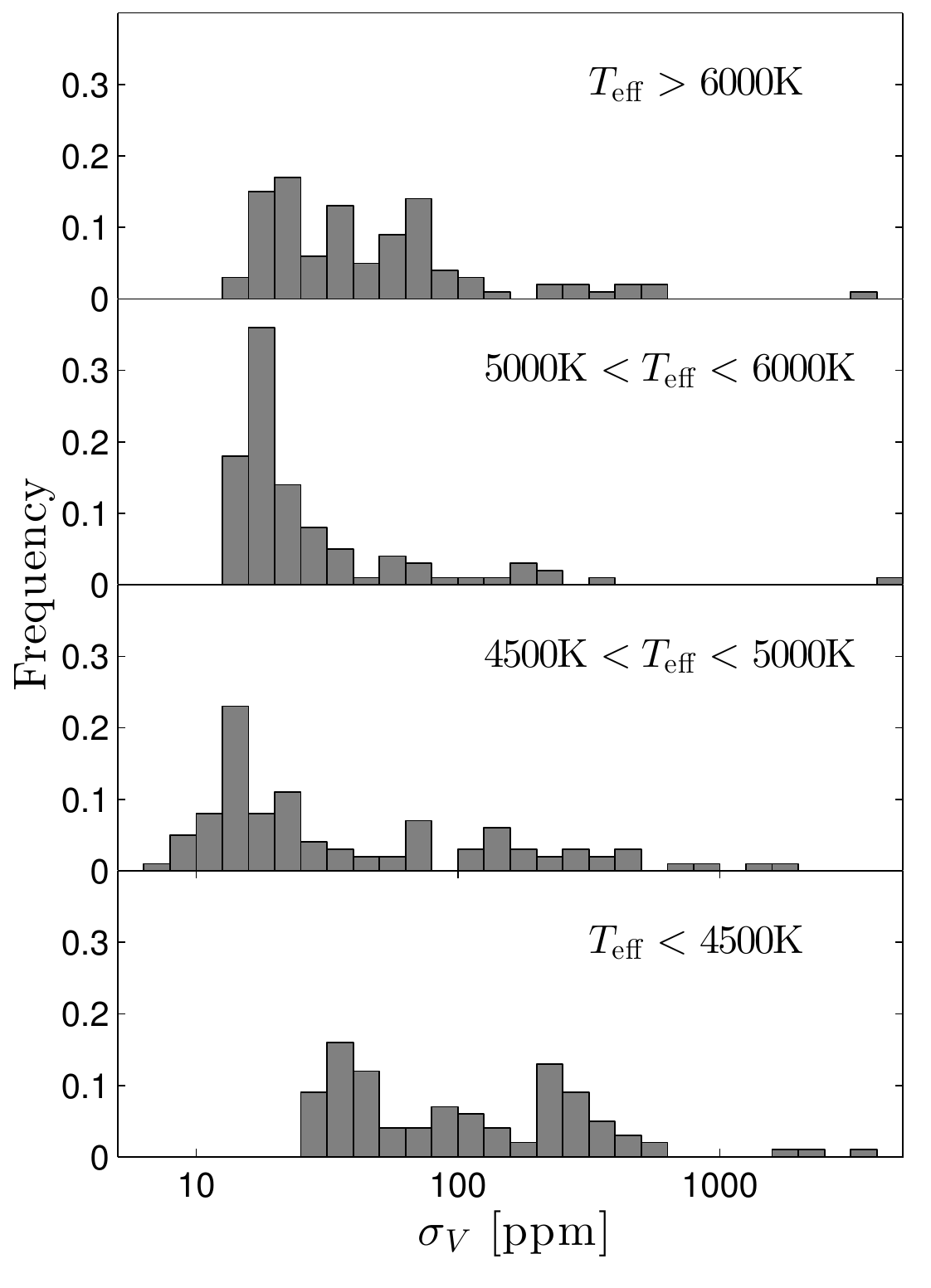}
\caption{The input distributions of the intrinsic stellar variability
  $\sigma_V$ per transit in parts per million (ppm). Each star in our
  catalog is assigned a variability statistic from these distributions
  according to its effective temperature. We calculate $\sigma_V$ from
  the 3-hour MDV statistic of \citet{basri} using Equation
  \ref{eq:mdv}.}
\label{fig:var}
\end{figure}

Figure \ref{fig:var} shows the resulting distributions of
variability. Each star in our simulated population is assigned a
variability index from a randomly-chosen member of the 100 stars in
the appropriate $\teff$ subsample. The variability of the
$\teff<4500$~K subsample is roughly 5 times greater than that of
solar-type stars. However, M dwarfs are the faintest stars that
\tess{} will observe, so instrumental noise and background will
dominate the photometric error of these targets.

Since the photometric variations associated with stellar variability
exhibit strong correlations on short timescales, we assume that the
level of noise due to intrinsic variability is independent of transit
duration: we do not adjust it according to $t^{-1/2}$ as would be the
case for white noise. However, we do assume that stellar variations
are independent from one transit to the next, so the noise
contribution from stellar variability scales with the number of
transits as $N^{-1/2}$.  In summary, the standard deviation in the
relative flux due to stellar variability, after phase-folding all of
the transits together, is taken to be
\begin{equation}
\label{eq:mdv}
\sigma_V = \frac{\rm MDV(3\;hr)}{\sqrt{2}}N^{-1/2}.
\end{equation}

\section{ECLIPSING SYSTEMS}
\label{sec:eclip}

We next assign planets to the simulated stars, and we identify the
transiting planets as well as the eclipsing binaries. We then
calculate the properties of the transits and eclipses relevant to
their detection and follow-up.

\begin{figure*}[ht]
\epsscale{1.1}
\plotone{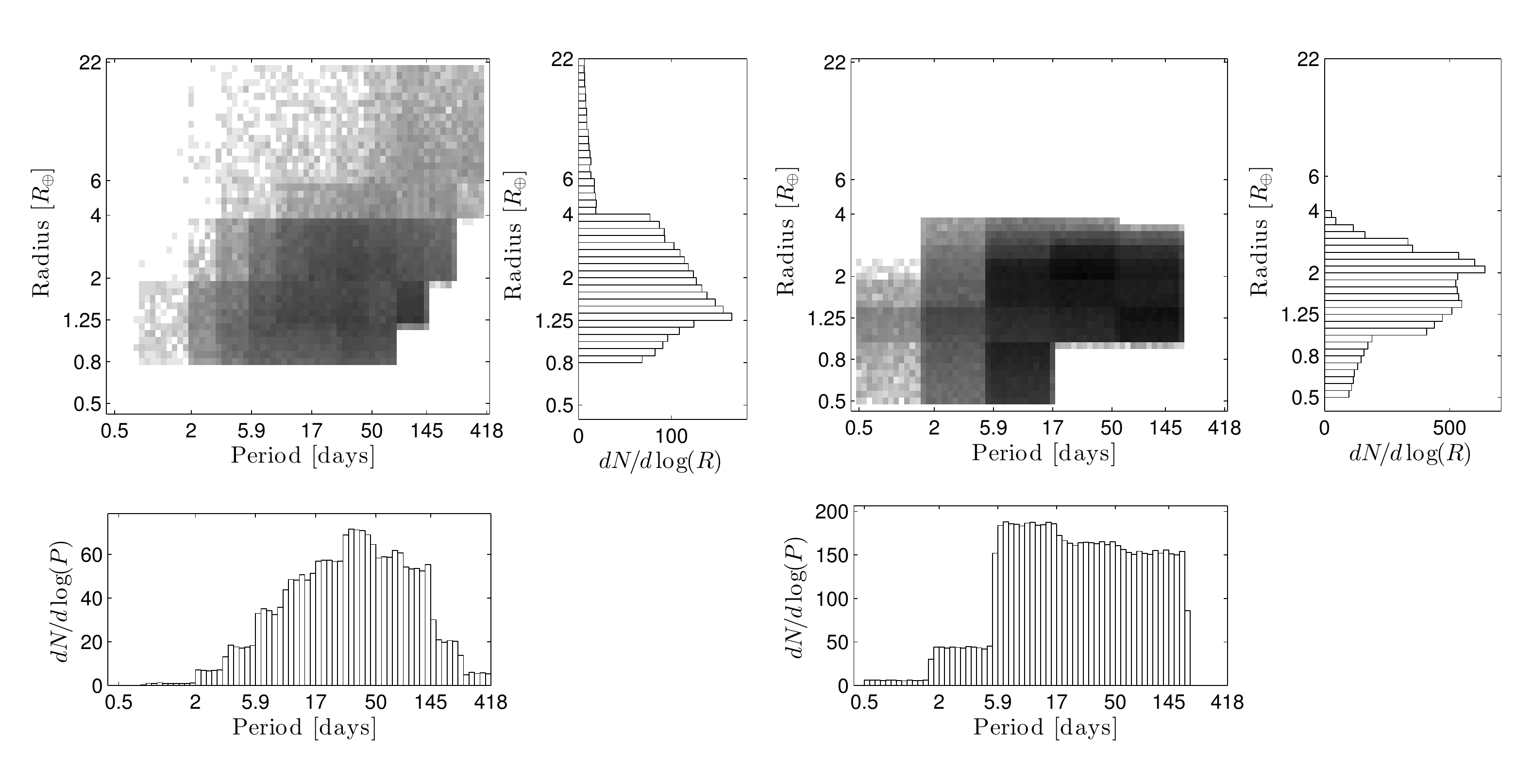}
\caption{The input distributions of planet occurrence in the
  period--radius plane. \textit{Left.}---For stars with $\teff >
  4000$~K, we use the planet occurrence rates reported by
  \citet{fressin2013}.  \textit{Right.}---For stars with $\teff <
  4000$~K, we use the planet occurrence rates reported by
  \citet{dressing2015}.}
\label{fig:occur}
\end{figure*}

\subsection{Planets}
\label{sec:planetin}

The planet assignments are based on several recent studies of
\kepler{} data. The \kepler{} sample has high completeness for the
planetary periods ($P\lesssim 20$ days) and radii ($R_p \gtrsim R_\oplus$) 
that are most relevant to \tess{}.

For FGK stars, we adopt the planet occurrence rates
from \citet{fressin2013}. For $\teff < 4000$~K, we adopt the
occurrence rates from \citet{dressing2015}, who updated the results
that were originally presented by \citet{dressing2013}. We note that
\citet{dressing2015} corrected their planet occurrence rates for 
astrophysical false positives by using the false-positive rates 
presented by \citet{fressin2013} as a function of the apparent planet size.

In both cases, the published results are provided as a matrix of
occurrence rates and uncertainties for bins of planetary radius and
period.  The incompleteness of the \kepler{} sample is considered for
each bin.  Because the bins are relatively coarse, we allow the radius
and period of a given planet to vary randomly within the limits of
each bin. Periods are assigned from a uniform distribution in $\log
P$. (We omit planets for which the selected period would place the
orbital distance within $2~R_\star$, on the grounds that tidal forces
would destroy any such planets.)

For the smallest radius bin examined
by \citet{fressin2013}, we choose the planet radius from a uniform
distribution between 0.8--1.25~$R_\oplus$.
For the larger-radius bins, we choose the planet radius within each bin according
to the distribution
\begin{equation}
\label{eq:power-law-radius}
\frac{dN}{dR_p} \propto R_p^{-1.7}.
\end{equation}
These intra-bin distributions were chosen {\it ad hoc} to provide a
relatively smooth function in the radius--period plane.  Likewise,
when applying the occurrence rates from \citet{dressing2015}, for the
smallest radius bin we choose the planet radius from a uniform
distribution between 0.5--1.0~$R_{\oplus}$. For the bin extending from
1.0--1.5)~$R_{\oplus}$, we chose the planet radius from a distribution
with a power-law index of $-1$.  For $R_p>1.5R_{\oplus}$ we use a
power-law index of $-1.7$. The maximum planet size in the
\citet{fressin2013} matrix is 22~$R_{\oplus}$, and the maximum planet
size in the \citet{dressing2015} matrix is 4~$R_{\oplus}$. The final
distributions are illustrated in Figure \ref{fig:occur}.

\begin{figure}[htb]
\epsscale{1.1}
\plotone{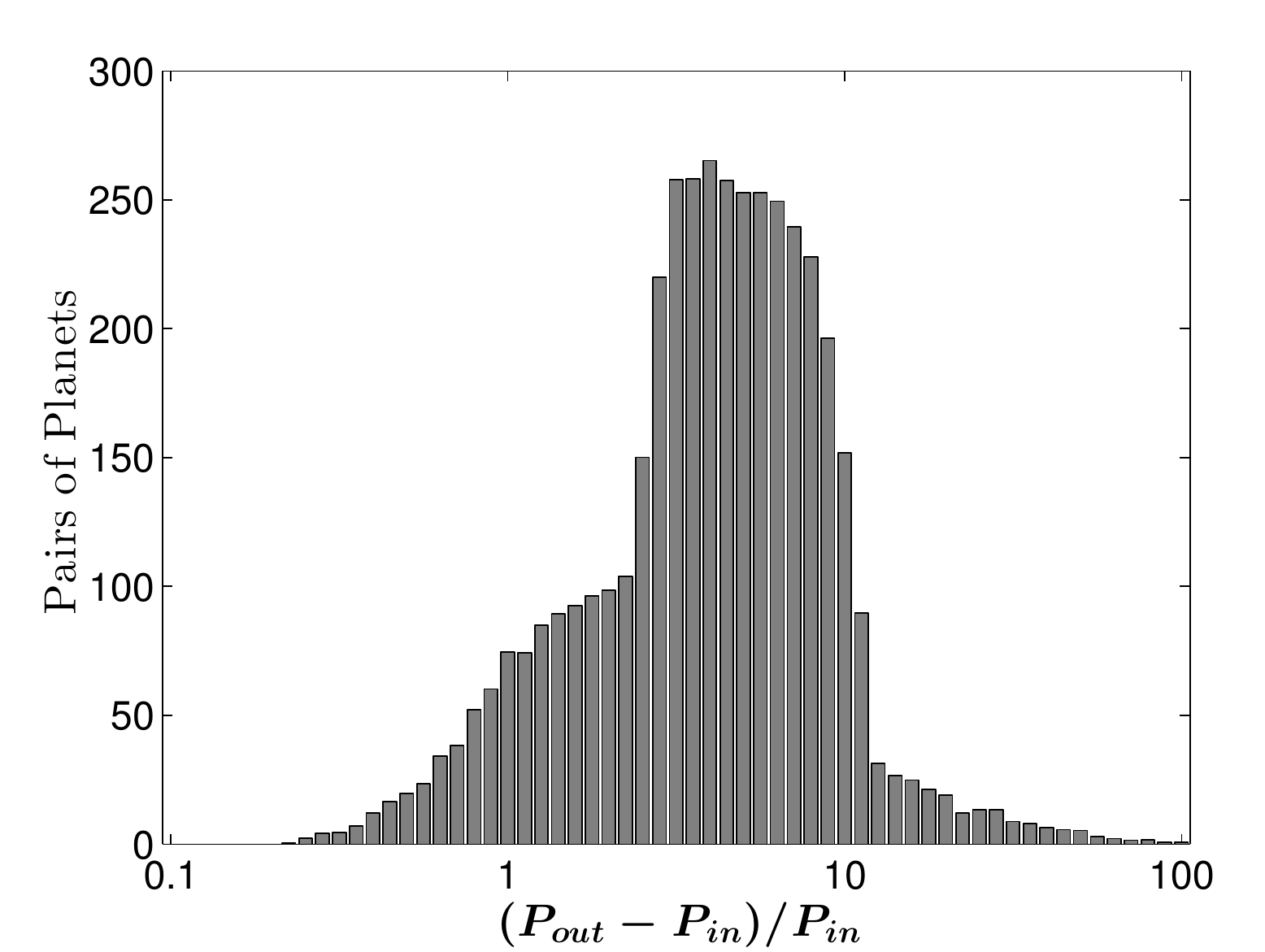}
\caption{The distribution in the relative period difference for
  multi-planet systems. In systems with more than two planets, the
  minimum period difference is counted. All systems with at
  least one transiting member and an apparent magnitude of $I_C<12$
  are counted.}
\label{fig:pr}
\end{figure}

We allow our simulation to assign more than one planet to a given star
with independent probability. The only exceptions are (1) we require
the periods of adjacent planetary orbits to have ratios of at least
1.2, and (2) planets around a star with a binary companion cannot have
orbital periods that are within a factor of 5 of the binary orbital
period. The result is that \multfgk{} of the transiting systems around
FGK stars and \multm{} of those around M stars are multiple-planet
systems. Figure~\ref{fig:pr} shows the resulting distribution of
period ratios. The orbits of multi-planet systems are assumed to be
perfectly coplanar, both for simplicity and from the evidence for low
mutual inclinations in compact multi-planet systems
\citep{fabrycky2014,figueira}.

As a sanity check, we compare the proportion of planets in
multi-transiting systems in our simulated stellar population to the
proportion of multi-transiting \kepler{} candidates. In our
simulation, \tmultfgk{} of planets around FGK stars and \tmultm{} of
planets around M stars reside in multi-transiting systems. Out of the
4,178 \kepler{} objects of interest, 41$\%$ are in
multi-transiting systems.

For simplicity, we assume that all planetary orbits are circular.
The orbital inclinations $i$ are assigned randomly from a uniform
distribution in $\cos i$. We identify the transiting
systems as those with $|b|<1$, where
\begin{equation}
b = \frac{a \cos i }{R_\star}
\end{equation}
is the transit impact parameter.

We then calculate the properties of the planets and their transits and
occultations. The transit duration $\Theta$ is given by Eqns.~(18) and (19) of
\citet{winnbook} in terms of the mean stellar density $\rho_\star$:
\begin{equation}
\label{eq:duration}
\Theta = 13\;\mathrm{hr}
\left(\frac{P}{365\;\mathrm{days}}\right)^{1/3}\left(\frac{\rho_{\star}}{\rho_{\odot}}\right)^{-1/3}
\sqrt{1-b^2}.
\end{equation}
The depth of the transit $\delta_1$ is given by $(R_p/R_\star)^2$.
The depth of the occultation (secondary eclipse) is found by
estimating the effective temperature of the planet ($T_p$)
and then computing the photon flux $\Gamma_p$ within the \tess{}
bandpass from a blackbody of radius $R_p$.
The photon flux from the planet is then divided by the combined photon flux from the planet and the star:
\begin{equation}
\delta_2 = \frac{\Gamma_p}{\Gamma_p + \Gamma_{\star}}.
\end{equation}
The equilibrium planetary temperature $T_p$
is determined by assuming radiative equilibrium with an albedo of zero and 
isotropic radiation (from a recirculating atmosphere), giving
\begin{equation}
\label{eq:tplanet}
T_p = \teff \sqrt{\frac{R_\star}{2a}}.
\end{equation}
We also keep track of the relative insolation of the planet $S/S_\oplus$, defined as
\begin{equation}
\frac{S}{S_\oplus} =
\left(\frac{a}{1~{\rm AU}}\right)^{-2}
\left(\frac{R_\star}{R_\odot}\right)^2
\left(\frac{\teff}{5777~{\rm K}}\right)^4.
\end{equation}

\subsection{Eclipsing Binaries}
\label{sec:ebin}

We identify the eclipsing binaries 
by computing the impact parameters $b_1$ and $b_2$ of the primary
and secondary eclipses, respectively:
\begin{equation}
b_{1,2} = \frac{a \cos i}{R_{1,2}}\left(\frac{1-e^2}{1 \pm e\sin\omega} \right)
\end{equation}
(see Eqns.~7-8 of \citealt{winnbook}).
Non-grazing primary eclipses are identified with the criterion
\begin{equation}
b_1R_1 < R_1-R_2,
\end{equation}
while grazing primary eclipses have larger impact parameters:
\begin{equation}
R_1 - R_2 < b_1R_1 < R_1 + R_2.
\end{equation}
The eclipse depth of non-grazing primary eclipses is given by
\begin{equation}
\delta_1 = \left( \frac{R_2}{R_1} \right)^2 \frac{\Gamma_1}{\Gamma_1 + \Gamma_2}
\end{equation}
where $\Gamma_1$ and $\Gamma_2$ are the photon fluxes from each
star. In the event that $R_2 > R_1$, the area ratio is set equal to
unity; in that case, the primary undergoes a total eclipse.  We neglect
limb-darkening in these calculations for simplicity.  Secondary
eclipses are identified and quantified in a similar manner.

For grazing eclipses, the area ratio $(R_2/R_1)^2$ is replaced with
the overlap area of two uniform disks with the appropriate separation
of their centers, given by Eqns.~(2.14-5) of \citet{kopal}. The durations
and timing of eclipses are calculated from Eqns.~(14-16) of
\citet{winnbook}.

We discard eclipsing binaries when the assigned parameters imply
$a<R_1$ or $a<R_2$. We also exclude systems where $a$ is less than the
Roche limit $a_R$ for either star, assuming they are tidally locked:
\begin{equation}
a_{R1,2} = R_{2,1}\left(3\frac{M_{1,2}}{M_{2,1}}\right)^{1/3}.
\end{equation}

\begin{figure}[htb]
\epsscale{1.0}
\plotone{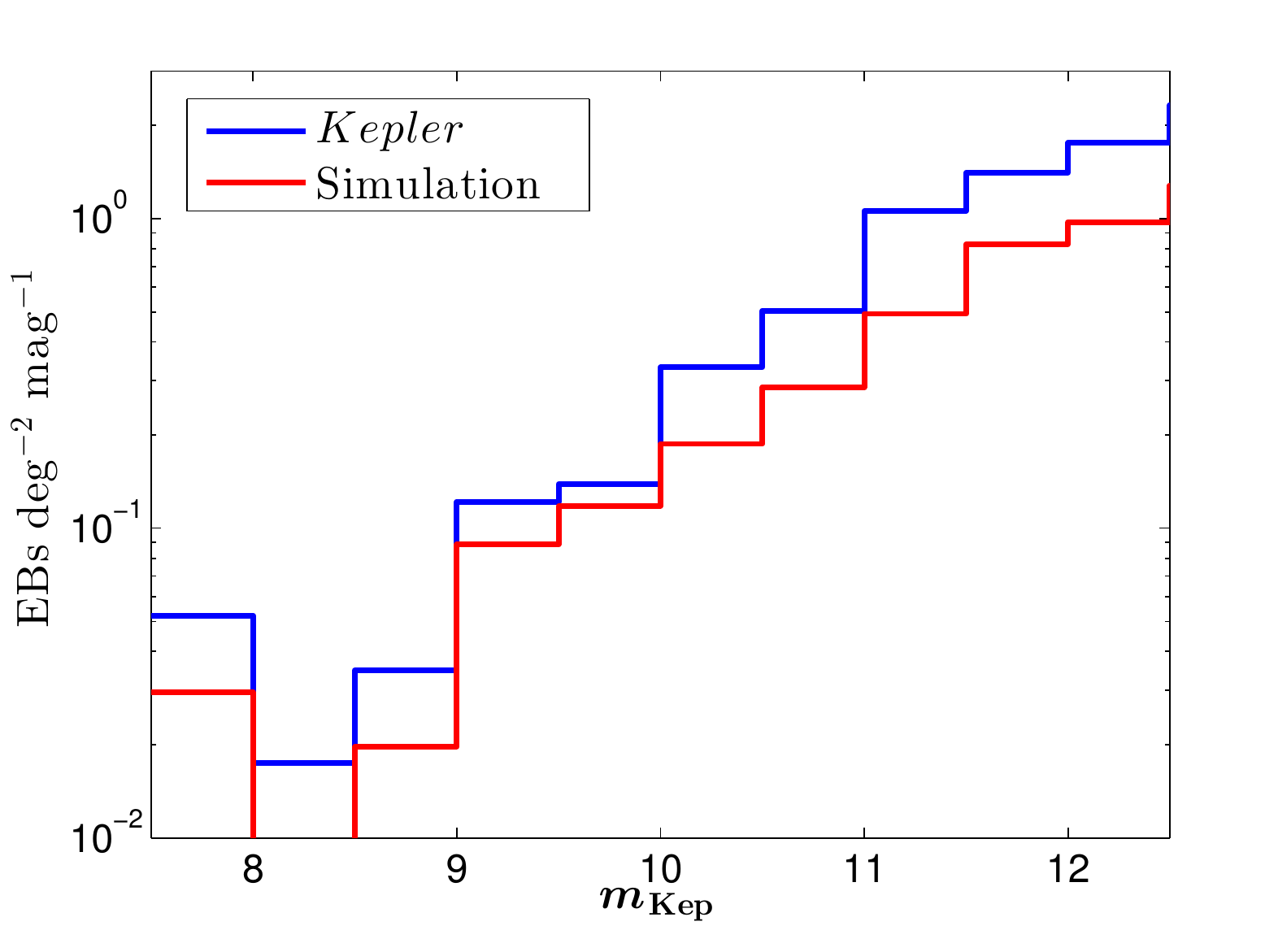}
\caption{Surface density of eclipsing binaries as a function of
  limiting magnitude in the \kepler{} bandpass. The blue curve
  represent actual observations by \citet{slawson}. The red curve is
  from our simulated stellar population in the vicinity of the
  \kepler{} field. All eclipsing systems with $0.5<P<50$ days are
  shown.}
\label{fig:kep_ebs}
\end{figure}

For primaries with $I_C<12$, our simulated stellar population has
\nebin{} eclipsing binaries over the $95\%$ of the sky that is covered
by the simulation. Another \nhebin{} systems contain eclipsing pairs
in a hierarchical system. As another sanity check, we compare the
simulated density of eclipsing systems on the sky to the catalog of
eclipsing binaries in the \kepler{} field. We use Version~2 of the
compilation\footnote{{\tt http://keplerebs.villanova.edu/v2}} from
\citet{prsa} and \citet{slawson} to plot the density of eclipsing
binaries as a function of apparent system magnitude in
Figure~\ref{fig:kep_ebs}. Within the range of $0.5<P<50$ days, this
catalog contains 1.85 EBs deg$^{-2}$ with $m_{\rm Kep}<12$. A 203 deg$^2$
subsample of our TRILEGAL catalog, taken from 15 HEALPix tiles and
centered on galactic coordinates $l=76^{\circ}$ and $b=13.4^{\circ}$
for similarity to the \kepler{} field, contains 1.04 EBs deg$^{-2}$
with $Kp<12$. This disparity suggests that our model of the
eclipsing-binary population could have systematic errors of nearly
80\%, at least for the relatively low galactic latitude of the {\it
  Kepler} field, where the TRILEGAL simulation loses accuracy, and the
steep increase in the stellar surface density makes it difficult to
accurately match the simulation results to the \kepler{} field.

\section{BEST STARS FOR TRANSIT DETECTION}
\label{sec:best}

Now that planets have been assigned to all of the stars with $K_s<15$,
it is interesting to explore the population of nearby transiting
planets independently from how they might be detected by \tess{} or
other surveys. This helps to set expectations for the brightest
systems that can reasonably be expected to exist with any desired set
of characteristics.

\begin{figure}[htb]
\epsscale{1.1}
\plotone{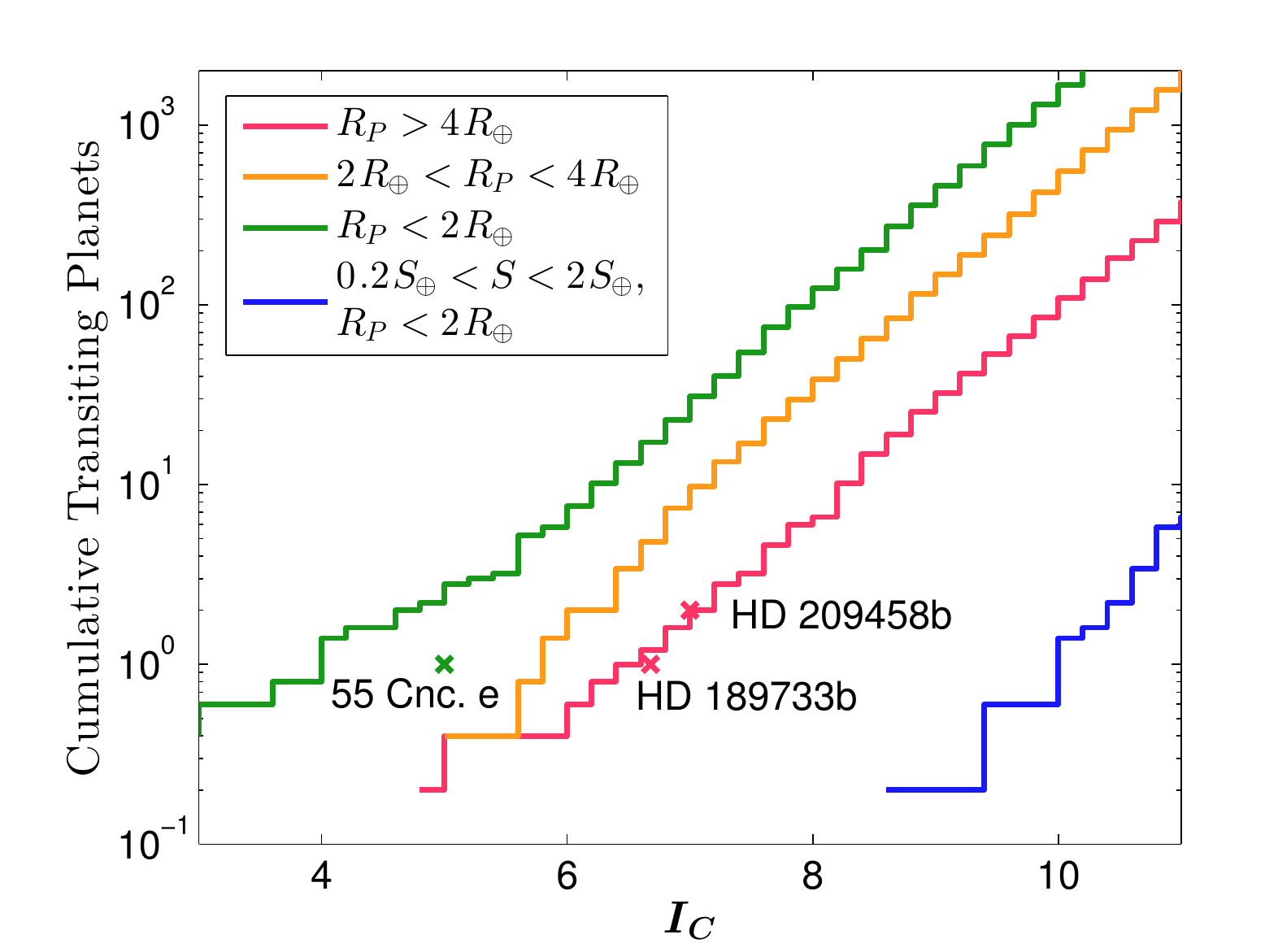}
\caption{Expected number of transiting planets that exist, regardless
  of detectability, over the 95\% of the sky covered by the
  simulation. The cumulative number of transiting planets is plotted
  as a function of the limiting apparent $I_C$ magnitude of the host
  star. The mean of five realizations is shown. We count all planets
  having orbital periods between 0.5-20~days and host stars with
  effective temperatures 2000-7000~K and radii 0.08-1.5~$R_\odot$. The
  planet populations are categorized by radius ranges as shown in the
  figure. Also marked are the apparent magnitudes of a few well-known
  systems with very bright host stars; their locations relative to the
  simulated cumulative distributions suggest that these systems are
  among the very brightest that exist on the sky.}
\label{fig:countplans}
\end{figure}

First, we identify the brightest stars with transiting planets.
Figure~\ref{fig:countplans} shows the cumulative number of transiting
planets as a function of the limiting apparent magnitude of the host
star. This is equal to the total number of planets that would be
detected in a 95\% complete magnitude-limited survey (since our HEALPix 
tiles cover this fraction of the sky). We include the stars with
effective temperatures between 2000 and 7000~K
and $R_{\star}<1.5 R_{\odot}$ that host planets with periods $<$20 days.
To reduce the statistical error, we combine the outcomes of 5 trials.

The brightest star with a transiting planet of size 0.8-2~$R_{\oplus}$
has an apparent magnitude $I_C = \spethicone$. The tenth brightest
such star has $I_C = \spethicten$. For transiting planets of size
2-4~$R_\oplus$, the brightest host star has $I_C = \smnepicone$ and
tenth brightest has $I_C = \smnepicten$. One must look deeper in order
to find potentially habitable planets with periods shorter than 20
days; if we require $0.8<R_p/R_{\oplus}<2$ and $0.2 < S/S_\oplus < 2$,
the brightest host star has $I_C = \hzicone$ and the tenth brightest
has $I_C = \hzicten$. (While there is also an outer limit to the HZ,
we do not impose a lower limit on $S$ since transit surveys are biased
toward close-in planets.)

In reality, the brightest host stars could be brighter or fainter than
the expected magnitudes.  In Figure~\ref{fig:countplans} we also show
the brightest known transiting systems for some of the
categories. Their agreement with the simulated cumulative
distributions suggest that some of the very brightest transiting
systems have already been discovered.

\section{INSTRUMENT MODEL}
\label{sec:inst}

Now that the simulated population of transiting planets and eclipsing
binaries has been generated, the next step is to calculate the
signal-to-noise ratio (SNR) of the transits and eclipses when they are
observed by \tess{}. The signal is the fractional loss of light during 
a transit or an eclipse ($\delta$), and the noise ($\sigma$) is calculated 
over the duration of each event. The noise is the quadrature sum of all 
the foreseeable instrumental and astrophysical components.

Evaluation of the SNR is partly based on the parameters of the cameras
already described in Section~\ref{sec:pre}.  We also need to describe
how well the \tess{} cameras can concentrate the light from a star
into a small number of pixels.  The same description will be used to
evaluate the contribution of light from neighboring stars that is also
collected in the photometric aperture.

Our approach is to create small synthetic images of each transiting or
eclipsing star, as described below. These images are then used to
determine the optimal photometric aperture and the SNR of the
photometric variations.

The synthetic images are also used to study the problem of background
eclipsing binaries.  Transit-like events that are apparent in the
total signal measured from the photometric aperture could be due to
the eclipse of any star within the aperture.  With only the
photometric signal, there is no way to determine which star is
eclipsing. If the timeseries of the $x$ and $y$ coordinates of the
flux-weighted center of light (the ``centroid'') is also examined,
then in some cases, one can determine which star is undergoing
eclipses.  As shown in Section~\ref{sec:centroid}, background
eclipsing binaries tend to produce larger centroid shifts during
eclipses than transiting planets.  The synthetic images allow us to
calculate the centroid during and outside of transits and eclipses.

\subsection{Pixel response function}
\label{sec:prf}

The synthetic images are constructed from the {\it pixel response
  function} (PRF), which describes the fraction of light from a star
that is collected by a given pixel. It is calculated by numerically
integrating the {\it point-spread function} (PSF) over the boundaries
of pixels. The {\it photometric aperture} for a star is the collection
of pixels over which the electron counts are summed to create the
photometric signal; they are selected to maximize the photometric SNR
of the target star.  Throughout this study, we assume that the pixel
values are simply summed without any weighting factors.

The \tess{} lens uses seven elements with two aspheres to deliver a
tight PSF over a large focal plane and over a wide bandpass. Due to
off-axis and chromatic aberrations, the \tess{} PSF must be described
as a function of field angle and wavelength. We calculate the PSF at
four field angles from the center (0$^{\circ}$) to the corner
(17$^{\circ}$) of the field of view. Chromatic aberrations arise both
from the refractive elements of the \tess{} camera and from the
deep-depletion CCDs absorbing redder photons deeper in the silicon. We
calculate the PSF for nine wavelengths, evenly spaced by 50~nm,
between 625 and 1025~nm.  These wavelengths are shown with dashed
lines in Figure \ref{fig:filt}. These wavelengths also correspond to a
set of bandpass filters that will be used in the laboratory to measure 
the performance of each flight \tess{} camera.

The \tess{} lens has been modeled with the Zemax ray-tracing
software. We use the Zemax model to trace 250,000 simulated rays
through the camera optics for each field angle and wavelength. The
model is set to the predicted operating temperature of $-75^{\circ}$C.
Rays are propagated through the optics and then into the silicon of
the CCD. A probabilistic model is used to determine the depth of
travel in the silicon before the photons are converted to electrons.
Finally, the diffusion of the electrons within the remaining depth of
silicon is modeled to arrive at the PSF.

Pointing errors from the spacecraft will effectively enlarge the PSF
because the 2~sec exposures are summed into 2~min stacks
without compensating for these errors. The spacecraft manufacturer
(Orbital Sciences) has provided a simulated time series of spacecraft
pointing errors from a model of the spacecraft attitude
control system. Using two minutes of this time series, we offset the
PSF according to the pointing error and then stack the resulting time
series of PSFs. The root-mean-squared (rms) amplitude of the pointing 
error is $\approx1\arcsec$, which is small in comparison to the pixel 
size and the full width half-maximum of the PSF. Thus, the impact of pointing errors on
short timescales turns out to be minor. Long-term drifts in the pointing
of the cameras will also introduce photometric errors, but this effect
is budgeted in the systematic error described in Section~\ref{sec:instnoise}.

Limits in the manufacturing precision of \tess{} cameras will also
increase the size of the PSF from its ideal value. In a Monte Carlo
simulation drawing from the tolerances prescribed in the optical
design, the the fraction of the flux captured by the brightest pixel
in the PRF is reduced by $\lesssim3\%$ in 80$\%$ of cases. To capture
this effect, we simply increase the size of the PSF by $\approx3\%$ to
achieve the same reduction.

Even after considering jitter and manufacturing errors, 
the PSF is still under-sampled by the 15~$\mu$m pixels of the \tess{} CCDs. 
Therefore, we must recalculate the PRF for a given offset and
orientation between the PSF and the pixel boundaries. We numerically
integrate the PSF over a grid of $16\times 16$ pixels to arrive at
the PRF. We do so over a $10\times 10$ grid of sub-pixel centroid
offsets and two different azimuthal orientations (0$^{\circ}$ and 45$^{\circ}$) 
with respect to the pixel boundaries. For the corner PSF (at a field angle
of 17$^{\circ}$), only the 45$^{\circ}$ azimuth angle is considered.

We can also view the PRF in terms of the cumulative fraction of light
collected by a given number of pixels. In Figure \ref{fig:prf}, we
average over all of the centroid offsets and both azimuthal
angles. For clarity, only three of the field angles and three values
of $\teff$ are shown. There is little change in the PRF across the
range of $\teff$, but the PRF degrades significantly at the
corners of the field.

\subsection{Synthetic images}
\label{sec:image}

For each target star with eclipses or transits, we create a synthetic
image in the following manner. First, we determine the appropriate PRF
based on the star's color and location in the camera field. We
calculate the field angle from its ecliptic coordinates and the
direction in which the relevant \tess{} camera is pointed. We randomly
assign an offset between the star and the nearest pixel center, and we
randomly assign an azimuthal orientation of either 0$^\circ$ or
45$^\circ$. We then look up the nine wavelength-dependent PRFs for the
appropriate field angle, centroid offset, and azimuthal angle. The
nine PRFs are summed with weights according to the stellar effective
temperature.

The weight of a given PRF is proportional to the stellar photon flux
integrated over the wavelengths that the PRF represents.  Outside of
the main simulation, we considered a Vega-normalized stellar template
spectrum of each spectral type from the \citet{pickles} library.  We
multiplied each template spectrum by the spectral response function of
the \tess{} camera, and we integrated the photon flux for each of the
nine PRF bandpasses. Next, we fitted a polynomial function to the
relationship between the stellar effective temperatures and the photon
flux in each bandpass.  During the simulation, the polynomial
functions are used to quickly calculate the appropriate PRF weights as
a function of stellar effective temperature.

\begin{figure}[ht]
\epsscale{1.2}
\plotone{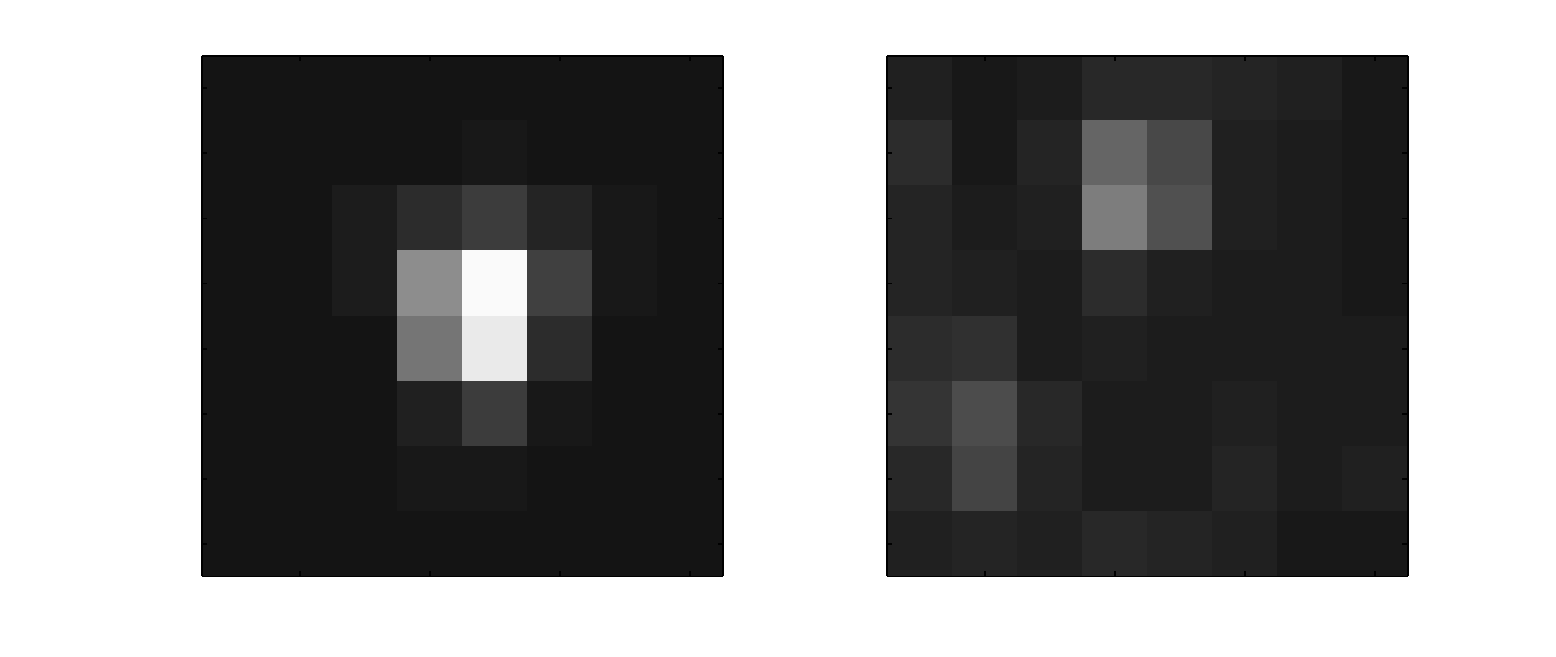}
\caption{Synthetic images produced from the pixel-response function (PRF). {\it
    Left.}---A target star. The PRFs computed for 9 wavelengths have
  been stacked to form a single image. The weight of each PRF in the
  sum depends on the the stellar effective temperature. {\it
    Right.}---Fainter stars in the vicinity of the target star. We
  sum the flux from neighboring stars, with PRFs weighted
  according to the $\teff$ of each star, in the same fashion as the
  target stars.}
\label{fig:imgs}
\end{figure}

\begin{figure}[ht]
\epsscale{1.0}
\plotone{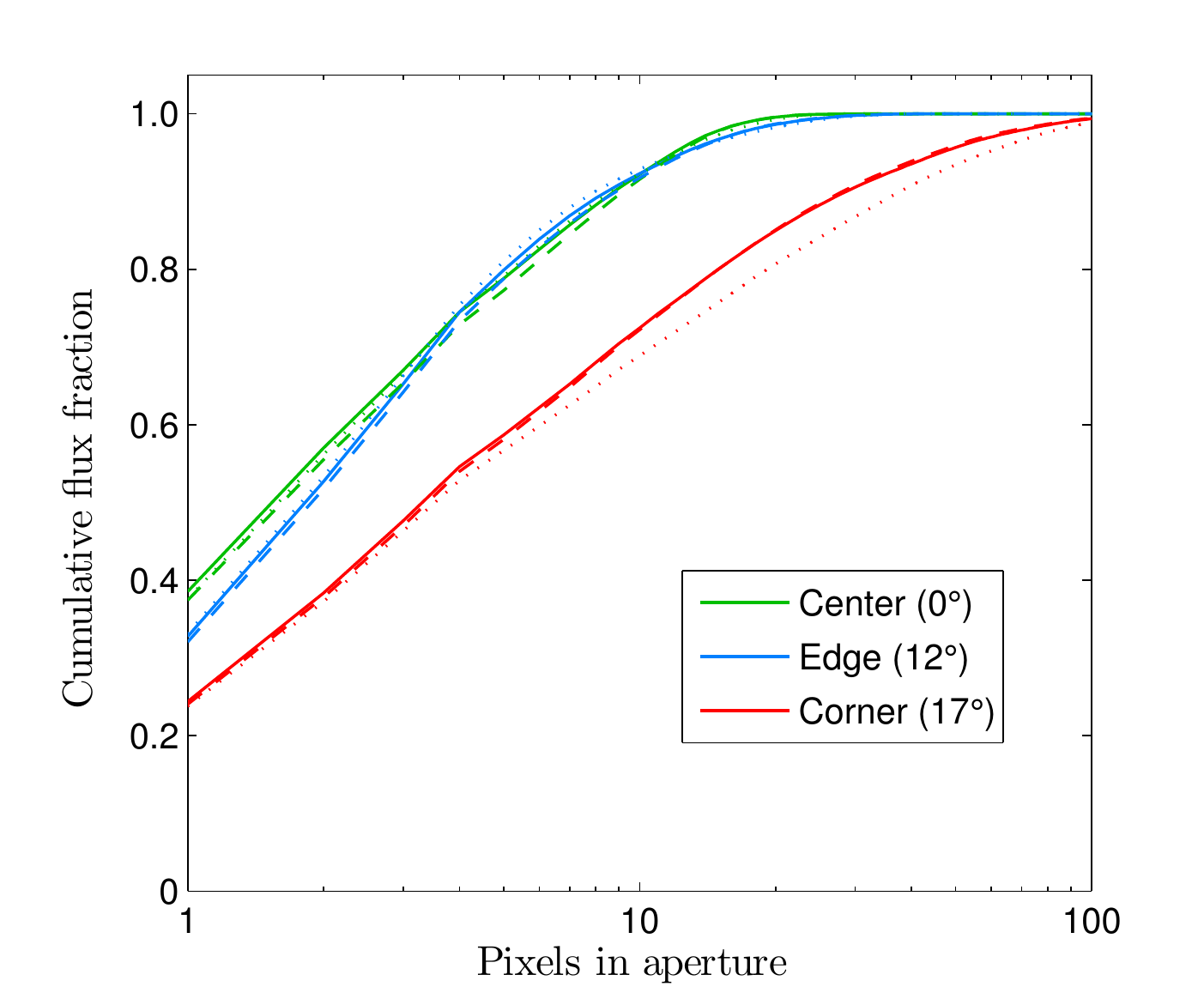}
\caption{The \tess{} pixel response function (PRF) after sorting and
  summing to show the cumulative fraction of light collected for a
  given number of pixels in the photometric aperture. We show this
  fraction for three field angles and three values of stellar
  effective temperature.  The dotted line is for $\teff = 3000$~K, the
  solid line is for 5000~K, and the dashed line is for 7000~K.  These
  temperatures span most of the range of the \tess{} target stars}
\label{fig:prf}
\end{figure}

Once the PRFs are summed, the result is a synthetic $16\times 16$
-pixel image of each target star. We only consider the central
$8\times 8$ pixels when determining the optimal photometric aperture;
the left panel of Figure \ref{fig:imgs} shows an example.

After synthesizing the image of each eclipsing or transiting target
star, a separate $16\times 16$ image is synthesized of all the
relevant neighboring stars and companion stars.  The neighboring stars
are drawn from all three star catalogs described in Section
\ref{sec:catalog}. The stars are assumed to be uniformly distributed
across each HEALPix tile, allowing us to randomly generate the
distances between the target star and the neighboring stars. Stars
from the target catalog are added to the synthesized image if they are
within a radius of 6~pixels from the target star. Stars from the
intermediate catalog are added if they are within 4~pixels, and stars
from the faint catalog are added if they are within 2~pixels. The
synthesized images are created in the same manner as described above:
by weighting, shifting, and summing the PRFs associated with each
star. The right panel of Figure \ref{fig:imgs} shows an example.

Synthetic images are also created for the eclipsing binary systems
drawn from the intermediate catalog, but a slightly different approach
is taken.  For each eclipsing binary, we search for any target stars
within 6~pixels.  If any are found, the brightest is added to the list
of target stars with apparent transits or eclipses. Separate synthetic
images are created for the target star, the eclipsing binary, and the
non-eclipsing neighboring stars.  Hierarchical binaries are treated in
a similar fashion; the non-eclipsing component is treated as the
target star, and a separate synthetic image is created for the
eclipsing pair so that its apparent depth can be diluted.  While this
approach may appear to strongly depend upon the somewhat arbitrary
magnitude limits adopted for the different catalogs, this is not
really the case. Both the eclipsing binaries from the target catalog
and the background eclipsing binaries from the intermediate catalog
end up being diluted by neighboring stars drawn from all of the
catalogs.

\subsection{Determination of optimal aperture}
\label{sec:optimal_aperture}

For each target star that is associated with an eclipse or transit
(whether it is due to the target star itself or a blended eclipsing
binary), we select the pixels that provide the optimal photometric
aperture from the central $8\times 8$ pixels of its synthetic
image. Starting with the three brightest pixels in the PRF, we add
pixels in order decreasing brightness one at a time. At each step, we
sum the flux of the pixels from the synthetic image of target star and
from the synthetic image of the neighboring stars.  We also consider
the read noise and zodiacal noise, which are discussed in Section
\ref{sec:noise}.  As the number of pixels in the photometric aperture
increases, more photons are collected from the target star, and more
noise is accumulated from the readout, sky background, and neighboring
stars.  The optimal photometric aperture maximizes the SNR of the
target star even if the eclipse is produced by a blended binary.  We
assume that the data will be analyzed with prior knowledge of the
locations of neighboring stars (but no prior knowledge of whether they
eclipse).

Once the optimal aperture is determined, we calculate the dilution
parameter $D$, which is the factor by which the true eclipse or transit depth is
reduced by blending with other stars in the photometric aperture.
Specifically, the dilution parameter is defined as the ratio of the
total flux in the aperture from the neighboring stars ($\Gamma_N$) and
target star ($\Gamma_T$) to the flux from the target star:
\begin{equation}
D = \frac{\Gamma_N+\Gamma_T}{\Gamma_T}.
\end{equation}
For blended binaries and hierarchical systems, the denominator is replaced with the
flux from the binary $\Gamma_B$, and the target star becomes a source of dilution:
\begin{equation}
D = \frac{\Gamma_N+\Gamma_T+\Gamma_B}{\Gamma_B}.
\end{equation}
With this definition, $D=1$ signifies an isolated system,
and in general, $D>1$. This parameter is later reported for all detected eclipses
under the ``Dil.'' column of Table~\ref{tbl:cat}.

\subsection{Noise Model}
\label{sec:noise}

The photometric noise model includes the photon-counting noise from
all of the stars in the photometric aperture, photon-counting noise
from zodiacal light, stellar variability, and instrumental
noise. Stellar variability and background stars are randomly assigned
from distributions, while the other noise terms are more deterministic
in nature. Figure \ref{fig:noise} shows the relative photometric noise
as a function of apparent magnitude and also breaks down the
contributions from the deterministic sources of noise.  Each
subsection below describes the noise terms in more detail.

\subsubsection{Zodiacal Light}
\label{sec:zodi}

Although \tess{} avoids the telluric sky background by observing from
space, it its still affected by the zodiacal light (ZL) and its
associated photon-counting noise. Our model of the zodiacal flux
is based on the spectrum measured by the Space Telescope Imaging
Spectrograph on the {\it Hubble Space Telescope}.\footnote{\tiny {\tt
    http://www.stsci.edu/hst/stis/performance/background/skybg.html}}
We multiply this ZL spectrum by the \tess{} spectral
response function and integrate over wavelength.
This gives the photon flux of
\begin{equation}
\label{eq:zl-count-rate}
2.56\times 10^{-3}~10^{-0.4 (V-22.8)}~{\rm ph~s}^{-1}~{\rm cm}^{-2}~{\rm arcsec}^{-2},
\end{equation}
where $V$ is the $V$-band surface brightness of the ZL in
mag~arcsec$^{-2}$. For \tess{}, the pixel scale is \pixsz{} and the effective
collecting area is \aeff{}.
To model the spatial dependence of $V$, we fit
the tabulated values of $V$ as a function of helio-ecliptic
coordinates\footnote{\tiny {\tt http://www.stsci.edu/hst/stis/documents/handbooks/currentIHB/c06\_exptime6.html\#689570}}
with a function
\begin{equation}
\label{eq:vband-sb-zl}
V = V_{\rm max} - \Delta V \left( \frac{b-90^\circ}{90^\circ} \right)^2
\end{equation}
where $b$ is the ecliptic latitude and $V_{\rm max}$ and $\Delta V$
are free parameters.  Because \tess{} will generally be pointed in the
anti-solar direction (near helio-ecliptic longitude $l \approx
180^{\circ}$), and because $V$ depends more strongly on latitude than
longitude in that region, we only fitted to the data with 
$l \geq 120^\circ$ and weighted the points in proportion to 
$(l-90^\circ)^2$. The least-squares best-fit has $V_{\rm max} = 23.345$ mag
and $\Delta V = 1.148$ mag.

Based on these results, we find that the zodiacal light collected in a
2~sec image ranges from 95-270~$e^-$~pix$^{-1}$ depending on
ecliptic latitude. The photon-counting noise associated with this
signal varies from 10-16~$e^-$~pix$^{-1}$ RMS, as mentioned in
Section~\ref{sec:pre}.

\subsubsection{Instrumental Noise}
\label{sec:instnoise}

The read noise of the CCDs is assumed to be 10~$e^-$~pix$^{-1}$ RMS in
each 2~sec exposure, which is near or below the level of
photon-counting noise from the ZL. Both the read noise and ZL noise
grow in proportion to the square root of the number of pixels used in
the photometric aperture.

Our noise model for \tess{} cameras also includes a systematic error
term of 60~ppm~hr$^{1/2}$. This is an engineering requirement on the
design rather than an estimate of a particular known source of
error. We assume that the systematic error is uncorrelated and scales
with the total observing time as $t^{-1/2}$. Under these assumptions,
the systematic error grows larger than 60~ppm for timescales shorter
than one hour, which is probably unrealistic; however, this issue is
not very relevant to our calculations because such timescales are
shorter than the typical durations of transits and eclipses.

It is thought that the systematic error of the \tess{} cameras will
primarily stem from pointing errors that couple to the photometry
through non-uniformity in the pixel response.  These pointing errors
come from the attitude control system, velocity aberration, thermal
effects, and mechanical flexure.  In addition, long-term drifts in the
camera electronics can contribute to the systematic error.  The data
reduction pipeline will use the same co-trending techniques that were
used by the \kepler{} mission to mitigate these effects, but the exact
level of residual error that \tess{} will be able to achieve is
unknown at this time.

\subsubsection{Saturation}

\begin{figure}[ht]
\epsscale{1.1}
\plotone{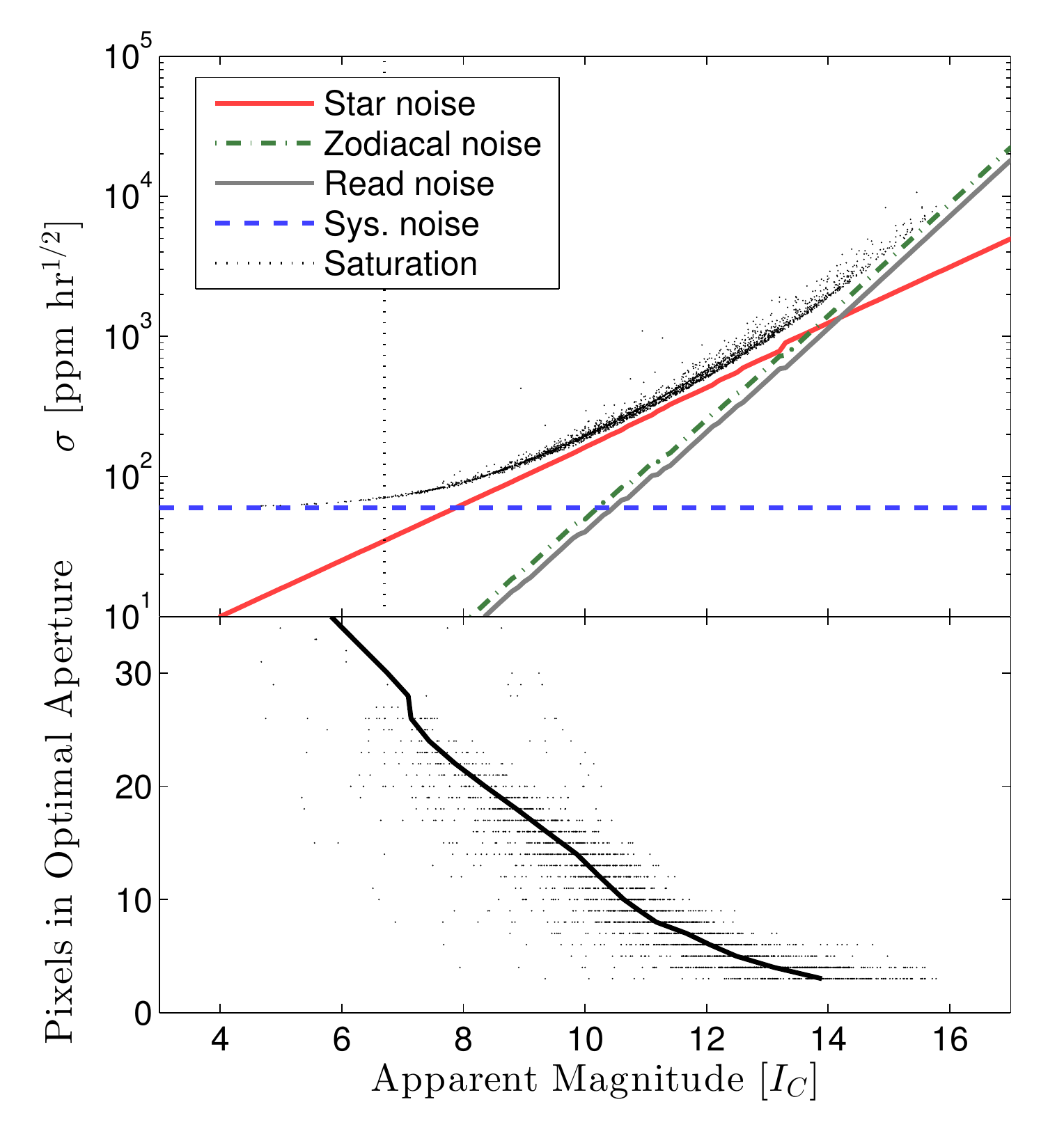}
\caption{Noise model for \tess{} photometry. \textit{Top.}---Expected
  standard deviation of measurements of relative flux, as a function
  of apparent magnitude, based on 1~hour of data.  For the brightest
  stars, the precision is limited by the systematic noise floor of 60
  ppm.  For the faintest stars, the precision is limited by noise from
  the zodiacal light (shown here for an ecliptic latitude of
  30$^{\circ}$). Over the range $I_C\approx 8$-13, the photon-counting
  noise from the star is the dominant source of uncertainty.
  \textit{Bottom.}---The number of pixels in the optimal photometric
  aperture, chosen to maximize the SNR. The scatter in the simulated
  noise performance and number of pixels is due to the random
  assignment of contaminating stars and centroid offsets in the PRF.}
\label{fig:noise}
\end{figure}

Stars with $T \lesssim \satmag{}$ will saturate the innermost pixels
of the PRF during the 2-second exposures. For reference, this
saturation magnitude is identified with a dotted line in
Figure~\ref{fig:noise}.  These saturated stars represent \satpct{} of
the target stars. As was the case with the \kepler{} CCDs, the \tess{}
CCDs are designed to conserve the charge that bleeds from saturated
pixels, and do not use anti-blooming structures. Since photometry of
saturated stars with \kepler{} has achieved the photon-counting limit
\citep{gilliland}, we assume that the systematic error is the same for
the saturated stars and the unsaturated stars.

While large photometric apertures will be needed to collect all of the
charge that bleeds from saturated stars, the read and zodiacal noise
are not important since the photometric precision will be dominated by
photon-counting noise and systematic errors. Because the photometric
precision will not depend strongly upon the number of pixels used in
the photometric aperture, we do not model the saturated stars
differently in our simulation.

\subsubsection{Cosmic Rays}
\label{sec:cr}

Typical back-illuminated CCDs have depletion depths of 10-50~$\mu$m.
In contrast, the \tess{} CCDs have a 100 $\mu$m depletion depth. This
is desirable to enhance the quantum efficiency at long wavelengths,
but it also makes the detectors more susceptible to cosmic rays (CRs)
since the pixel volume is larger and the maximum amount of charge
collected per event can be larger.

To assess the effect of cosmic rays, we consider a typical cosmic ray
flux of 5~events~s$^{-1}$~cm$^{-2}$ and minimally-ionizing events that
deposit 100~$e^-$~$\mu$m$^{-1}$ within silicon. Each pixel has an
optical exposure time of 2~sec. The accumulated images also spend an
average of 1~sec in the frame-store region of the CCD, where they are
still vulnerable to cosmic rays. Given these parameters, for each
2~min stack of values from one pixel, there is a 10\% chance of
experiencing a cosmic ray event with an energy deposition above the
combined read and zodiacal noise of 110~$e^-$.  The distribution in
the energy deposition values has a peak near 1500~$e^-$, which is
comparable to the photon-counting noise of bright stars observed with
2~min cadence. Electrons from cosmic rays will therefore add
significantly to the photometric noise, but will not be easily
detected in the 2~min or 30~min data products.

Cosmic rays are far more conspicuous in the 2~sec images.  Therefore,
it is probably best to remove the contaminated pixel values before
they are combined into the 2~min and 30~min stacks. The Data Handling
Unit on \tess{} will apply a digital filter that rejects outlier values 
during the stacking process either periodically or adaptively. 
A possible side-effect of this filter, depending on the algorithm used, 
is a reduction in the signal-to-noise ratio to the degree that 
uncontaminated data is also rejected in the absence of cosmic rays.

The exact algorithm that will be used to mitigate cosmic-ray noise is
still being studied. For the present simulations we have budgeted for
a 3$\%$ loss in the SNR. In the simulation code, we simply raise the
detection threshold (described in Section \ref{sec:threshold}) by
3$\%$ to compensate for the reduced SNR, and we assume that there are
no other residual effects from cosmic rays.

\subsection{Duration of observations}
\label{sec:obs}

The SNR of transits or eclipses will depend critically on how long the
star is observed.  Figure \ref{fig:npoint} is a sky map showing the
number of times that \tess{} will point at a given location as a
function of ecliptic coordinates. As noted above, the simulations
assign coordinates to each star through a uniform random distribution
across the HEALPix tile to which it belongs.  The star's ecliptic
coordinates are then converted to $x$ and $y$ pixel coordinates for
each \tess{} pointing. We tally the number of pointings for which the
target falls within the field-of-view of a \tess{} camera. The total
amount of observing time is calculated as the total duration of all
consecutive pointings.

The duty cycle of observations must also be considered. At each
orbital perigee, \tess{} interrupts observations in order to transmit
data to Earth and perform other housekeeping operations.  This takes
approximately 0.6~days. We model this interruption in the simulation,
so each 13.6-day spacecraft orbit actually results in 13.0 days of
data.

The presence of the Earth or Moon in the field-of-view of any camera
will also prohibit observations. We do not model this effect since
predicting their presence depends upon the specific launch date of
\tess{}. However, our simulations do show that if observations are
interrupted near \tess{}'s orbital apogee in addition to its perigee,
then the planet yields are approximately proportional to the duty
cycle of observations.

\subsection{Detection}
\label{sec:threshold}

The model for the detection process is highly simplified: we adopt a
threshold for the signal-to-noise ratio, and we declare a signal to be
detected if the total SNR exceeds the threshold.  In other words, the
detection probability is modeled as a step function of the computed
SNR.  (The matched-filter technqiues of the \tess{} pipeline probably
have a smoother profile, such as a standard error function
[\citealt{jenkins1996}]).  For transiting planets, all of the observed
transits contribute to the total SNR. For eclipsing binaries, we allow
both the primary and secondary eclipses to contribute to the total
SNR.

The choice of an appropriate SNR threshold was discussed in detail by
\citet{jenkins2002} in the context of the \kepler{} mission. Their
criterion was that the threshold should be sufficiently high to
prevent more than one ``detection'' from being a purely statistical
fluke after analyzing all of the data from the entire mission.  We
adopt the same criterion here. Since the number of astrophysical false
positives is at least several hundred (as discussed below), this
criterion allows statistical false positives to be essentially
ignored.

To determine the appropriate threshold, we use a separate Monte Carlo
simulation of the transit search. We produce $2\times10^5$ light
curves containing uncorrelated, Gaussian noise and analyze them for
transits in a similar manner as will be done with real data. Then, we
find the SNR threshold that results in approximately one statistical
false positive. Each light curve consists of 38,880 points,
representing two \pointdur{}-day \tess{} pointings with 2-minute
sampling. We chose a timeseries length of two pointings rather than
one to account for the stars observed with overlapping pointings.

To search for transits, we scan through a grid of trial periods, times
of transit, and transit durations. At each grid point, we identify the
data points belonging to the candidate transit intervals. 
The SNR is computed as the mean of the in-transit data
values divided by the uncertainty in the mean.

The grid of transit durations $\Theta$ starts with 28~min (14 samples) and
each successive grid point is longer by 4~min (2 samples). The
grid of periods $P$ is the range of periods that are compatible with
the transit duration.  The periods are calculated by inverting
Eqn.~(\ref{eq:duration}):
\begin{equation}
P = \left(\mathrm{365\;days}\right)\left(\frac{\Theta}{\mathrm{78\;min}}\right)^3\frac{\rho_{*}}{\rho_{\odot}}\left(1-b^2\right)^{-3/2}
\end{equation}
We allow $P$ to vary over a sufficient range to include plausible stellar densities $\rho_\star/\rho_\odot$ from 0.5 to 100.
The fractional step size in the period $\Delta P/P$ is then
$3\Delta\Theta/\Theta$,
which has a minimum value of 0.43 for the shortest periods. We consider orbital periods ranging from 1.7~hr
(which is below the period corresponding to Roche limit)
to \pointdur{}~days (half of the nominal observing interval). 
The transit phase is stepped from zero to the
orbital period in increments of one-half the transit duration.

\begin{figure}[htb]
\epsscale{1.0}
\plotone{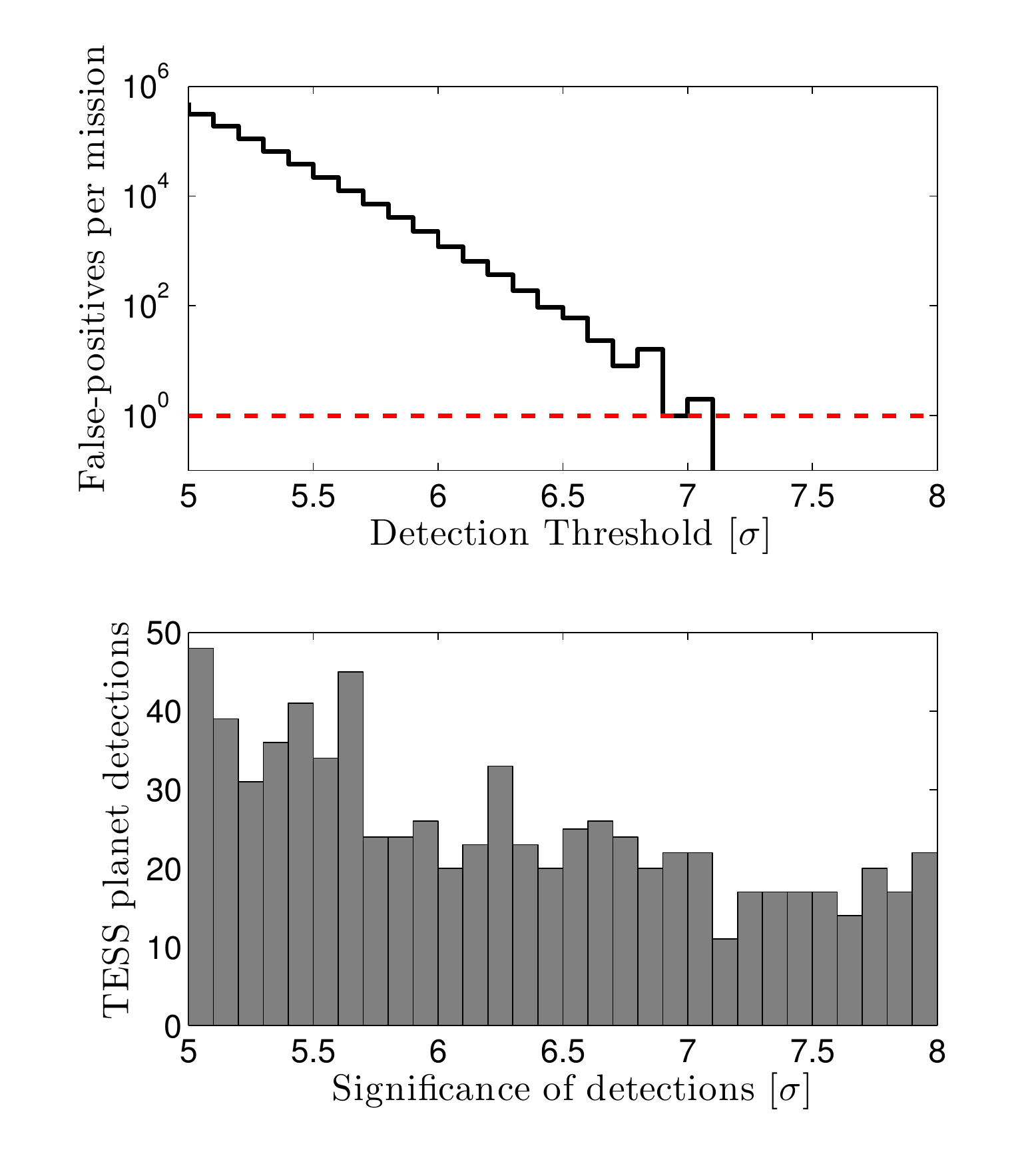}
\caption{Determination of the SNR threshold. {\it Top.}---The statistical
  false-positive rate for the \tess{} mission as a function of the
  detection threshold. We do not want more than one statistical
  false positive to occur (red dashed line), which dictates a
  threshold of \snrthresh. {\it Bottom.}---The SNR distribution of
  transits near the threshold from the full \tess{} simulation
  (presented in Section \ref{sec:planetout}).
  The small slope of this distribution near \snrthresh{} suggests that the
  planet yield is not extremely sensitive to the detection process or threshold.}
\label{fig:snr}
\end{figure}

Figure~\ref{fig:snr} shows how the number of false-positive detections
scales with the detection threshold. We find that a SNR of
\snrthresh{} produces approximately one statistical false positive
within the library of $2\times 10^5$ light curves. By coincidence,
this is equal to the SNR threshold of 7.1 that was calculated for
\kepler{} mission by \citet{jenkins2002}. \tess{} searches twice as
many stars as the $10^5$ considered in the \kepler{} study, and over a
larger dynamic range in period; \kepler{} searches for planets with
longer periods using longer intervals of data.

To account for the expected reduction in SNR due to the cosmic-ray
rejection algorithm (see Section \ref{sec:cr}), we adopt a slightly
higher threshold of \snrcr{} in this paper. In addition, we only
consider a transit or eclipse to be detected if two or more events are
observed. We also record the single events that exceed the SNR
threshold, but we do not count them as ``detections'' in the tallies
and the discussion that follows. The planets detected with a single
transit generally have longer periods than the multiple-transit
detections.  It is also worth noting that \tess{} may detect some
single transits from the population of planets with periods longer
than a year, which we have not simulated at all, because our sources
for planet occurrence rates do not extend to such long periods. The
single-transit detections may represent interesting opportunities to
study the properties of more distant planets.  However, they will
require additional ground-based follow-up observations to determine
the orbital period and discriminate against astrophysical or
statistical false positives.

\subsection{Selection of target stars}
\label{sec:target_selection}

From the \nbrightcat{} stars in the $K_s<15$ catalog, we must select
the $2\times 10^5$ target stars for which pixel data will be saved and
transmitted with 2~min time sampling. In our simulation, the target
stars are chosen according to the prospects for detecting the transits
of small planets, which depend chiefly on stellar radius and apparent
magnitude.

In the simulation, we have complete knowledge of the properties of
each star, which makes it straightforward to determine whether a
fiducial transiting planet with a given radius and period could be
detected with \tess{}. We adopt an orbital period of 20 days; for each
\pointdur{}-day pointing that \tess{} spends observing a star, we
assume that 2 transits are observed. The stellar radius and mass are
used to calculate the transit duration with a 20-day period, thereby
determining the total exposure time during transits. Then, we use the
simplified noise model from Section \ref{sec:maglim} that considers
the read noise and photon-counting noise of the star and zodiacal
light. We then check to see if the fiducial transiting planet would be
detectable with a signal-to-noise ratio exceeding of \snrcr{}.

The number of stars meeting this detection criterion depends strongly
on the radius of the fiducial planet. Starting from small values, we
increase the radius until the number of stars for which the planet
would be detectable is $2\times 10^5$. This is achieved for $R_p =
2.25~R_{\oplus}$. Through this procedure, the target star catalog is
approximately complete for planets smaller than 2.25~$R_{\oplus}$ with
orbital periods shorter than 20~days. There is a higher density of
target stars assigned near the ecliptic poles due to the longer
duration of \tess{} observations in those regions.

In selecting the target stars, we do not assume prior knowledge of
whether a star is part of a multiple-star system. If it is, we assume
that all components of the system fall within a single photometric
aperture, and they are all observed at the 2~min cadence.

\begin{figure}[ht]
\epsscale{1.0}
\plotone{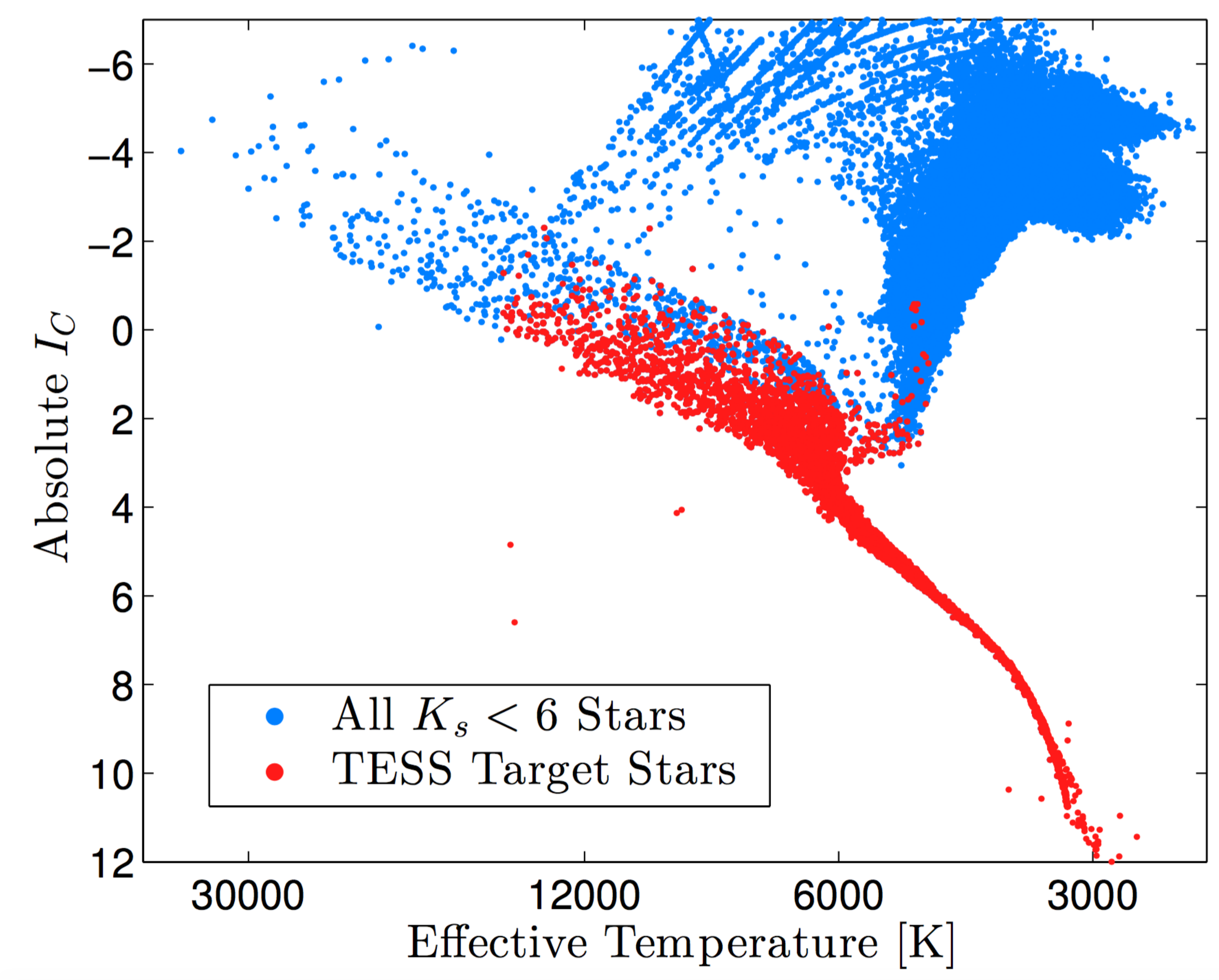}
\caption{Selection of the $2\times 10^5$ target stars on a
  Hertzsprung-Russell diagram. To reduce the number of plotted points
  to a manageable number, the blue points represent only those
  simulated stars with apparent $K_s<6$, and the red points are a
  random selection of 1$\%$ of the target stars. Nearly all
  main-sequence dwarfs smaller than the Sun are selected as target
  stars; a decreasing fraction of larger stars are selected.}
\label{fig:hr}
\end{figure}

Figure~\ref{fig:hr} illustrates the selection of the target stars on a
Hertzsprung-Russell diagram. For clarity, we show a magnitude-limited
subsample ($K_s<6$) of our ``bright'' catalog as well as a
randomly-selected subsample of the $2\times 10^5$ target stars.
Nearly all main-sequence stars with $\teff{}<6000$~K are selected as
target stars. Stars that are larger than the Sun are only included if
they have a sufficiently bright apparent magnitude.  White dwarfs
could also be interesting targets for \tess{}, but we do not include
them in our simulation because the occurrence rates of planets around
white dwarfs is unknown.

\begin{figure}[ht]
\epsscale{1.0}
\plotone{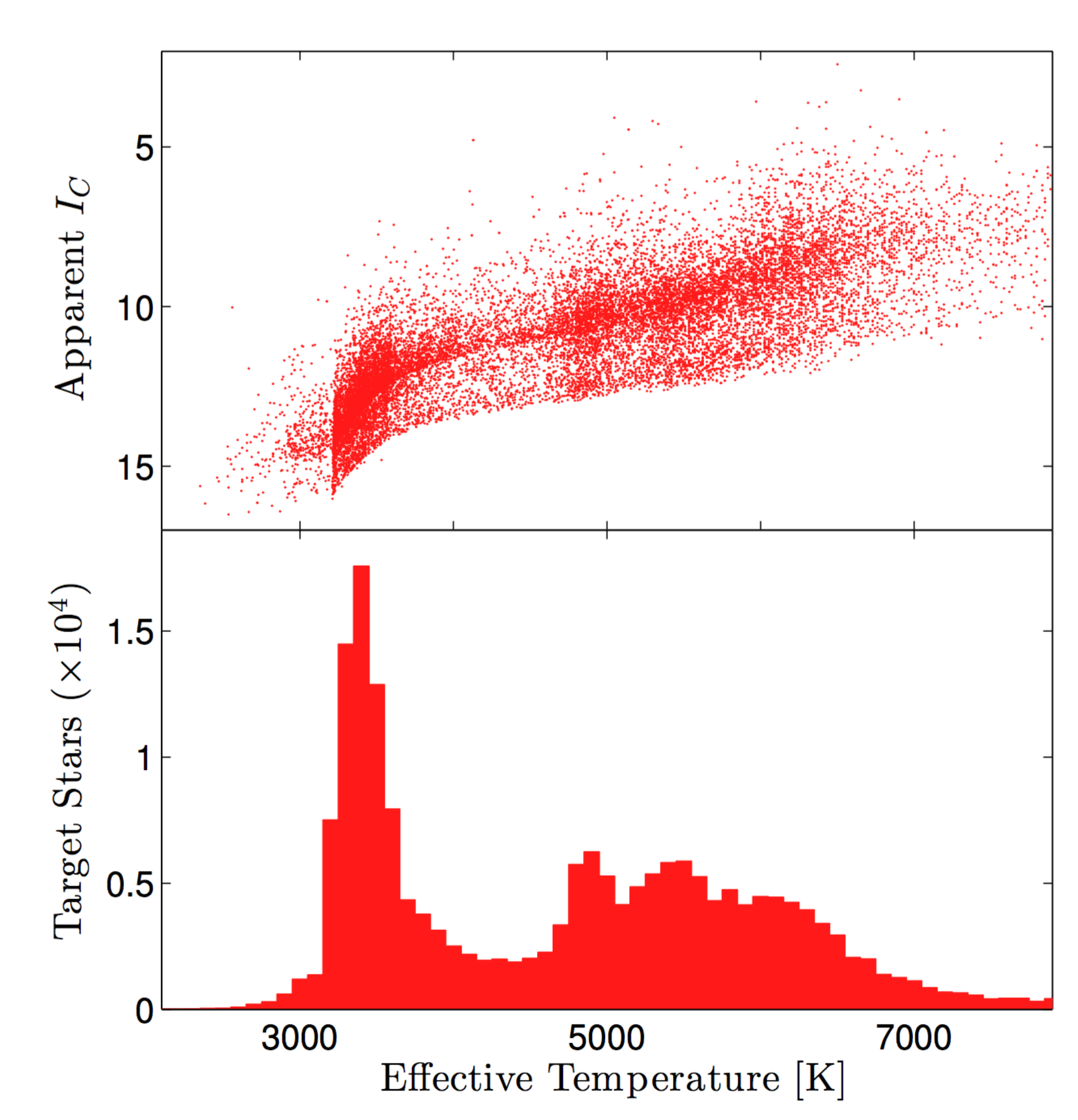}
\caption{The distributions of apparent $I_C$ magnitude and effective
  temperature of the the \tess{} target stars. To reduce the number of
  plotted points to a manageable number, the top panel shows a random
  subset of 10$\%$ of the target stars.}
\label{fig:icteff}
\end{figure}

Figure~\ref{fig:icteff} shows the distribution of target stars as a
function of effective temperature, along with their {\it apparent}
$I_C$ magnitudes. The distribution in effective temperature of the
target stars is bimodal, with a sharp peak near 3400~K and a broader
peak near 5500~K.

In reality, it will not be quite as straightforward to select the
target stars for \tess{}. While proper-motion surveys (e.g.,
\citealt{lspm}) can readily distinguish red giants from dwarf stars,
it is much more difficult to distinguish dwarfs from subgiant stars
\citep{tdc}.  Ultimately, the selection of the \tess{} target stars
may rely on parallaxes from the ongoing {\it Gaia} mission
\citep{gaia2001}. Errors in selecting the target stars might be
mitigated by simply observing a larger number of stars at 2~min
cadence. There is also the possibility of detecting transits in the
full-frame images, which is described below.

\subsection{Full-frame images}

\tess{} will record and downlink a continuous sequence of full-frame
images (FFIs) with an effective integration time of 30~min or
shorter. Transiting planets can still be detected with 30~min
sampling, but the longer integration time of the FFIs reduces the
sensitivity to events with a short duration. Our simulation estimates
the yield of transiting planets from the FFIs in the following
fashion.

First, we identify all the transiting or eclipsing stars that are not
among the pre-selected $2\times 10^5$ target stars. We assign to each
system a random phase between the beginning of a 30~min window and the
beginning of an eclipse.  Next, we calculate the number of 30~min data
points that are required to cover the transit or eclipse duration. The
data points at the beginning and end of the series are omitted if they
do not increase the signal-to-noise.  Finally, we compute the
effective depth of the transit or eclipse by averaging over all of the
30~min data points spanning the event.  This step can reduce the depth
because some of the data points include time outside of the transit or
eclipse.

For transits with durations shorter than 1 hour, the 30-minute
integration time of the FFIs causes the apparent transit duration to
be lengthened and the apparent transit depth to become more shallow.
However, the depths and durations of transits with longer durations
are largely unaffected. The effects of time averaging on the
uncertainties in transit parameters derived from light-curve fitting
have been analyzed by \citet{kipping2010} and \citet{price2014}.

Our calculated detection threshold of \snrcr{} only ensures that no
more than one statistical false positive is detected among the
$2\times 10^5$ target stars.  Since many more stars can be searched
for transits in the FFIs, the number of statistical false-positives
will be much greater than one if the same threshold is adopted.

\begin{figure*}[htb]
\epsscale{1.0}
\plotone{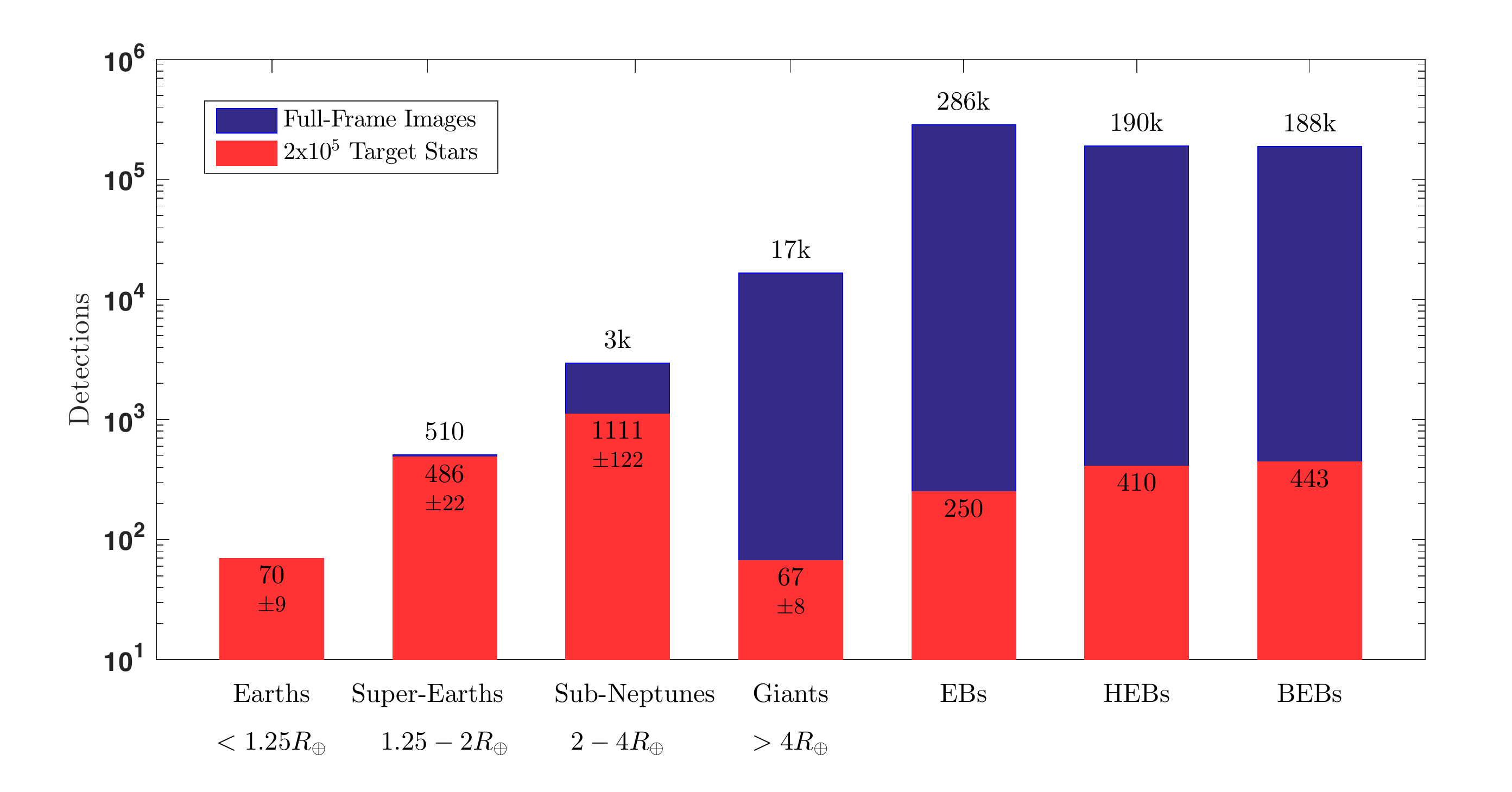}
\caption{Mean numbers of planets and eclipsing binaries that are
  detected in the \tess{} simulation. Results are shown for the
  $2\times 10^5$ target stars that are observed with 2~min time
  sampling as well as stars in the full-frame images that are observed
  with 30~min sampling. The quoted uncertainties are based on the
  statistical uncertainties due to Poisson fluctuations and the
  uncertainties in the planet occurrence rates. For eclipsing
  binaries, there may be additional systematic uncertainties as high
  as $\approx$50$\%$ (see the text).}
\label{fig:baryield}
\end{figure*}

\begin{figure*}[bht]
\epsscale{0.70}
\plotone{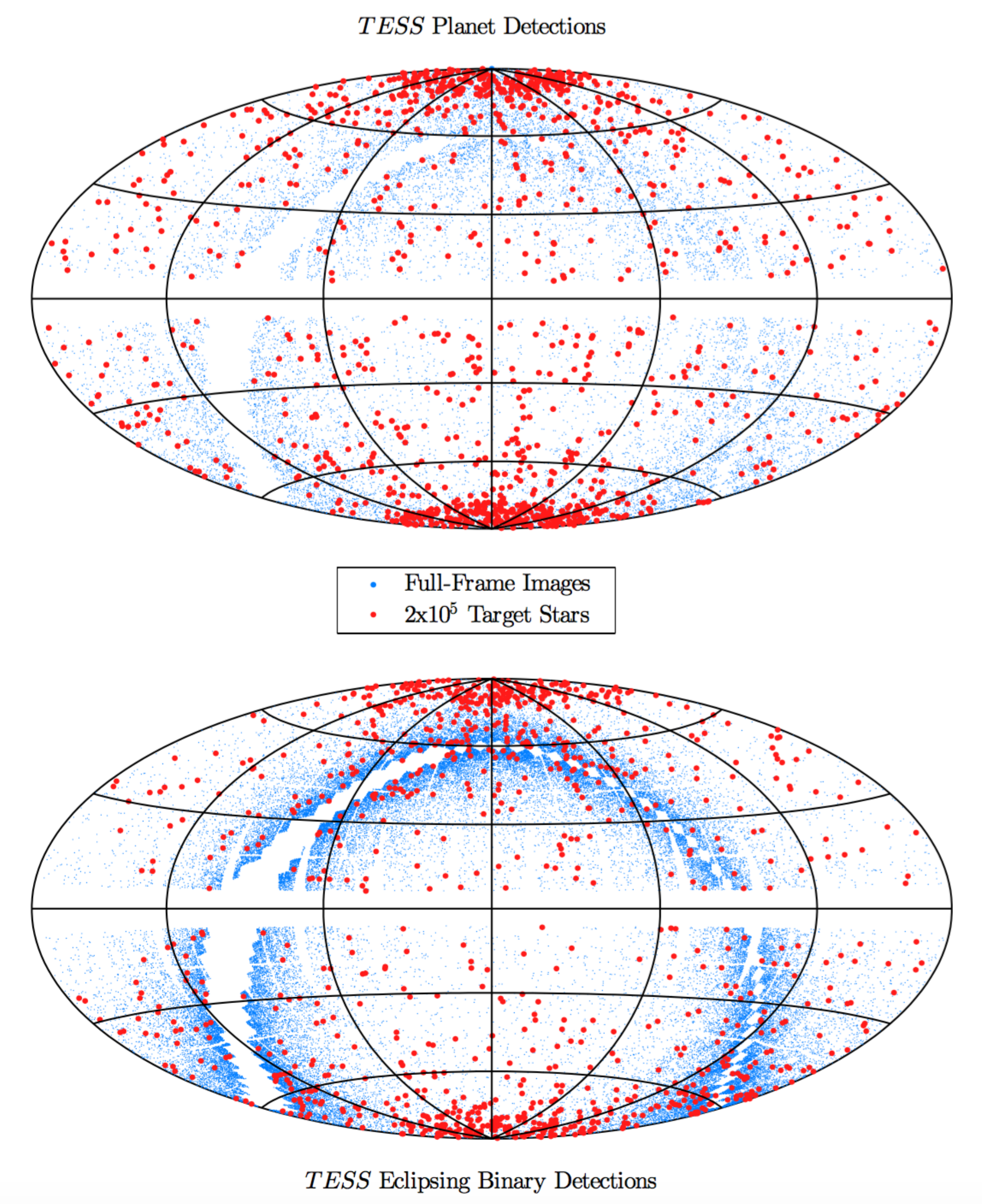}
\caption{Sky maps of the simulated \tess{} detections in
  equal-area projections of ecliptic coordinates.  The lines of
  latitude are spaced by 30$^{\circ}$, and the lines of longitude are
  spaced by 60$^{\circ}$. {\it Top.}---Planet detections. Red points
  represent planets detected around target stars (2~min cadence). Blue
  points represent planets detected around stars that are only
  observed in the full-frame images (30~min cadence).  Note the
  enhancement in the planet yield near the ecliptic poles, which
  \tess{} observes for the longest duration.  Note also that the inner
  6$^{\circ}$ of the ecliptic is not observed. {\it
    Bottom.}---Astrophysical false positive detections, using the same
  color scheme.  For clarity, only 10$\%$ of the false positives
  detected in the full-frame images are shown. (All other categories show 
  100$\%$ of the detections from one trial.) Note the enhancement in
  the detection rate near the galactic plane, which is stronger for
  false positives than for planets.}
\label{fig:map}
\end{figure*}

\section{SURVEY YIELD}
\label{sec:yield}

Having calculated the SNR for each eclipsing or transiting system, we
determine that a system is ``detected'' if the SNR~$\geq \snrcr{}$ in
the phase-folded light curve and at least 2 transits or eclipse events
are observed.  We thereby produce a simulated catalog of detected
planets and false positives.

Figure~\ref{fig:map} is a sky map in ecliptic coordinates of the
simulated detections from one trial. Figure~\ref{fig:baryield} shows
the tallies for each class of planet and false positive. 
For the $2\times 10^5$ target stars, the yields we show are the average
over five trials of the \tess{} mission; for the full-frame images, 
the yields are reported from a single trial since the computation 
time is much longer for this case.

The uncertainties that are printed in Figure~\ref{fig:baryield} (for 
planets transiting the $2 \times 10^5$ target stars) are based on the two
primary sources of statistical uncertainty: the Poisson fluctuations
in the number of detected planets and the statistical uncertainties
in the planet occurrence rates (which are partly due to Poisson
fluctuations in the {\it Kepler} sample of detected planets). We
propagate the uncertainties in the occurrence rates by running 100
trials of the simulation. In each trial, the occurrence rates were
perturbed by adding random Gaussian deviates to the quoted occurrence
rate with the standard deviation set to the quoted uncertainty in the
occurrence rate. In this way, the standard deviation in the number of
planet detections across the 100 trials is essentially the quadrature
sum of the Poisson fluctuations and the uncertainties propagated from
the input occurrence rates. Poisson fluctuations are dominant for the
categories of planets where the mean number of detected planets is
small, such as habitable-zone planets.

The preceding calculations do not take into account systematic
uncertainties. Among the sources of systematic uncertainty are the
models of galactic structure and extinction,
the stellar luminosity function, the stellar mass-radius-luminosity
relations, and any bias in the planet occurrence rates. It is beyond
the scope of this work to gauge the uncertainties in all of these
inputs and the resulting impact on the planet yield. We can, however,
make some general comments.  We expect that the uncertainties in
galactic structure and extinction will only be significant near the
galactic plane, where it will be more difficult for {\it TESS}
to detect planets due to crowding.  Regarding the stellar luminosity
function, it seems plausible that there are residual biases at the
level of $\approx$10\%, given that we found it necessary to adjust the
model luminosity function by $\approx$30\% across all absolute
magnitudes to match the various sets of observational inputs.  When
coupled with uncertainties in the stellar mass-radius-luminosity
relations, we would guess that the net impact on the planet detection
statistics is at the level of $\approx$30\%.  Regarding biases in the
planet occurrence rates upon which our simulation is based, it seems plausible that
they are of the same order as the reported statistical errors, which
have a median of $\approx$40\% across all planetary sizes and periods.
Therefore, the systematic uncertainties in the number of planet
detections could be as large as 50\%.

The number of planet detections from the full-frame images is
sufficiently large that the systematic uncertainties almost certainly
dominate over the statistical uncertainties, and therefore, the
results should probably be valid to within a factor of two.  For the
same reason, we have not reported statistical uncertainties for the
yields of astrophysical false positives. In addition to the systematic
uncertainties mentioned above, there are additional uncertainties
arising from the models for the stellar multiplicity fraction, mass
ratio distribution, and eccentricity/period distributions. Our
comparison to the Kepler eclipsing binary catalog indicates that for
low galactic latitudes these uncertainties are of order of 80$\%$ (see
Figure~\ref{fig:kep_ebs}).

\subsection{Transiting Planets}
\label{sec:planetout}

\begin{figure}[ht]
\epsscale{1.1}
\plotone{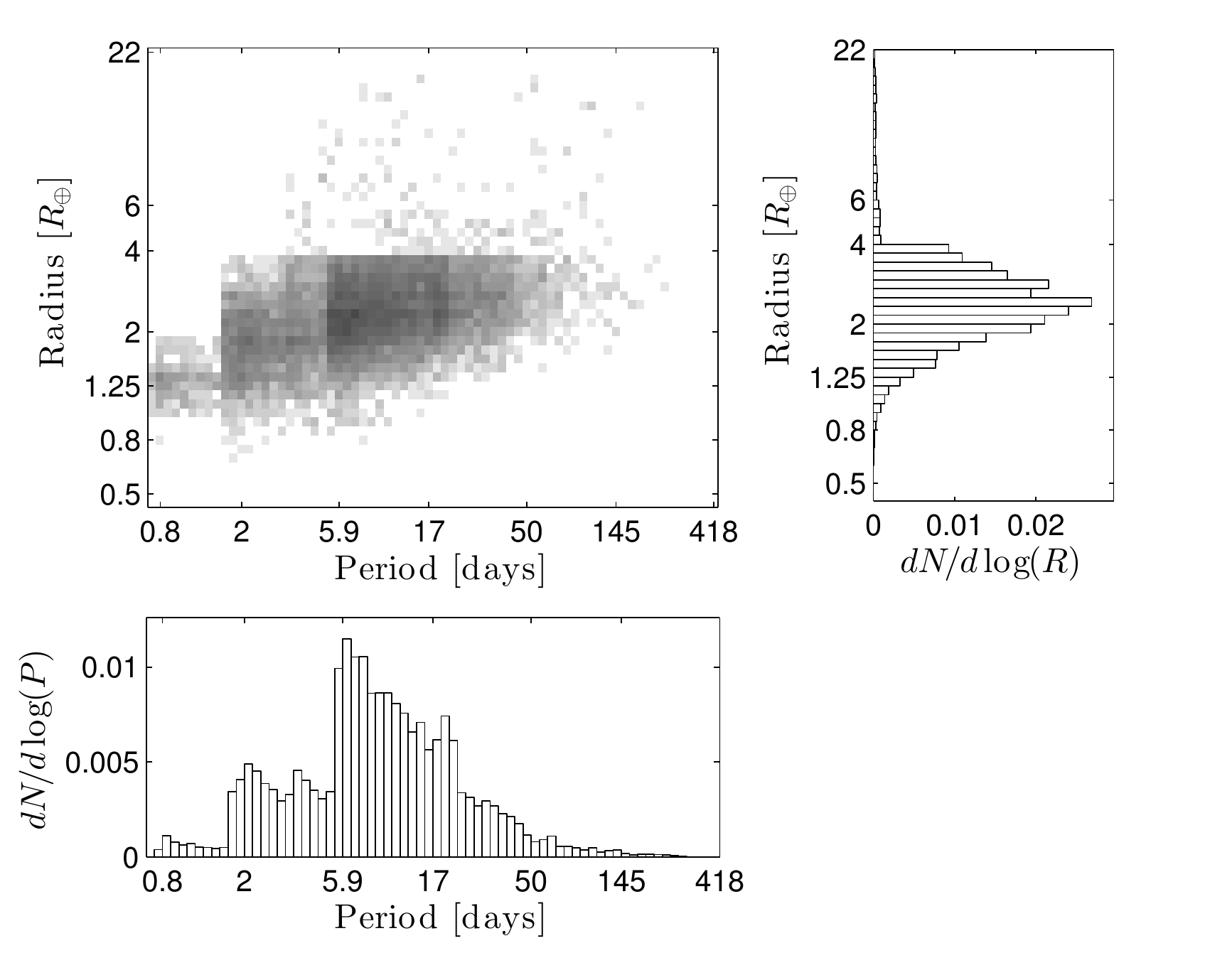}
\caption{The distribution of detected planets on the period--radius
  plane.  The shading of the 2-d histogram is the same as in
  Figure~\ref{fig:occur}.  The sawtooth patterns in the radius and
  period histograms are an artefact of the planet occurrence rates
  having coarse bin sizes in radius and period combined with the
  sensitivity of \tess{} favoring planets with larger radii and
  shorter periods.}
\label{fig:detplan}
\end{figure}

{\it Total number of detections.}---Based on five trials with the
$2\times 10^5$ target stars, we expect \tess{} to find \neartherr{}
planets smaller than 1.25~$R_{\oplus}$, \nspetherr{} planets in the
range 1.25-2~$R_\oplus$, \nsmneperr{} planets in the range
2-4~$R_{\oplus}$, and \ngiantpserr{} planets larger than 4~$R_\oplus$.
Table \ref{tbl:cat} presents the catalog of planets from one of these
five trials. Figure \ref{fig:detplan} shows the distribution of
detected planets plotted on the radius-period plane, in the same
fashion that the input planet occurrence rates were plotted in Figure
\ref{fig:occur}.

The top panel of Figure \ref{fig:map} maps the simulated planet
detections in ecliptic coordinates. Detections among the target stars
(red points) are enhanced in the vicinity of the ecliptic poles
because of the overlapping pointings they receive.
Apart from that conspicuous feature, the detections are nearly
uniformly distributed across the sky. The detections from stars that
are only observed in the full-frame images (blue dots) show a
strong enhancement near the galactic plane. This is due to the vast 
number of faint and distant stars around which giant planets can be 
detected.

\begin{figure}[ht]
\epsscale{1.1}
\plotone{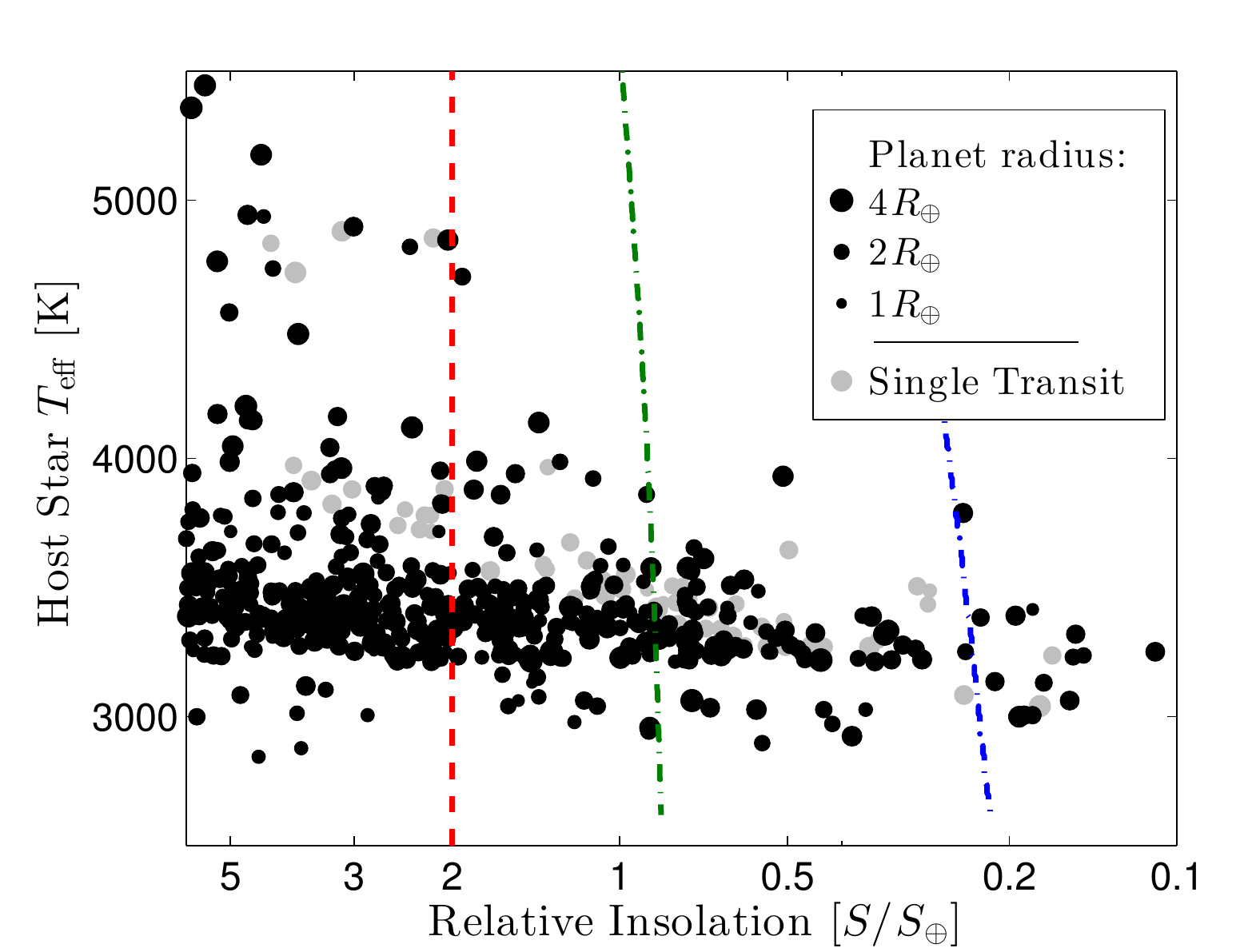}
\caption{Small planets in and near the habitable zone from one trial
  plotted against stellar effective temperature and relative
  insolation $S/S_{\oplus}$.  The dash-dot lines show the inner
  (green) and outer (blue) edges for the HZ defined in
  \citet{kopparapu13}. The vertical red dashed line indicates
  $S/S_\oplus = 2$.  The gray points represent planets for which only
  a single transit is detected. (We note that this is the only figure
  in this paper that includes single-transit detections.)}
\label{fig:hz}
\end{figure}

{\it Habitable-zone planets.}---Of the \nsmplas{} planets smaller than
2~$R_{\oplus}$, a subset of \nhzerr{} have a relative insolation on
the range $0.2 < S/S_\oplus < 2$ and are therefore near the habitable
zone. We also expect a smaller subset of \nkopphz{} to be within the
more restricted zone defined by \citet{kopparapu13}. This definition
of the habitable zone extends approximately from $0.2 < S/S_\oplus <
1$, with the exact bounds depending on stellar effective temperature.
Figure~\ref{fig:hz} shows the distribution of $S/S_\oplus$ and $\teff$
for the simulated detections in the vicinity of the habitable
zone. Because the sensitivity of \tess{} favors short periods, the
potentially-habitable planets must orbit low-mass, cool stars with
$\teff \lesssim 4000$~K.  Furthermore, the yield of such planets
depends strongly upon the definition of the inner edge of the
habitable zone, but much less so upon the outer edge.

{\it Small planets with measurable masses.}---The smallest planets
will be of particular interest for mass measurement since there are
presently very few small (and potentially rocky) planets with measured
masses and sizes. Among the \nearths{} simulated planets smaller than
1.25~$R_\oplus$, the median period is 2.1 days, and the median stellar
effective temperature is 3450~K. The median $I_C$ magnitude is 11.6.

\begin{figure*}[htb]
\epsscale{0.8}
\plotone{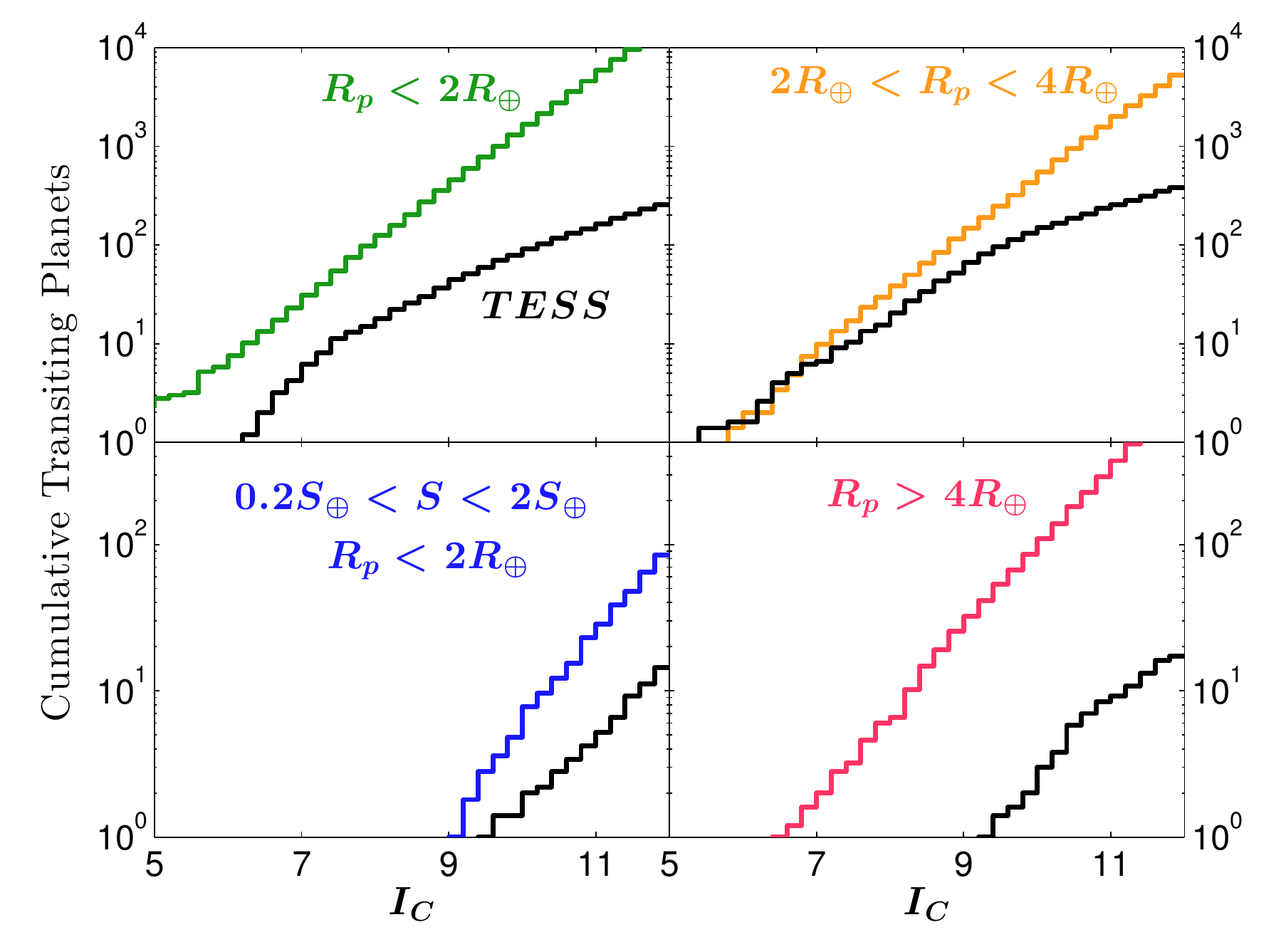}
\caption{Completeness of the \tess{} survey. For each category of
  planet, we plot the cumulative number of transiting planets as a
  function of the limiting apparent magnitude of the host star.  Only
  planets with $P<20$ days and host stars with $\teff<7000$~K and
  $R_\star<1.5 R_\odot$ are considered. The colored lines show the
  distributions for all transiting planets in the simulation; the
  black lines show the simulated \tess{} detections. The completeness
  is partly limited from the selection of the $2 \times 10^5$ target
  stars, which is evident for $R_p>4R_{\oplus}$ planets.}
\label{fig:complete}
\end{figure*}

{\it Survey completeness.}---The degree of completeness of the \tess{}
survey can be assessed by comparing the simulated planet detections
against the total number of transiting planets on the sky (as
discussed in Section \ref{sec:planetin}). Plotted in Figure
\ref{fig:complete} are the cumulative numbers of transiting planets as
a function of the limiting apparent magnitude of the host star. We
make the comparison for short-period planets around Sun-like and
smaller stars for planets of different sizes as well as small planets
near the HZ. For planets with $R_p<2 R_\oplus$, the completeness of
the \tess{} survey is limited by instrumental noise. For planets with
$R_p>4 R_\oplus$, the completeness is limited by the maximum number of
target stars ($2 \times 10^5$).

{\it Diluting flux.}---Whenever the photometric aperture contains flux
from neighboring stars, the measured transit depth will be smaller
than it would be if the star were observed in isolation. If this
effect is not taken into account (by using observations with higher
angular resolution), then the planet's radius will be underestimated.
The source of the ``diluting flux'' can be a star that is
gravitationally bound to the target star, or it can be one or more
completely unrelated stars along the same line-of-sight. 
In our simulation, we find that \underadten{} of detected planets suffer 
dilution by more than $>21\%$, making them vulnerable to radius
underestimation by $>10\%$. For \underadtwy{} of planets, 
the radii could be underestimated by $>20\%$.  
We note that we do not consider cases of underestimated
planet sizes to be ``false positives'', in contrast to
\citet{fressin2013}. Those authors considered the detection of
transits with significant dilution to be a false positive because
they were concerned with determining the occurrence rates of planets
as a function of planet radius.

A separate scenario in which the transit depth can be diluted is when
the transiting planet is actually orbiting a background star rather
than the target star. Simulating these background transiting planets
is a more computationally challenging problem which we conducted
separately from the main simulations. We generated planets around the
background stars represented by in ``faint'' star catalog and
simulated the detection of the transiting planets blended with
target stars. We found this type of transit detection to be very
rare. Of the 2$\times10^5$ target stars, we find that only $\sim$1
planet transiting a background star will be detectable with
\tess{}. In the 30-minute full-frame images, approximately 70 such
planets might be detected.  The transit depths of these planets must
be very deep to overcome the diluting flux of the brighter target
star. In the simulations, the median radius of blended transiting
planets is 17$R_{\oplus}$. Our conclusion is in agreement with those
of \citet{fressin2013}, who found that transits of background stars
are a less important source of detections than transits of planets
around gravitationally bound companion stars (see their Figure~10).

{\it Single-transit detections.}---In a few notable cases,
the SNR of a transit exceeds the threshold of $\snrcr$, but only a
single transit is observed. We expect \nsingles{} such planets to be
detected with one transit. These are not counted as detections in the
tallies given above, but they are included in Figure~\ref{fig:hz} as
gray points.  These planets have longer periods and lower equilibrium
temperatures than the rest of the \tess{} sample. There may even be
additional single-transit detections from planets with orbital periods
exceeding one year, which we have not modeled at all. Although the
periods will not be well-constrained using \tess{} data alone, and the
probability of a ``detection'' being a statistical fluke is higher, it
may still be worthwhile to conduct follow-up observations of these
stars.  The single-transit detections have a median planet size of
$\sim$3~$R_\oplus$, a median orbital period of $\sim$30~days, and a
median insolation of 1.9~$S_\oplus$.

\subsection{False positives}
\label{sec:ebout}

Among the $2\times 10^5$ target stars, \tess{} detects \nebtot{}
eclipsing binary systems along with the transiting planets.  The
uncertainty in this figure is based only on the Poisson fluctuations;
we acknowledge that the true uncertainty is likely to be significantly
larger. Based on our comparison with the {\it Kepler} eclipsing binary
catalog (see Section~\ref{sec:ebin}), the uncertainty may be as large
as 80\% for relatively low galactic latitudes.

The false-positives can be divided into the following cases:
\begin{enumerate}
\item Eclipsing Binary (EB): The target star is an eclipsing binary
  with grazing eclipses. There are \nebs{} detections of EBs.
\item Hierarchical Eclipsing Binary (HEB): The target star is a triple or quadruple
  system in which one pair of stars is eclipsing. There are \nhebs{}
  detections of HEBs.
\item Background Eclipsing Binary (BEB): The target star is blended
  with a background eclipsing binary. There are \nbebs{} detections of
  BEBs.
\end{enumerate}
These tallies are also illustrated in Figure \ref{fig:baryield}.
The bottom panel of Figure~\ref{fig:map} shows a sky map of the
astrophysical false positives in the same coordinate system as the
top panel. The surface density of false positives is a much stronger
function of galactic coordinates than the density of planet
detections, for binary eclipses are deeper than planetary transits
and can be detected out to greater distances. The period and depth
distributions of the eclipsing binary population is discussed in
Section \ref{sec:pd}.

\section{DISTINGUISHING FALSE POSITIVES FROM PLANETS}
\label{sec:fp}

Experience has shown that the success of a transit survey depends
crucially on the ability to distinguish transiting planets from
astrophysical false positives. Our simulations suggest that for
\tess{}, the number of astrophysical false positives will be
comparable to the number of transiting planet detections. In many
cases, it will be necessary (or at least desirable) to undertake
ground-based follow-up observations to provide a definitive
classification.

However, there will also be useful clues within the \tess{} data that
a candidate is actually an eclipsing binary, even before any follow-up
observations are undertaken. These clues are: (1) ellipsoidal
variations, (2) secondary eclipses, (3) lengthy ingress and egress
durations, or (4) centroid motion associated with the eclipse
events. In this section, we investigate the prospects for using these
four characteristics to identify false positives with \tess{} data
alone. Specifically, we determine the number of cases, summarized in
Table \ref{tbl:fp}, for which any of these characteristics can be
measured with an SNR of 5 or greater.  This statistic indicates that
the information will be available to help make the distinction between
planet and false positive. The next step would be to combine all the
measurable characteristics in a self-consistent manner and attempt to
arrive at a definitive classification. This is a complex process which
we have not attempted to model here.

\subsection{Ellipsoidal Variations}

\begin{figure}[htb]
\epsscale{1.0}
\plotone{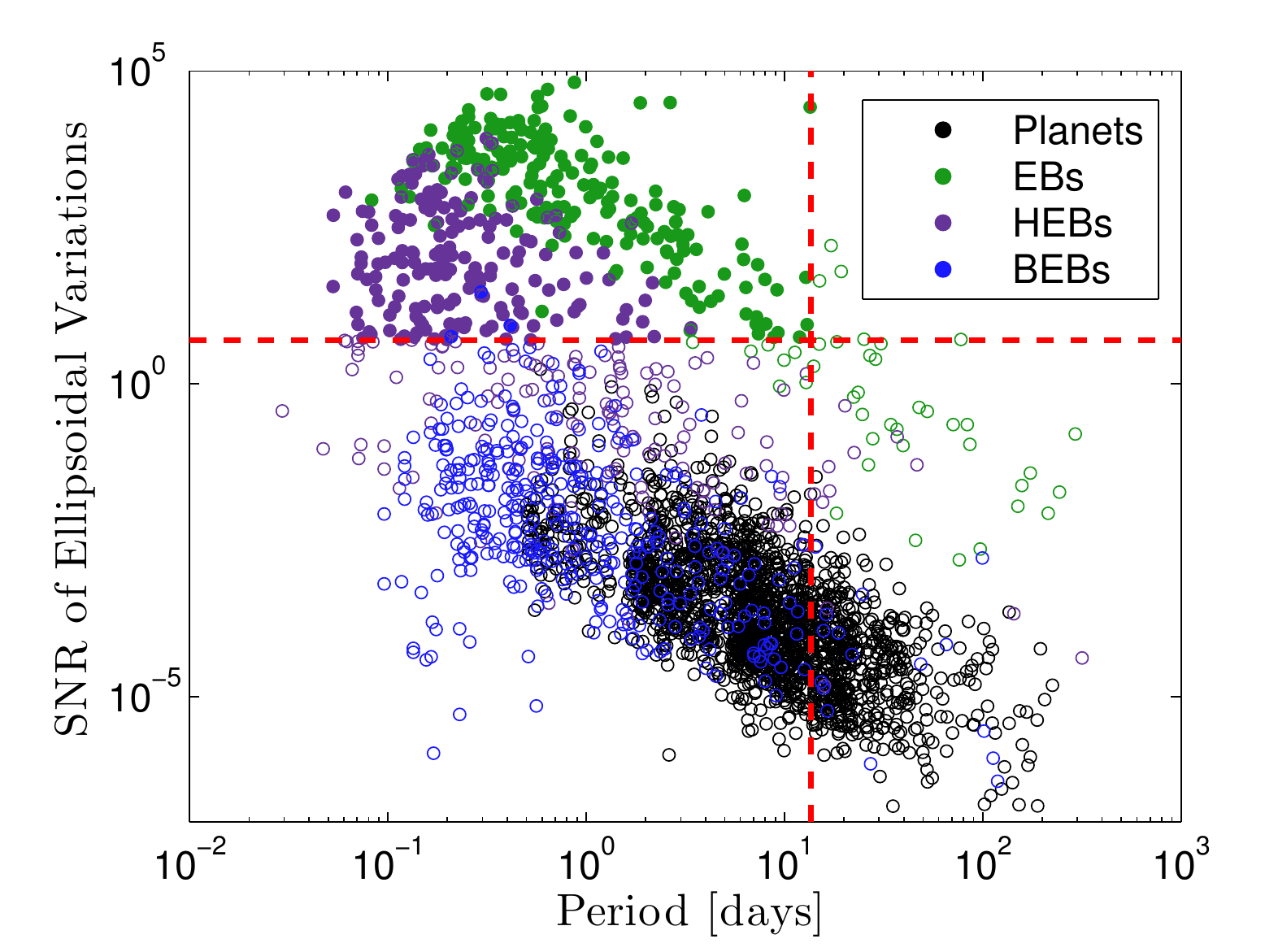}
\caption{Ellipsoidal variations of the primary star among the
  simulated \tess{} detections. Short-period systems give larger
  ellipsoidal variations. We consider the variations to be detectable
  if the semi-amplitude is greater than 10~ppm and the SNR exceeds 5
  (horizontal dashed line). We also require the orbital period of the 
  system to be shorter than the orbital period of \tess{} (vertical dashed
  line) due to systematic errors. A significant number of EBs and HEBs 
  can be identified on this basis. Only a small number of BEBs, 
  and zero planets, give rise to
  detectable ellipsoidal variations.}
\label{fig:ellipsoidal}
\end{figure}

The members of a close binary exert strong tidal gravitational forces
on one another, causing their photospheres to deform into ellipsoids.
These deformations lead to ellipsoidal variations in the light
curve. A model for these photometric variations was presented by
\citet{morris}. \citet{mazeh} gave a simple expression for the
dominant component, which has a period equal to half of the orbital
period, and a semi-amplitude
\begin{equation}
\frac{\Delta\Gamma_1}{\Gamma_1} = 0.15~\frac{(15+u_1)(1+\tau)}{(3-u_1)}q\left(\frac{R_{1}}{a}\right)^2\sin^2 i,
\end{equation}
where $R_1$ is the primary radius, $a$ is the orbital distance, $i$ is
the orbital inclination, $u_1$ is the linear limb-darkening
coefficient, $\tau$ is the gravity-darkening coefficient, and $q$ is
the mass ratio.  To estimate the amplitude of this effect for our
simulated \tess{} detections, we adopt an appropriate value of $u_1$ for each
star using the tables of \citet{claret2012} and \citet{claret2013},
which come from the PHOENIX stellar models. For gravity darkening,
we use a value of $\tau=0.32$ for all stars, which is thought to be
appropriate for stars with convective envelopes \citep{lucy1967}.

The formal detection limits for ellipsoidal variations are quite low
because the signal is present throughout the entire light curve
rather than being confined to eclipses of a narrower duration.
 Since the period and phase are fixed from the
observed eclipses, we model the detection of the ellipsoidal
variations as a cross-correlation of the light curve with a cosine function of the
appropriate period.
If the fractional uncertainty in flux of each data point is
$\sigma$, and the total number of data points is $N$,
then the SNR of ellipsoidal variations is
\begin{equation}
\mathrm{SNR}_{\rm EV} =
\frac{\Delta\Gamma_1}{\Gamma_1}\frac{\sqrt{N}}{D\sigma\sqrt{2}}.
\end{equation}
Here, $D$ denotes the dilution of the target star in the photometric
aperture, which is defined in Section \ref{sec:optimal_aperture}. Due
to this factor, ellipsoidal variations from BEBs are more difficult to
detect since their eclipses are usually more diluted than EBs and
HEBs. The factor of $\sqrt{2}$ arises from the RMS value of a cosine
function.

It seems likely that correlated noise will prevent the detection limit
from averaging down to extremely low values as the duration of
observations is extended. Somewhat arbitrarily, we require the
semi-amplitude of the ellipsoidal variations to exceed 10~ppm, in
addition to the criterion SNR$_{\rm EV} > 5$, to be counted as
``detectable.''  We also require that the orbital period of the
binary, which is twice the period of ellipsoidal variations, is
shorter than one spacecraft orbit (13.6 days) out of concern that
thermal or other variations of the satellite will induce systematic
errors with a similar frequency. Under these detection constraints,
shown in Figure~\ref{fig:ellipsoidal}, ellipsoidal variations are
detected for \nebellip{} of the eclipsing binaries in the simulation.
The majority of these are grazing-eclipse binaries rather than HEBs or
BEBs.  The results are summarized in the second column of Table
\ref{tbl:fp}.

\subsection{Secondary Eclipse Detection}

Another key difference between eclipsing binaries and transiting
planets is that the secondary star in a binary is more luminous than a
planetary companion. This distinction is somewhat blurred when
comparing brown-dwarf and hot-Jupiter companions but is quite clear
between ordinary stars and lower-mass planets. If the two stars in a
binary have nearly the same surface brightness, then the depths of the
primary and secondary eclipses will be indistinguishable. In this
case, the system might appear to be a planet with an orbital
period equal to half of the true orbital period of the
binary. However, if the surface brightnesses of the stars differ and
both eclipses are detected with a sufficiently high SNR, then the
secondary eclipse can be distinguished from the primary eclipse and
the system can be confidently classified as an eclipsing binary.

\begin{figure}[htb]
\epsscale{1.0}
\plotone{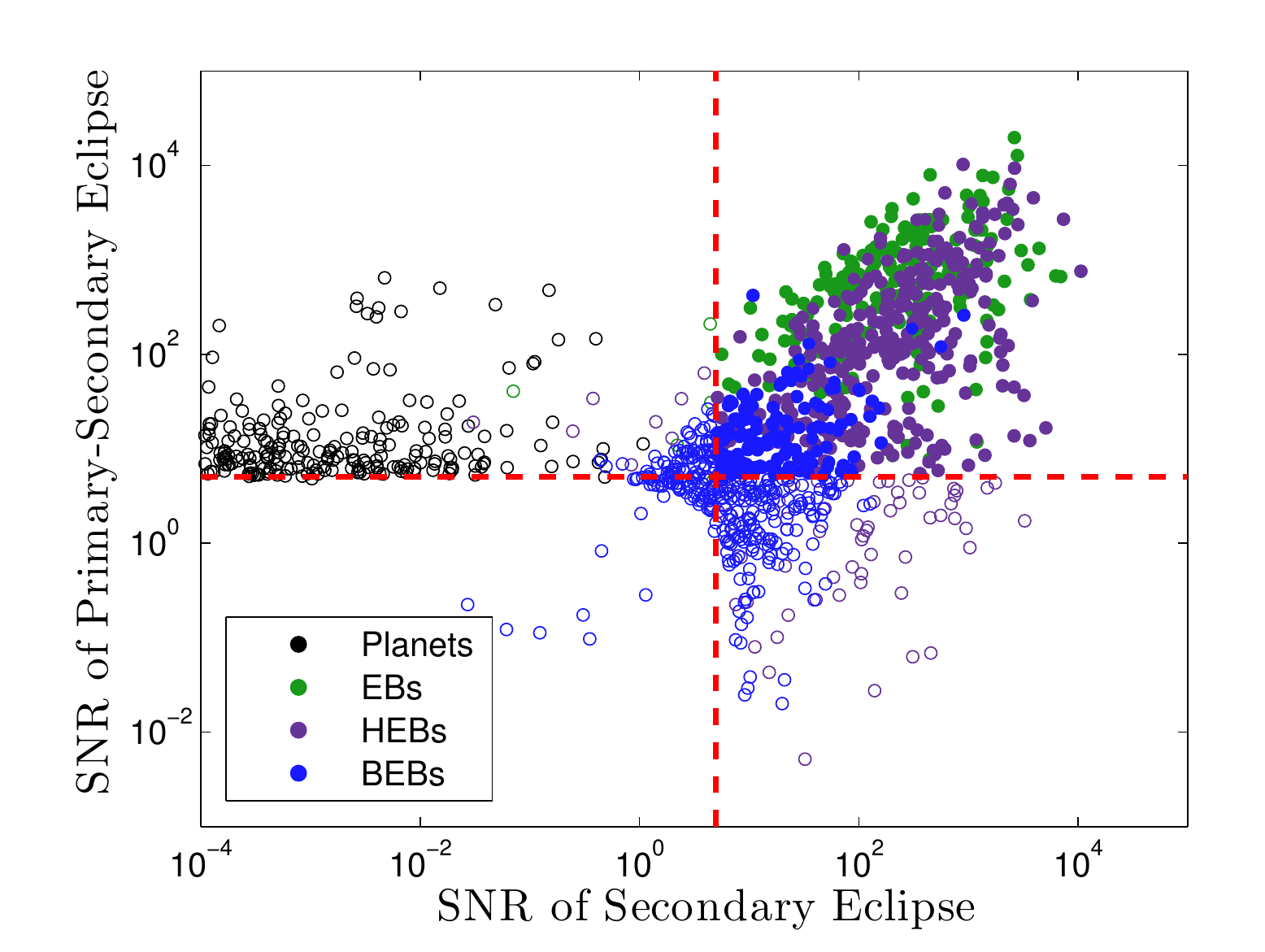}
\caption{Distinguishing secondary eclipses from primary eclipses based
  on \tess{} photometry. The vertical dashed red line shows where the
  secondary eclipses can be detected at SNR$_2>5$. The horizontal dashed
  red line shows where the difference in eclipse depths can be measured
  with SNR$_{1-2}>5$. Points in the upper-right quadrant of the plot meet 
  conditions, so the secondary eclipse can be distinguished from the primary eclipse. For \nebsec{} of the eclipsing binaries that \tess{} detects in the
  simulation, it is possible to classify them as false positives from the \tess{} data alone.}
\label{fig:dep}
\end{figure}

To estimate the number of cases for which the primary and secondary
eclipses are distinguishable, we identify the simulated
systems for which signal-to-noise of the secondary eclipses, SNR$_2$, is $>5$, 
and the SNR in the difference between the primary and secondary eclipse
depths, SNR$_{1-2}$, is also $>5$. The latter quantity is calculated as
\begin{equation}
\mathrm{SNR}_{1-2} = \frac{\delta_1-\delta_2}{\sqrt{\sigma_1^2 + \sigma_2^2}},
\end{equation}
where $\delta_{1,2}$ denote the depths of the eclipses and $\sigma_{1,2}$ denote the noise in the relative flux over the observed duration of each eclipse.
Figure \ref{fig:dep} shows the detectability of secondary eclipses by plotting SNR$_{1-2}$ versus SNR$_2$. The secondary eclipse can be distinguished from the primary eclipse for the systems that lie in the upper-right quadrant of the plot.

The results are also summarized in the third column of Table~\ref{tbl:fp}. 
A majority of the false positives have detectable secondary eclipses that are distinguishable in depth from the primary eclipses. 
The notable exceptions include the HEBs in which the eclipsing pair consists of
equal-mass stars ($q\approx 1$).  In such cases, $\delta_1 \approx
\delta_2$ and it is impossible to distinguish between primary and
secondary eclipses. For the BEBs, the difficulty is that the eclipse
depths are often strongly diluted and the secondary eclipses are not
detectable. Most planets are too small and faint to produce detectable
secondary eclipses in the \tess{} bandpass.  In the simulations, the
fraction of detected planets with detectable secondary eclipses is
only \nplasec{}.

\subsection{Ingress and Egress Detection}

Eclipsing binaries can also be distinguished from transiting planets
based on the more prolonged ingress and egress phases of stellar
eclipses. As above, we adopt an SNR threshold of 5 for the
ingress/egress phases to be detectable. The average ``signal'' during
ingress and egress is half the maximum eclipse depth, and the
``noise'' is calculated for the combined durations of ingress and
egress. In order to ensure that the ingress/egress can be temporally
resolved, we require the duration of the ingress or egress to be more
than twice as long as the duration of an individual data sample (2~min
for the target stars and 30~min for the rest of the stars).

Since transiting planets generally have ingrees or egress phases
lasting a few minutes, \tess{} will only be able to detect the
ingress/egress for a small fraction ($\approx$10\%) of transiting
planets observed with 2~min sampling. Only large planets observed in
the 30~min. FFIs would have resolvable ingress/egress.  However, the
the ingress/egress phases of eclipsing binaries are more readily
detectable.

We note that detection of the ingress/egress alone does not classify a
signal as an eclipsing binary. One would next examine the period and
shape of the eclipse signals to determine whether the radius of the
eclipsing body is consistent with the observed depth.

\begin{figure}[htb]
\epsscale{1.0}
\plotone{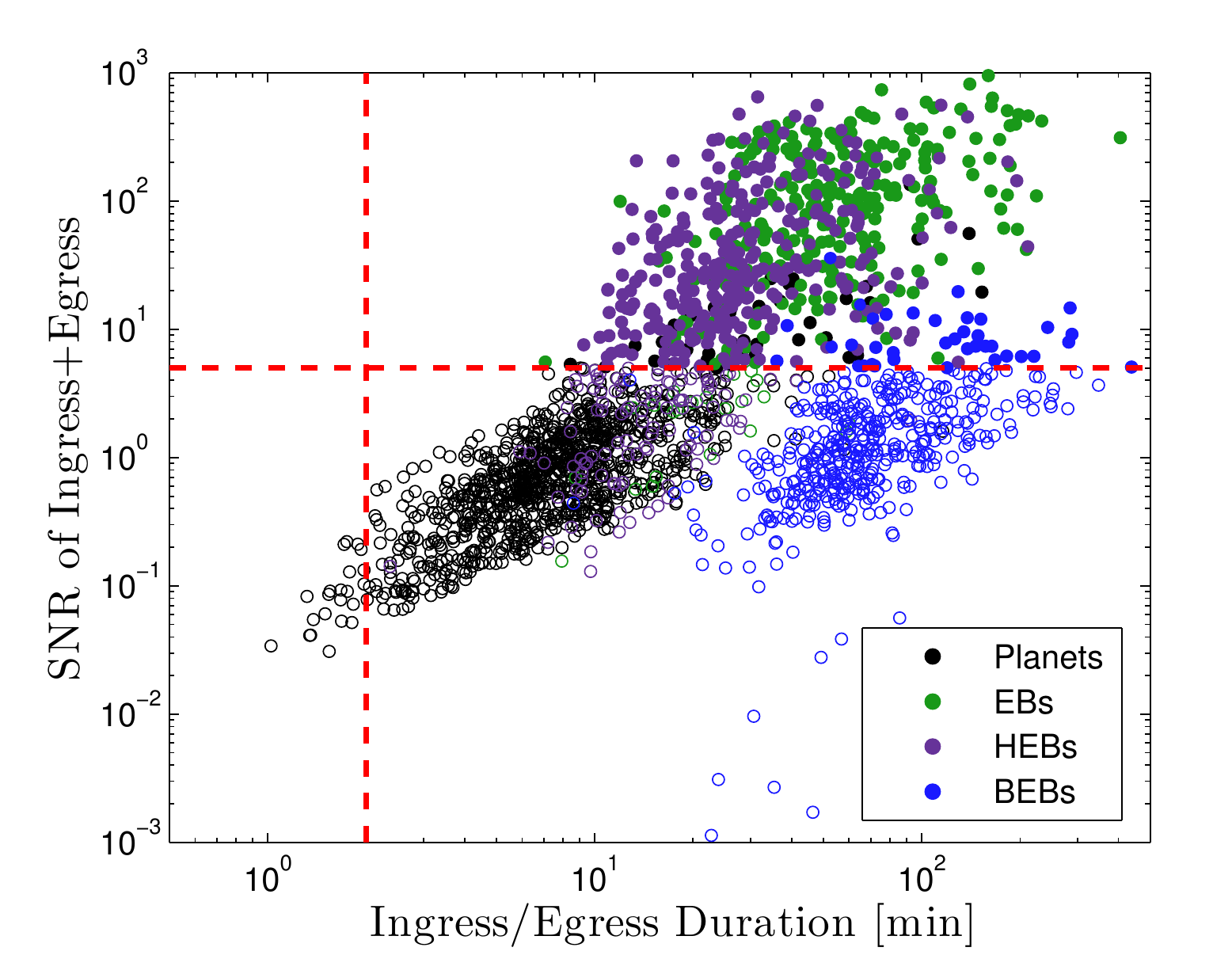}
\caption{Detectability of the ingress and egress phases of eclipses
   observed with \tess{}. We require the time-averaged ingress/egress depth 
   (half of the full depth) must be detectable with SNR > 5 from data 
   obtained during ingress/egress (horizontal dashed line).
   Also, we require the ingress/egress duration to be longer than the 
   2 min averaging time of each sample (vertical dashed line). 
   Filled circles represent systems for which the ingress/egress
   are detectable according to these criteria.}
\label{fig:gress}
\end{figure}

Figure \ref{fig:gress} illustrates the detection of ingress/egress for
planets and false positives. The fourth column of Table~\ref{tbl:fp}
summarizes the results. Approximately 70\% of the eclipsing binary
systems that \tess{} detects among the target stars might be
classified as false positives by virtue of a lengthy ingress or egress
duration. For stars that are only observed at a 30~min cadence, this
method is not as effective.

\vfill
\subsection{Centroid Motion}
\label{sec:centroid}

Another diagnostic of false positives, particularly background
eclipsing binaries, is the centroid motion that accompanies the
photometric variations. If there are detectable shifts in the centroid
of the target star during transit or eclipse events, it is more likely
that the target is a blended eclipsing binary rather than a transiting
planet or an eclipse of the target star itself. Transits or eclipses
of the target star can still have significant centroid motion if
another bright star is blended with the target.

With real data, one could interpret the amplitude and direction of the
measured centroid shift using the known locations of neighboring stars
in order to determine the most likely source of the photometric
variations. This is a complicated process to simulate, so we simply
investigate the issue of the detecting the centroid shift. As verified
in our simulations, the systems with detectable centroid shifts are
much more likely to be false positives than transiting planets.

We simulate the detectability of centroid shifts by calculating the
two-dimensional centroid (center-of-light) of the target star, $C_x$
and $C_y$, within the $8\times 8$ synthetic images described in
\ref{sec:image}.  We calculate the centroids both during and outside
of the loss of light to find the magnitude and direction of the
centroid shift.  Next, we calculate the uncertainty in the centroid
$\sigma_{Cx}$ and $\sigma_{Cy}$, which stems from the photometric
noise of each pixel. If each pixel $(i,j)$ has coordinates $(x,y)$,
and its photometric noise relative to the total flux is denoted by
$\sigma_{i,j}$, then the noise propagates to the centroid measurement
uncertainty through
\begin{equation}
\sigma_{Cx}^2 = \sum\limits_i (x_i-C_x)^2 \sigma_{ij}^2 
\quad {\rm and} \quad
\sigma_{Cy}^2 = \sum\limits_j (y_j-C_y)^2 \sigma_{ij}^2.
\end{equation}
In an analogous fashion to determining the optimal photometric
aperture, we select the pixels that maximize the signal-to-noise ratio
of the centroid measurement. Finally, we project the $x$ and $y$
centroid uncertainties in the direction of the centroid shift. The
signal-to-noise ratio of the centroid measurement is the magnitude of
the centroid shift divided by the centroid uncertainty projected in
the direction of the centroid shift. We consider a centroid shift to
be detectable if the signal-to-noise is 5 or greater.

In practice, the centroid measurement uncertainty could be much larger
if the spacecraft jitter does not average down during the hour-long
timescales of transits and eclipses.  On the other hand, monotonic
drifts in the spacecraft pointing during a transit or eclipse are less
likely to impact the centroid measurement since the motion is common
to all stars.

\begin{figure}[htb]
\epsscale{1.0}
\plotone{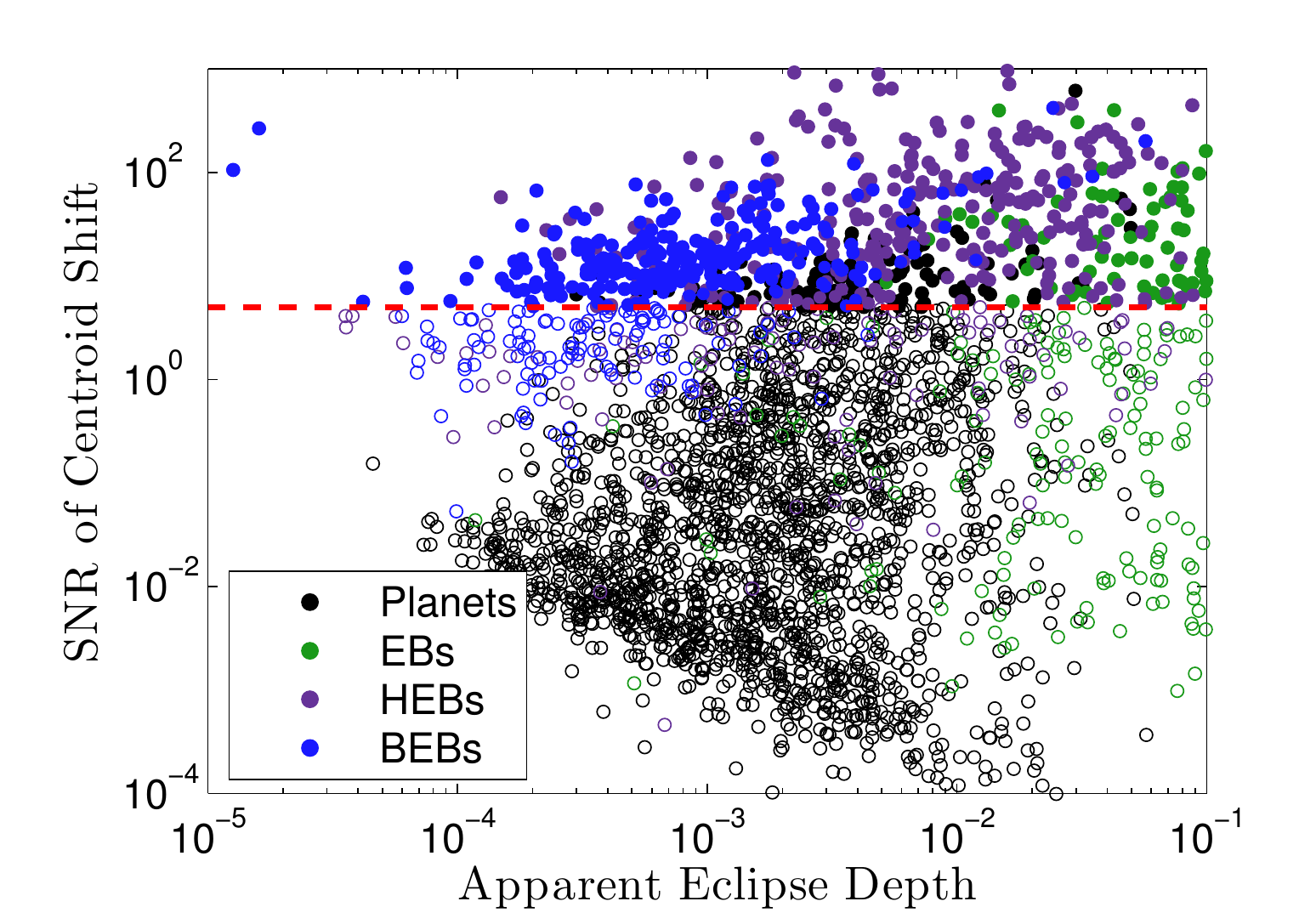}
\caption{Measurement of the shift in the centroid of the target star
  during eclipses for various types of detections. Eclipses from
  background binaries give the largest centroid shifts for a given
  depth. If the \tess{} data permits a measurement of the centroid
  shift with SNR~$>5$, we consider the shift to be detectable and plot
  it with a filled circle.}
\label{fig:centroid}
\end{figure}

We find that centroid shifts can be detected for \hbebcenpct{} of the
BEBs and HEBs. These results are illustrated in Figure
\ref{fig:centroid} and summarized in column 5 of
Table~\ref{tbl:fp}. The BEBs have a higher fraction of detectable
centroid shifts from the larger angular separations between the
eclipsing system and the target star. Only \plancenpct{} of planet
transits produce a detectable centroid shift.

\begin{deluxetable}{rccccccc}
\tabletypesize{\scriptsize}
\tablecolumns{6}
\tablewidth{0pt}
\tablecaption{Methods of distinguishing false positives from transiting planets.  \label{tbl:fp}}

\tablehead{
\colhead{} & 
\colhead{$N$\tablenotemark{a}} &
\colhead{Ellip.\tablenotemark{b}} &
\colhead{Sec. Ecl.\tablenotemark{b}} &
\colhead{In/Egress\tablenotemark{b}} &
\colhead{Centroid\tablenotemark{b}} & 
\colhead{Any\tablenotemark{c}} 
}

\startdata
EB     & 250 & 79.9 & 80.2 & 92.5 & 31.7 & 98.6 \\ 
HEB    & 410 & 43.1 & 73.4 & 74.5 & 71.2 & 93.0 \\ 
BEB    & 443 & 0.8 & 30.4 & 10.9 & 69.1 & 74.1 \\ 
\hline
\\
All FP & 1103 & 34.4 & 57.7 & 53.0 & 54.2 & 86.7 \\ 
\cutinhead{Planets\tablenotemark{d}}
$<4~R_{\oplus}$ &  1667 & 0.0 & 0.0 & 1.9 & 6.3 & 1.9 \\ 
$>4~R_{\oplus}$ &   67 & 0.0 & 0.3 & 40.7 & 9.6 & 40.7  
\enddata

\tablenotetext{a}{Mean number of each type of system that is detected.}

\tablenotetext{b}{The central four columns indicate the percentage of
  systems each with detectable ellipsoidal variations, secondary
  eclipses, ingress and egress, and centroid motion.}

\tablenotetext{c}{The percentage of systems for which at least one of
  these four characteristics is detectable.}

\tablenotetext{d}{Same, but restricted to planets larger or smaller
  than $4~R_{\oplus}$. For large planets the ingress/egress and the
  secondary eclipses are occasionally detectable.}

\end{deluxetable}

\break
\subsection{Imaging}
\label{sec:fpphot}

As shown in Table \ref{tbl:fp}, the simulations suggest that blended
eclipsing binaries are the type of false positive that is most
difficult to identify based only on \tess{} data. Assuming that all of
the false-positive tests described in the previous sections are
applied, approximately \nfpimg{} of the \nebtot{} false positives
would fail to be identified. The large majority (\bebimgpct{}) of
these more stubborn cases are BEBs.

If archival images or catalogs do not reveal a system in the vicinity
of a \tess{} target star that is consistent with any measurable
centroid motion, then additional imaging is needed.  An effective way
to identify these BEBs is through ground-based imaging with higher
angular resolution than the \tess{} cameras. A series of images
spanning an eclipse could reveal which star (if any) is the true
source of variations. Due to the large pixel scale of the \tess{}
optics, it will not be difficult to improve upon the angular
resolution with ground-based observations. Even modest contrast and a
well-sampled PSF can resolve many ambiguous cases.

\begin{figure}[htb]
\epsscale{1.0}
\plotone{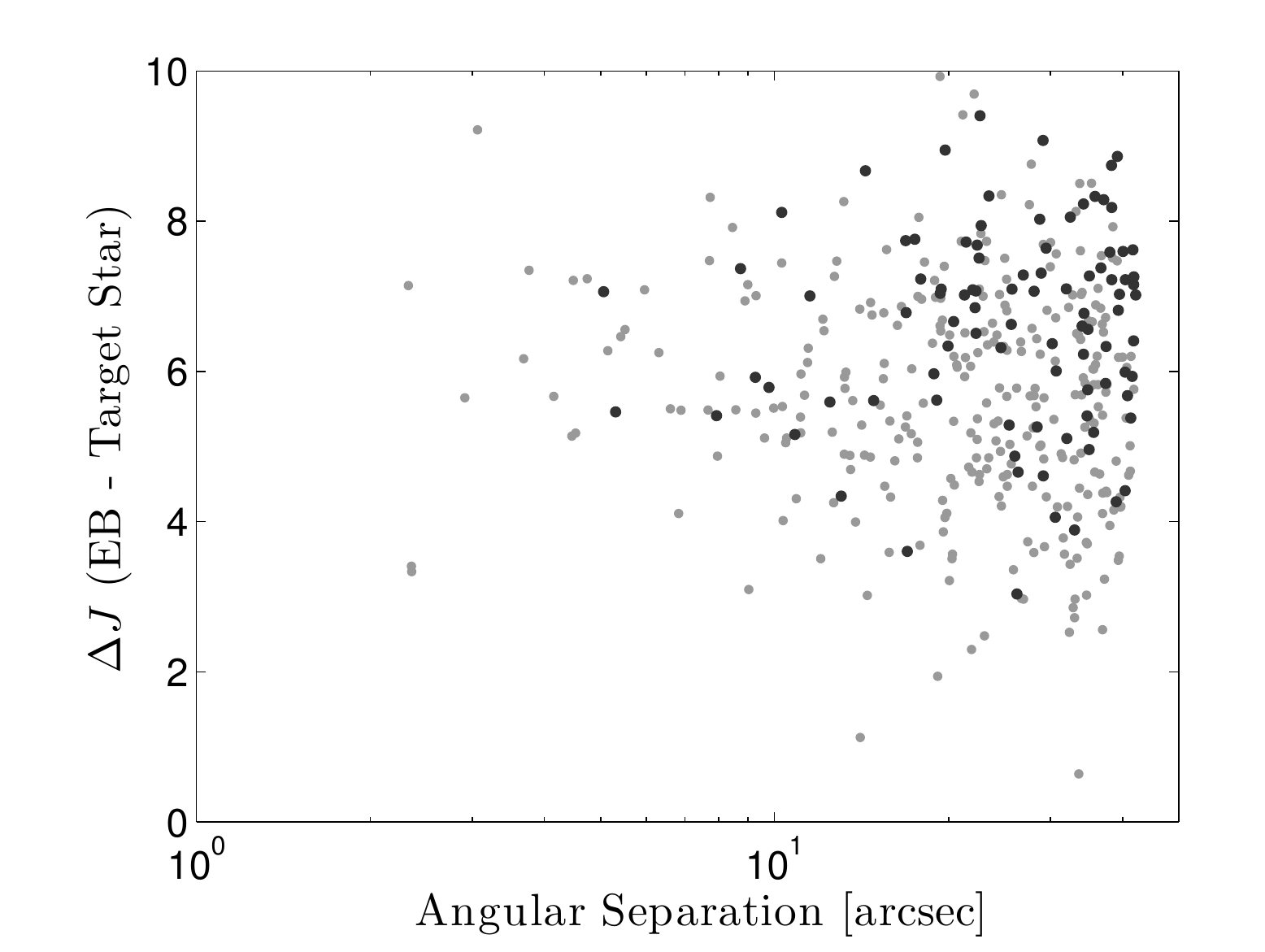}
\caption{Magnitude differences and angular separations between BEBs and 
the associated target star. Gray dots show the BEBs for which the TESS 
photometric data already provides some evidence that the source is a 
false positive through ellipsoidal variations, secondary eclipses, 
ingress/egress, or centroid motion. Black dots are the BEBs for
which none of those effects are detectable; ground-based images 
spanning an eclipse might be the most useful discriminant in such cases.}
\label{fig:contrast}
\end{figure}

Figure~\ref{fig:contrast} illustrates the requirements on angular
resolution and contrast. For each BEB, we have plotted the angular
separation and the $J$-band magnitude difference between the BEB and
the target star. Natural-seeing images with $1\arcsec$ resolution
would be sufficient to resolve all of the simulated BEBs. In more
difficult cases, adaptive optics might be necessary to enable high
contrast.

\begin{figure}[htb]
\epsscale{1.0}
\plotone{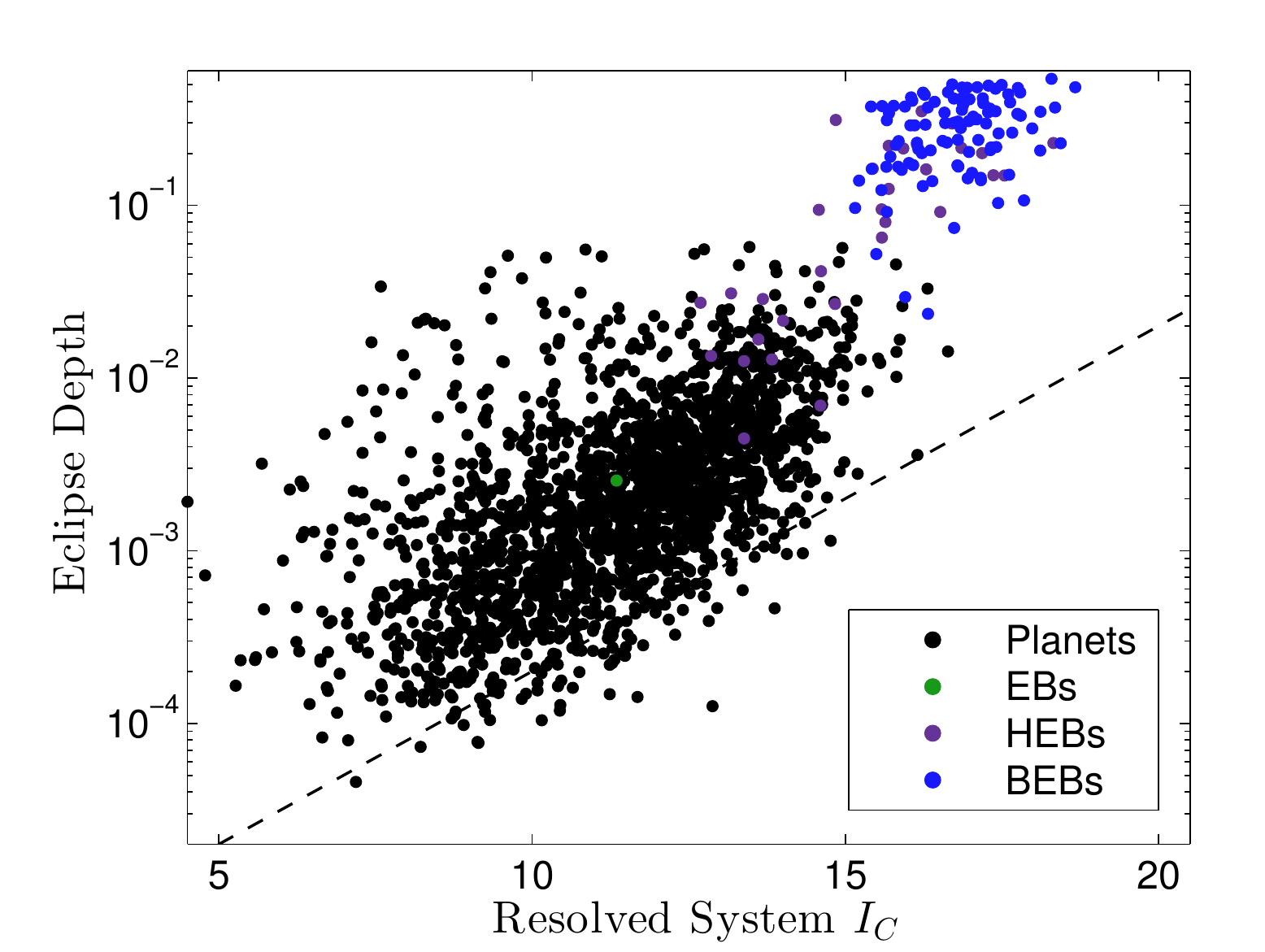}
\caption{Follow-up photometry of the \tess{} candidates, which are a
  mixture of planets and astrophysical false positives. We only show
  the false positives that cannot be ruled out from the \tess{}
  photometry, which are primarily BEBs. In order to show the
  photometric precision that is required to detect a transit or
  eclipse, we plot the depth against apparent magnitude. We assume
  that the BEBs are resolved from the target star (see Figure
  \ref{fig:contrast}), so the full eclipse depth and apparent
  magnitude of the binary are observable. An observation limited by
  photon-counting noise designed to detect most of the planets (dashed
  line) is sufficient to detect the eclipsing binaries as well.}
\label{fig:phot}
\end{figure}

Figure~\ref{fig:phot} shows the photometric requirements to detect
the planets as well as BEBs and other eclipsing systems for which the \tess{} photometry 
cannot distinguish whether the candidate is a false positive. We plot the 
eclipse depth against apparent system magnitude to indicate the photometric
precision that is required of the facilities performing these observations.

\subsection{Statistical Discrimination}
\label{sec:pd}

The false positives and transiting planets have significantly
different distributions of orbital period, eclipse/transit depth, and
galactic latitude.  Therefore, the likelihood that a given source is a
false positive can be estimated from the statistics of these
distributions in addition to the characteristics described above that
can be observed on a case-by-case basis.

\begin{figure*}[htb]
\epsscale{0.9}
\plotone{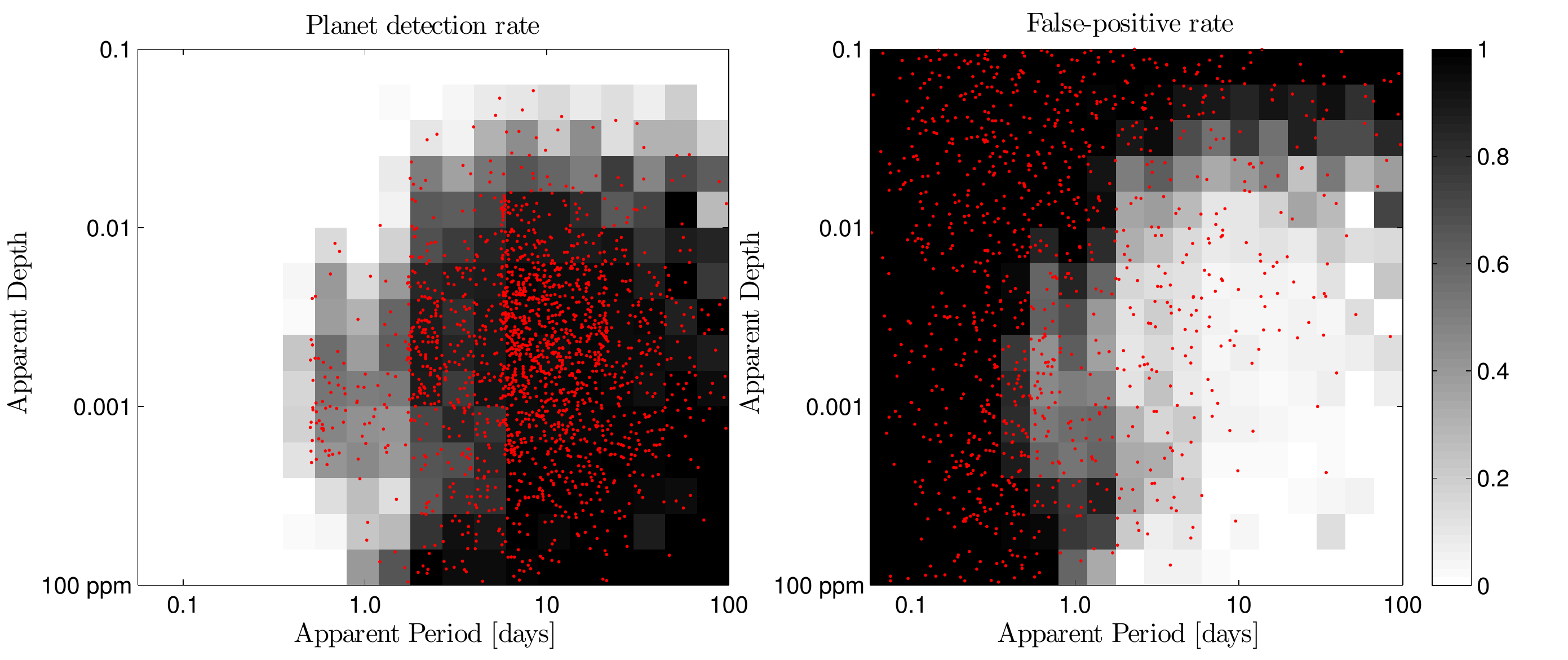}
\caption{The grayscale shows the likelihood that an eclipse observed 
  with \tess{} is a false positive or transiting planet based on 
  its apparent period and depth. \textit{Left.}---The fraction of detections
  from five trials that are transiting planets; the planets from one      
  trial are plotted as red dots. 
  \textit{Right.}---The fraction of all eclipses that are due to false    	
  positives; the red dots are individual false-positives.}
\label{fig:rates}
\end{figure*}

Figure \ref{fig:rates} shows the distributions of apparent period and
apparent depth of the eclipses caused by transiting planets and
false positives. Here, the ``apparent period'' is the period one would
be likely to infer from the \tess{} photometry; if the secondary
eclipse is detectable but not distinguishable from the primary
eclipse, one would conclude that the period is half of the true
orbital period. The ``apparent depth'' takes into account the dilution
of an eclipse from background stars or, in the case of BEBs, the
dilution from the target star.

These populations are seen to be quite distinct. Eclipsing binary
systems tend to have larger depths and shorter periods than
planets. Simply by omitting sources which have eclipse/transit depths
$>$5\% or periods $<$0.5~days, approximately \fppdpct{} of the false
positives among the target stars would be discarded. 

\begin{figure}[htb]
\epsscale{1.0}
\plotone{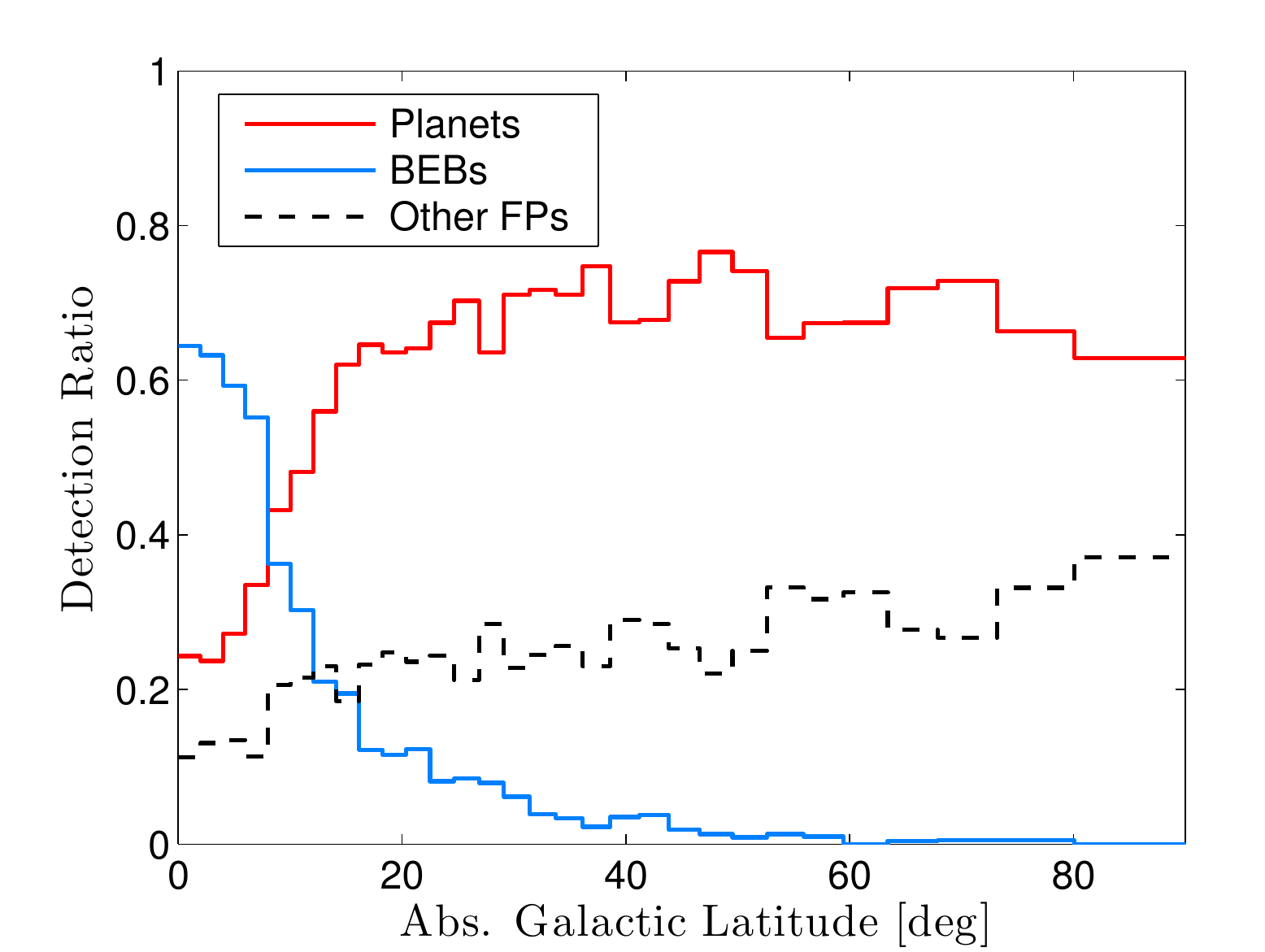}
\caption{The likelihood that an eclipse observed with \tess{} is a
  false positive or transiting planet as a function of galactic latitude.
  Planets tend to be detected at higher galactic latitude while
  background eclipsing binaries (BEBs) dominate detections at low
  galactic latitude. Here, we consider all eclipses with an apparent
  depth $<$10\%.}
\label{fig:glat}
\end{figure}

The galactic latitude $b$ of the target also has a strong influence on
the likelihood that a given source is a false
positive. Figure~\ref{fig:glat} shows the fraction of detections that
are due to planets, BEBs, and other false positives as a function of
galactic latitude. Only the events with apparent depth $<$10\% are
included in this plot. For $|b|<$ 10$^\circ$, the density of
background stars is very high, and any observed eclipse is far more
likely to be from a BEB than any other kind of eclipse. For $|b|>$
20$^\circ$, planets represent a majority over false positives. A
weaker dependence on galactic latitude is seen for grazing-eclipse
binaries and hierarchical eclipsing binaries.

\section{PROSPECTS FOR FOLLOW-UP OBSERVATIONS}
\label{sec:followup}

We now turn to the prospects for follow-up observations to
characterize the \tess{} transiting planets. As already discussed in
Section~\ref{sec:fpphot}, it is desirable to obtain transit light
curves of the planetary candidates with a higher signal-to-noise than
the \tess{} discovery.  The photometry could be carried out with
ground-based facilites or with upcoming space-based facilities such as
{\it CHEOPS} \citep{cheops}.  This data can be used to look for
transit timing variations and to improve our estimates of relative
plantary radii.

Constraining the absolute planetary radii of the \tess{} planets will
benefit from additional determinations of the radii of their host
stars.  Interferometric observations may be possible for the brightest
and nearest host stars. For this reason, we report the stellar radii
and distance moduli (in the ``DM'' column) of Table~\ref{tbl:cat},
allowing for estimation of angular diameters.

Asteroseismology can also be used to determine the radii of host stars if
finely-sampled, high-precision photometry is available. Such data could come
from the \tess{} data or the upcoming {\it PLATO} mission \citep{plato}.
There is discussion of having \tess{} record the pixel values of 
the most promising targets for asteroseismology with a time sampling
shorter than 2~min.

Next, we turn to the follow-up observations that \tess{} is designed
to enable: radial-velocity observations to measure a planet's mass and
spectroscopic observations to detect and characterize a planet's
atmosphere.

\subsection{Radial Velocity}
\label{sec:rv}

\begin{figure*}[htb]
\epsscale{0.9}
\plotone{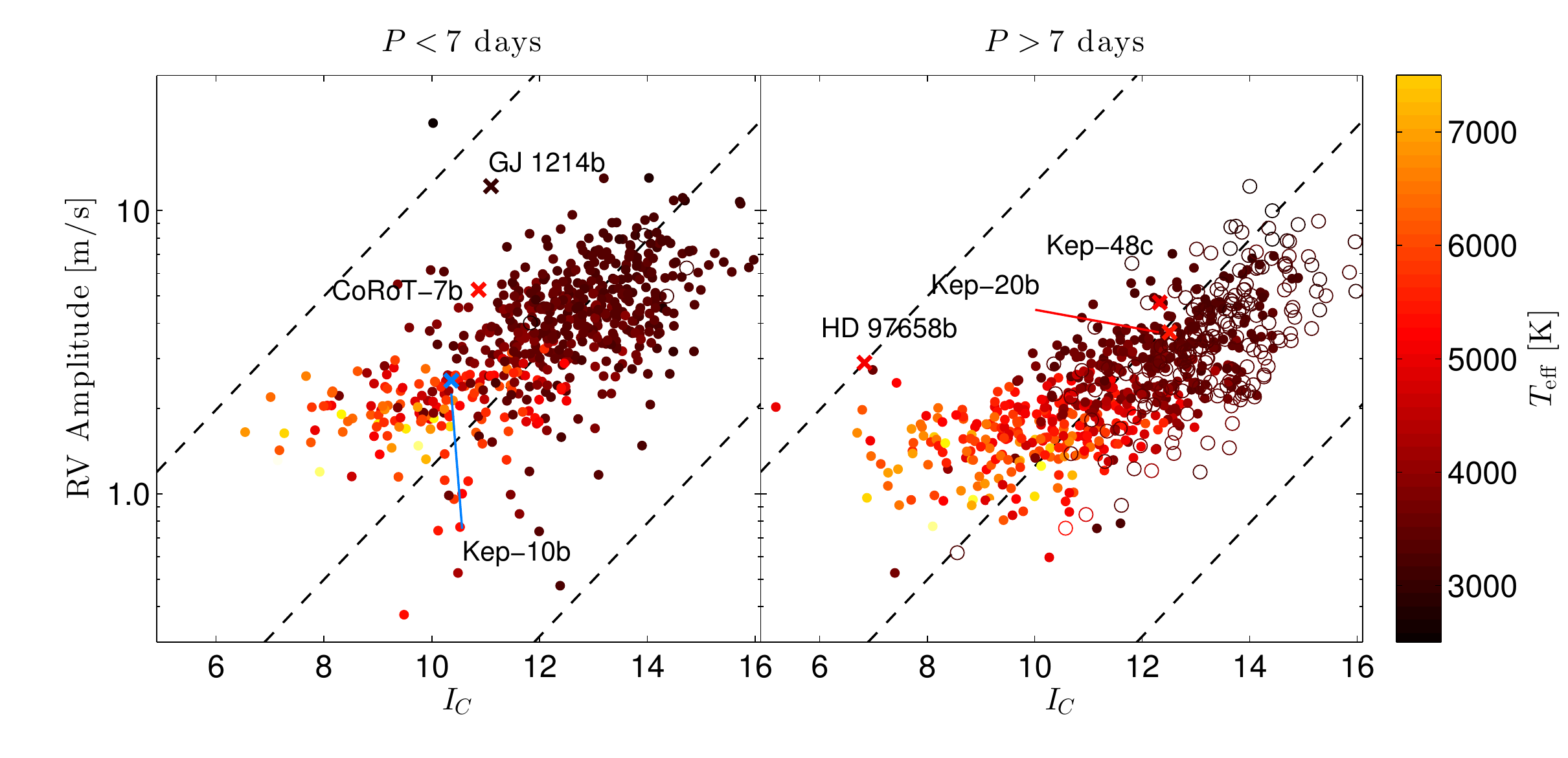}
\caption{Mass measurement of the \tess{} planets. The radial velocity
  semi-amplitude $K$ plotted against apparent magnitude for the
  \tess{} planets with $R_p < 3R_{\oplus}$. The sample is split at the
  median period of 7 days, and open symbols indicate planets near the
  habitable zone with an insolation $S<2S_{\oplus}$. We assume the
  mass-radius relation from \citet{weiss}. Several well-known
  exoplanets are also shown for context with $\times$ symbols: HD
  97658b \citep{dragomir2013}, CoRoT-7b \citep{hatzes2011}, GJ 1214b
  \citep{gj1214b}, Kepler-20b and Kepler-48c \citep{marcy2014}, and
  Kepler-10b \citep{dumusque2014}, which is plotted in blue for
  clarity.}
\label{fig:rv}
\end{figure*}

The \tess{} planets should be attractive targets for radial-velocity
observations because the host stars will be relatively bright
and their orbital periods will be relatively short. Both of these
factors facilitate precise Doppler spectroscopy. 
To evaluate the detectability of the Doppler signal
we assign masses to the simulated planets using the empirical mass-radius
relation provided by \citet{weiss}. For $R_p < 1.5~R_{\oplus}$, the
planet mass $M_p$ is calculated as
\begin{equation}
M_p =  M_{\oplus} \left[ 0.440 \left(\frac{R_p}{R_{\oplus}}\right)^3 + 0.614 \left(\frac{R_p}{R_{\oplus}}\right)^4 \right],
\end{equation}
and for $R_p \geq 1.5 R_{\oplus}$, the mass is calculated as
\begin{equation}
M_p = 2.69 M_{\oplus} \left(\frac{R_p}{R_{\oplus}} \right)^{0.93}.
\end{equation}
This simple one-to-one relationship between mass and radius is used
here for convenience. In reality, there is probably a distribution of
planet masses for a given planet radius (see, e.g.,
\citealt{rogers2014}).

From the masses calculated here, we then find the radial-velocity
semiamplitude $K$, which is reported in Table~\ref{tbl:cat}.
Figure~\ref{fig:rv} shows $K$ values of each planet detected in one 
trial as a function of the apparent magnitude of the host star. 
Because of the short periods, even planets smaller than
2~$R_\oplus$ will produce a radial-velocity semiamplitude $K$ close to
1~m~s$^{-1}$, putting them within reach of current and upcoming
spectrographs.

\subsection{Atmospheric Characterization}
\label{sec:atm}

\begin{figure*}[ht!]
\epsscale{0.55}
\plotone{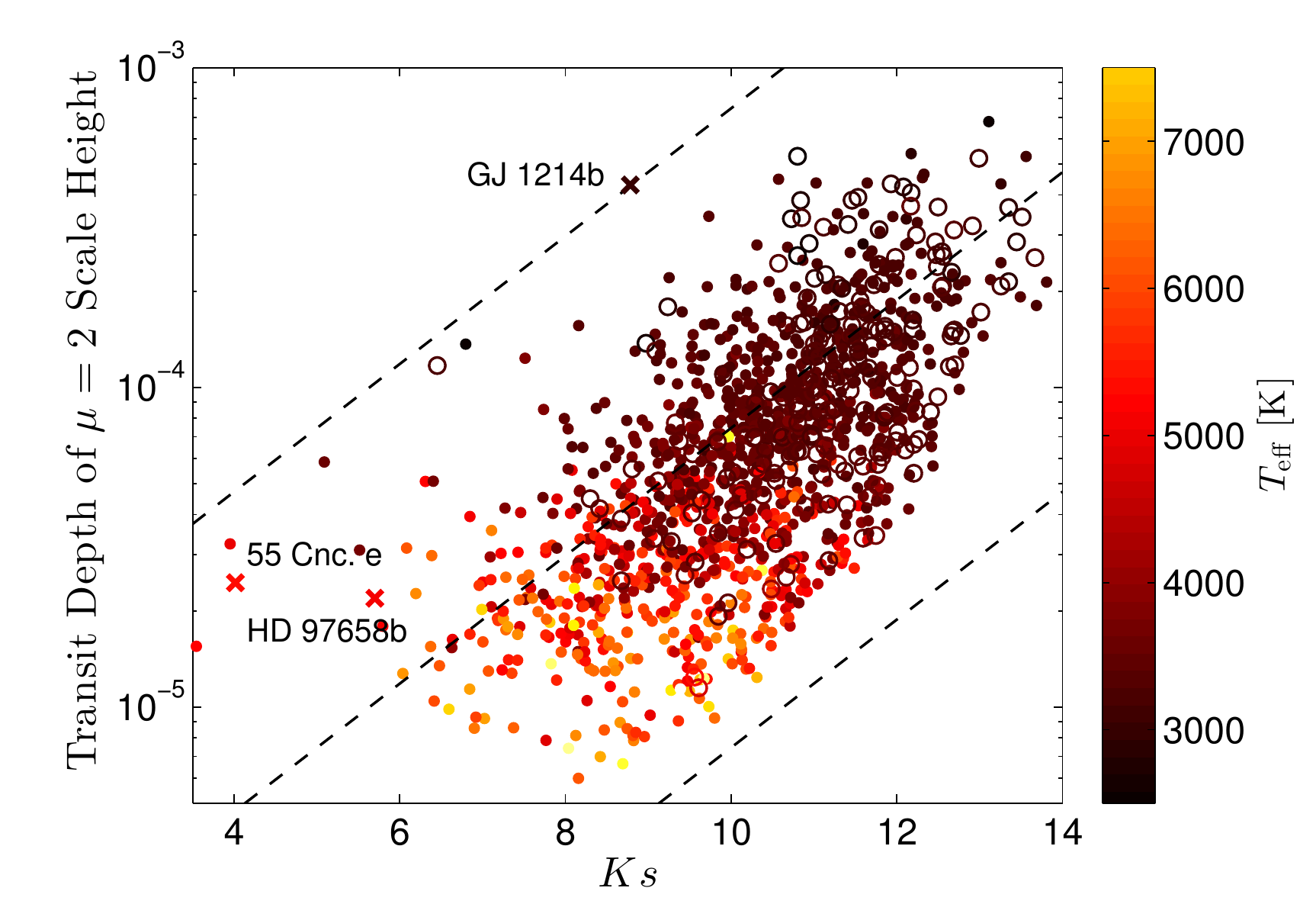}
\caption{Feasibility of transit spectroscopy of the \tess{} planets.
  The transit depth of one atmospheric scale height, assuming a pure
  H$_2$ atmosphere, is plotted against the apparent stellar $Ks$
  magnitude.  Atmospheric transit depths are lower by a factor of
  $\mu/2$ for other mean molecular weights.  The points are colored by
  stellar $\teff$, and open symbols indicate planets with an
  insolation $S < 2S_{\oplus}$.  The dashed lines indicate the
  relative photon-counting noise versus magnitude, spaced by decades.
  Planets with $R_p<3R_{\oplus}$ are shown in addition to GJ1214b
  \citep{gj1214b}, 55~Cancri~e \citep{winn55ce}, and HD97658b
  \citep{hd97658b}.}
\label{fig:atmos}
\end{figure*}

\begin{figure*}[ht!]
\epsscale{0.45}
\plotone{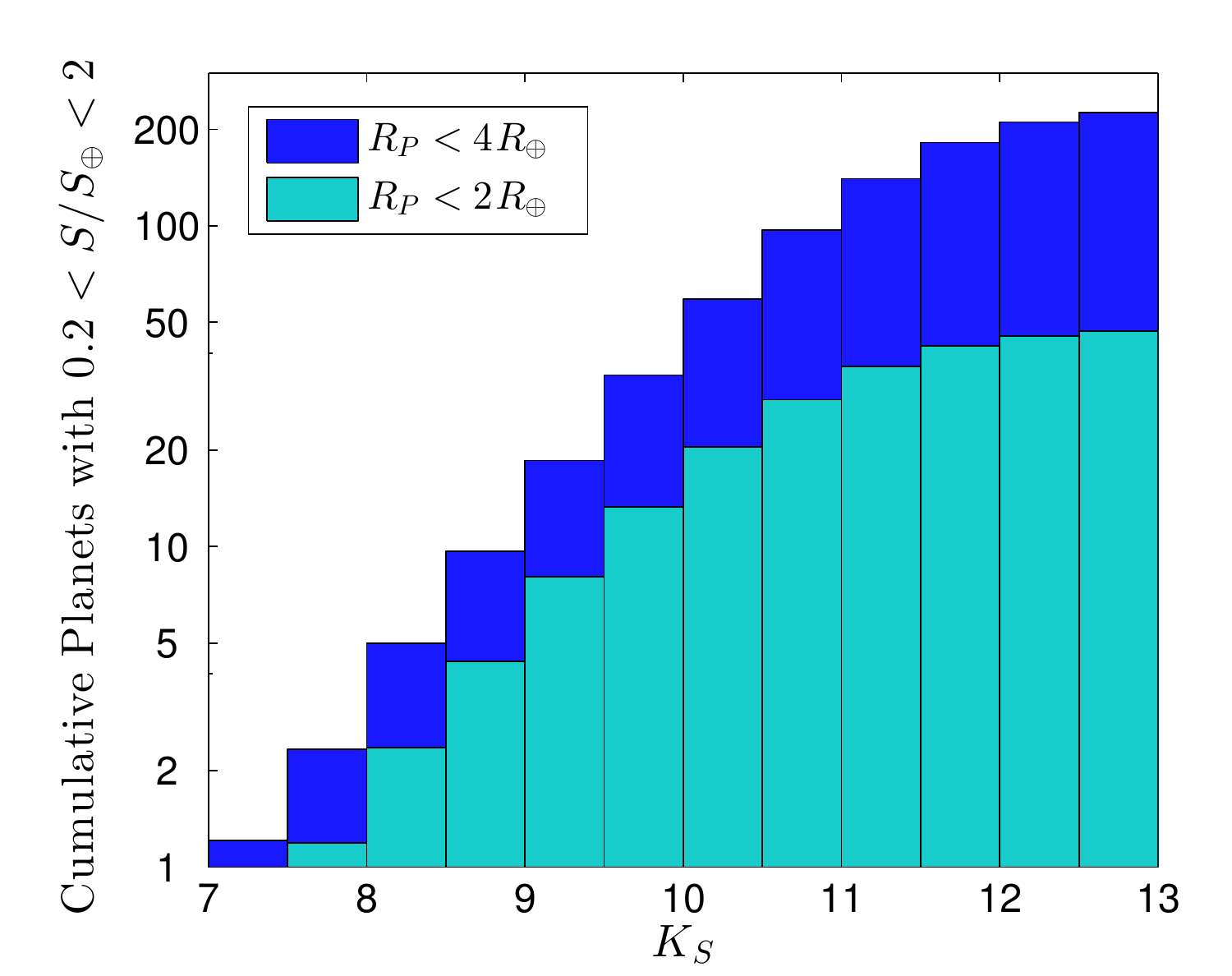}
\caption{The cumulative distribution of apparent $K_s$ magnitudes of 
the \tess{}-detected planets with $0.2 < S/S_{\oplus} < 2$.}
\label{fig:hzbright}
\end{figure*}

The composition of planetary atmospheres can be probed with transit
spectroscopy. Such measurements can be carried out with space-based or
balloon-based facilities, or even from ground-based facilities if the
resolution is high enough to separate telluric features from stellar
and planetary features. The enhanced sensitivity of \tess{} to
transiting planets near the ecliptic poles will provide numerous
targets for observations inside or near the continuous viewing zone of
the {\it James Webb Space Telescope}. The prospects for follow-up with
{\it JWST} have been detailed in \citet{deming2009} and elsewhere.
More specialized space missions, including {\it FINESSE}
\citep{finesse} and {\it EChO} \citep{echo}, have also been proposed
to perform transit spectroscopy.

Here, we use the simulation results to explore the relative difficulty
of transit spectroscopy of the \tess{} planets independent from the
facility that is used to observe them.  We compute a figure-of-merit
$\delta_H$, which is the fractional loss-of-light from an annulus
surrounding the planet (with radius $R_p$) and a thickness equal to
the scale height, $H$:
\begin{equation}
\delta_H = \frac{2HR_p}{R_\star^2}
\end{equation}
The scale height is calculated from
\begin{equation}
H = \frac{k_{\rm B} T_p R_p^2 }{G M_p \mu m_p },
\end{equation}
where $M_p$ is the planet mass and $m_p$ is the proton mass. We
calculate the temperature of the planet, $T_p$, assuming it is in
radiative equilibrium with zero albedo and isotropic re-radiation (see
Eqn.~\ref{eq:tplanet}).  We assume a mean molecular weight $\mu$ of
2~amu, which corresponds to an atmosphere consisting purely of H$_2$.
In any other case, the atmospheric transit depth $\delta_H$ is reduced
by a factor of $\mu/2$. An Earth-like atmosphere would have
$\mu=29$~amu, and a Venusian atmosphere would have $\mu=44$~amu.

Figure \ref{fig:atmos} shows $\delta_H$ for all of the detected
planets in the simulation as a function of the apparent magnitude of
the host star.  For a molecular species to be identifiable, one must
observe transits with a sensitivity on the order of $\delta_H$ both in
and out of the absorption bands of that species. The detection of
various species therefore depends on the depth of the absorption bands
and the spectral resolution used to observe them. The presence of
clouds and haze can reduce the observable thickness of the atmosphere.

Next, we look specifically at the number of planets with a relative
insolation $0.2<S/S_{\oplus}<2$, placing them within or near the
habitable zone. These planets are especially attractive targets for
atmospheric spectroscopy because they may have atmospheres similar to
that of the Earth, and may present ``biomarkers'' indicative of life.
Such observations are most feasible for the planets with the brightest
possible host stars.  For that reason, we show in
Figure~\ref{fig:hzbright} the cumulative distribution of apparent
$K_s$ magnitudes of stars hosting planets with $0.2<S/S_{\oplus}<2$.
With the statistical errors in the planet occurrence rates and Poisson
fluctuations in the number of detected planets, \nhzbright{} planets
with $0.2<S/S_{\oplus}<2$ and $SR<2R_{\oplus}$ have host stars
brighter than $K_s = 9$.

Of particular interest for atmospheric spectroscopy with {\it JWST}
are the planets that are located near the continuous-viewing zones of
{\it JWST}, which will be centered on the ecliptic poles.  A subset of
\njwsterr{} planets with $R_p < 2 R_{\oplus}$ and $0.2<
S/S_{\oplus}<2$ are found within 15$^{\circ}$ of the ecliptic poles.
The brightest stars hosting these planets have $K_s \approx$9.

\section{SUMMARY}

We have simulated the population of transiting planets and eclipsing
binaries across the sky, and we have identified the subset of those
systems that will be detectable by the \tess{} mission.  To do so, we
employed the TRILEGAL model of the galaxy to generate a catalog of
stars covering 95$\%$ of the sky. We adjusted the modelled properties
of those stars to align them with more recent observations and models
of low-mass stars, the stellar multiplicity fraction as a function of
mass, and the $J$-band luminosity function of the galactic disk.  We
then added planets to these stars using occurrence rates derived from
\kepler{}. Then, we modeled the process through which \tess{} will
observe those stars and estimated the signal-to-noise ratio of the
eclipse and transit events.

We report the statistical uncertainties in our tallies of detected 
planets arising from Poisson fluctuations and uncertainties in the planet 
occurrence rates. However, systematic errors in the occurrence rates, 
the luminosity function, and stellar properties are also significant. 
We also assumed that we can perfectly identify the $2\times 10^5$ best 
``target stars'' for \tess{} to observe at the 2-min cadence. 
In reality, it is difficult to select these stars since subgiants can 
masquerade as main-sequence dwarfs. Parallaxes
from {\it Gaia} could help determine the radii of \tess{} target stars more 
accurately, and examining the full-frame images will help find planets 
transiting the stars excluded from the 2-minute data.

The \tess{} planets will be attractive targets for follow-up
measurements of transit properties, radial velocity measurements, and
atmospheric transmission. Knowing the population of planets that
\tess{} will detect allows the estimation of the follow-up resources that 
are needed, and it informs the design of future instruments that will
observe the \tess{} planets. The simulations provide fine-grained
statistical samples of planets and their properties which may be of
interest to those who are planning follow-up observations or building
instruments to enable such observations. Table \ref{tbl:cat} presents
the results from one trial of the \tess{} mission. This catalog
contains all the detected transiting planets from among the $2\times
10^5$ target stars that are observed at a 2~min cadence.

We look forward to the occasion, perhaps within 5-6 years, when
\tess{} will have completed its primary mission and we are able to
replace this simulated catalog with the real \tess{} catalog. 
This collection of transiting exoplanets will represent the 
brightest and most favorable systems for further study.

\clearpage

\acknowledgments We are grateful to the referee, Scott Gaudi, for
providing constructive criticism that led to improvements in this
paper. We are also grateful to Luke Bouma and Hans Deeg for
scrutinizing our results carefully and bringing some errors to our
attention. We thank the members of the \tess{} Science Team for their
contributions to the mission. In particular, Jacob Bean, Tabetha
Boyajian, Eric Gaidos, Daniel Huber, Geoffrey Marcy, Roberto
Sanchis-Ojeda, and Keivan Stassun provided helpful comments on the
manuscript, and discussions with Jon Jenkins helped inform our
approach to the simulations. We acknowledge Anthony Smith, Kristin
Clark, and Michael Chrisp at MIT-Lincoln Laboratory for their
respective roles in the project management, systems engineering, and
optical design for the \tess{} payload. In addition, Barry Burke and
Vyshnavi Suntharalingam at MIT-LL provided useful input to the PSF
model. We thank Leo Girardi for adding the \tess{} bandpass to the
TRILEGAL simulation and for providing a perl script to facilitate the
queries.  We also thank Gibor Basri for sharing the stellar
variability data from {\it Kepler}.

This publication makes use of data products from the Two Micron All
Sky Survey, which is a joint project of the University of
Massachusetts and the Infrared Processing and Analysis
Center/California Institute of Technology, funded by the National
Aeronautics and Space Administration and the National Science
Foundation.


\begin{deluxetable*}{ccccccccccccccccc}
\tabletypesize{\scriptsize}
\tablecolumns{15}
\tablewidth{0pt}
\tablecaption{Catalog of simulated \tess{} detections. \label{tbl:cat}}
\tablehead{
\colhead{$\alpha$ [$^{\circ}$]} &
\colhead{$\delta$ [$^{\circ}$]} &
\colhead{$R_p$} &
\colhead{$P$ [days]} & 
\colhead{$S/S_{\oplus}$} & 
\colhead{$K$~[m~s$^{-1}$]} &
\colhead{$R_\star$~[$R_\odot$]} &
\colhead{$\teff$~[K]} & 
\colhead{$V$} &
\colhead{$I_C$} &
\colhead{$J$} &
\colhead{$K_s$} &
\colhead{DM} &
\colhead{Dil.} &
\colhead{$\log_{10}(\sigma_V)$} &
\colhead{SNR} &
\colhead{Mult.}
}

\startdata
 0.439 & 45.217 & 3.31 & 9.14 & 361.7 & 2.03 & 1.41 & 6531 & 8.47 & 7.97 & 7.63 & 7.41 & 5.00 & 1.00 & -4.87 & 16.8 &  1 \\ 
 0.480 & -66.204 & 2.19 & 14.20 & 2.1 & 3.11 & 0.32 & 3426 & 15.08 & 12.83 & 11.56 & 10.79 & 3.90 & 1.01 & -4.22 & 12.3 &  3 \\ 
 0.646 & 42.939 & 1.74 & 4.96 & 235.0 & 1.66 & 0.95 & 5546 & 10.12 & 9.35 & 8.81 & 8.41 & 4.95 & 1.00 & -4.64 &  7.5 &  0 \\ 
 0.924 & -26.065 & 1.48 & 2.16 & 1240.1 & 1.95 & 1.12 & 5984 & 8.06 & 7.42 & 6.98 & 6.67 & 3.65 & 1.00 & -4.50 &  8.4 &  0 \\ 
 1.314 & -24.954 & 2.29 & 9.75 & 5.9 & 2.95 & 0.42 & 3622 & 14.19 & 12.15 & 10.99 & 10.19 & 4.10 & 1.00 & -4.44 &  8.6 &  2 \\ 
 1.384 & 10.606 & 2.32 & 13.99 & 2.1 & 3.32 & 0.32 & 3425 & 15.04 & 12.79 & 11.52 & 10.74 & 3.85 & 1.07 & -3.50 &  7.8 &  2 \\ 
 1.783 & -71.931 & 3.29 & 8.42 & 4.5 & 5.23 & 0.34 & 3444 & 15.29 & 13.06 & 11.79 & 11.02 & 4.25 & 1.00 & -4.39 & 26.4 &  2 \\ 
 1.789 & -9.144 & 2.81 & 5.62 & 2.6 & 9.15 & 0.17 & 3228 & 15.05 & 12.52 & 11.13 & 10.36 & 1.85 & 2.28 & -4.41 & 28.7 &  3 \\ 
 1.948 & -16.995 & 17.15 & 1.34 & 10164.5 & 16.03 & 2.11 & 6668 & 7.51 & 7.05 & 6.72 & 6.53 & 5.00 & 1.00 & -4.42 & 457.4 &  2 \\ 
 2.172 & -15.533 & 4.80 & 17.14 & 341.2 & 2.10 & 2.11 & 6668 & 7.51 & 7.05 & 6.72 & 6.53 & 5.00 & 1.00 & -4.68 & 13.1 &  2 \\ 
 4.071 &  9.507 & 1.97 & 11.45 & 1.5 & 4.09 & 0.22 & 3300 & 14.96 & 12.55 & 11.21 & 10.44 & 2.65 & 1.00 & -4.50 & 17.7 &  2 \\ 
 4.634 & -23.500 & 4.71 & 5.17 & 116.3 & 4.60 & 0.80 & 5000 & 9.52 & 8.54 & 7.85 & 7.32 & 3.35 & 1.00 & -4.68 & 64.0 &  0 \\ 
 4.788 & 78.625 & 1.56 & 0.62 & 71.2 & 8.92 & 0.22 & 3283 & 15.94 & 13.50 & 12.14 & 11.38 & 3.50 & 1.00 & -4.43 & 16.9 &  2 \\ 
 5.322 & -55.554 & 2.85 & 18.16 & 2.4 & 3.01 & 0.41 & 3592 & 14.23 & 12.15 & 10.98 & 10.18 & 4.00 & 1.00 & -4.46 & 15.5 &  2 \\ 
 5.704 & 50.726 & 2.24 & 2.75 & 308.3 & 2.82 & 0.80 & 5188 & 11.28 & 10.40 & 9.77 & 9.28 & 5.35 & 1.01 & -4.67 & 10.6 &  2 \\ 
 5.951 & -28.675 & 3.75 & 17.75 & 4.3 & 3.46 & 0.50 & 3844 & 11.73 & 9.94 & 8.89 & 8.08 & 2.55 & 1.00 & -4.29 & 52.5 &  1 \\ 
 6.166 & 32.455 & 1.11 & 0.79 & 53.4 & 2.95 & 0.22 & 3304 & 14.81 & 12.41 & 11.08 & 10.32 & 2.65 & 1.00 & -3.50 &  8.5 &  2 \\ 
 6.521 & -4.048 & 1.53 & 7.51 & 4.8 & 2.78 & 0.32 & 3435 & 13.66 & 11.43 & 10.17 & 9.41 & 2.50 & 1.00 & -3.49 &  8.3 &  2 \\ 
 6.662 & -79.377 & 2.31 & 1.89 & 45.5 & 5.48 & 0.40 & 3551 & 14.42 & 12.30 & 11.10 & 10.31 & 4.00 & 1.00 & -3.49 & 19.9 &  0 \\ 
 7.592 & -79.924 & 7.72 & 3.92 & 320.7 & 7.41 & 0.91 & 5623 & 11.06 & 10.32 & 9.79 & 9.40 & 5.85 & 1.00 & -4.54 & 96.5 &  1 \\ 
 8.071 & -77.185 & 1.49 & 1.02 & 887.9 & 3.50 & 0.70 & 5030 & 8.00 & 7.05 & 6.40 & 5.85 & 1.60 & 1.00 & -4.74 & 40.7 &  1 \\ 
 8.396 & -53.966 & 2.38 & 5.59 & 6.8 & 4.55 & 0.31 & 3442 & 14.14 & 11.91 & 10.66 & 9.89 & 2.95 & 1.25 & -3.60 & 20.3 &  3 \\ 
 8.919 & 67.419 & 2.94 & 29.94 & 1.3 & 2.61 & 0.42 & 3611 & 13.33 & 11.30 & 10.14 & 9.35 & 3.20 & 1.09 & -3.69 & 18.1 &  1 \\ 
 9.843 & -11.832 & 2.82 & 6.28 & 62.9 & 2.98 & 0.69 & 4819 & 11.69 & 10.60 & 9.88 & 9.26 & 4.90 & 1.00 & -4.85 & 17.2 &  0 \\ 
10.467 & -40.605 & 1.71 & 15.14 & 1.3 & 2.83 & 0.26 & 3359 & 14.51 & 12.18 & 10.87 & 10.11 & 2.70 & 1.02 & -3.49 &  9.1 &  1 \\ 
10.551 & -27.297 & 3.60 & 5.99 & 283.9 & 2.96 & 1.04 & 5970 & 9.85 & 9.21 & 8.76 & 8.44 & 5.25 & 1.00 & -4.52 & 22.5 &  0 \\ 
11.067 & -52.752 & 2.40 & 17.18 & 0.9 & 4.35 & 0.22 & 3287 & 15.54 & 13.10 & 11.74 & 10.98 & 3.10 & 1.00 & -4.32 & 17.8 &  1 \\ 
11.145 & 29.347 & 2.43 & 8.66 & 13.3 & 2.80 & 0.53 & 3948 & 12.80 & 11.09 & 10.08 & 9.27 & 3.95 & 1.01 & -4.46 & 13.8 &  2 \\ 
11.145 & -49.044 & 1.51 & 8.12 & 6.4 & 2.27 & 0.39 & 3557 & 13.67 & 11.56 & 10.37 & 9.58 & 3.25 & 1.03 & -4.45 & 10.7 &  1 \\ 
11.207 & 37.355 & 2.64 & 1.98 & 28.9 & 7.02 & 0.32 & 3437 & 14.67 & 12.43 & 11.18 & 10.40 & 3.55 & 1.02 & -4.01 & 22.6 &  3 \\ 
11.547 & -44.670 & 5.26 & 39.43 & 56.6 & 1.90 & 1.47 & 6577 & 8.65 & 8.16 & 7.82 & 7.62 & 5.30 & 1.00 & -4.01 & 14.5 &  0 \\ 
11.909 & -67.746 & 3.67 & 4.78 & 216.5 & 3.69 & 0.84 & 5598 & 11.69 & 10.94 & 10.40 & 10.00 & 6.25 & 1.01 & -3.77 & 20.4 &  0 \\ 
12.015 & 74.529 & 3.85 & 1.85 & 11.8 & 17.24 & 0.17 & 3225 & 16.89 & 14.36 & 12.95 & 12.18 & 3.75 & 1.10 & -3.49 & 45.8 &  3 \\ 
12.085 & -51.662 & 11.96 & 40.02 & 16.1 & 4.97 & 0.99 & 5598 & 10.31 & 9.55 & 9.02 & 8.63 & 5.25 & 1.01 & -4.74 & 169.9 &  0 \\ 
12.261 & -60.310 & 2.12 & 10.35 & 3.4 & 3.19 & 0.33 & 3467 & 13.88 & 11.68 & 10.44 & 9.67 & 2.90 & 1.00 & -3.53 & 13.3 &  2 \\ 
12.320 & 75.640 & 6.92 & 5.25 & 513.2 & 5.25 & 1.20 & 6295 & 10.12 & 9.57 & 9.18 & 8.91 & 6.10 & 1.54 & -4.12 & 64.3 &  1 \\ 
12.410 & -10.212 & 2.97 & 6.84 & 2.2 & 8.51 & 0.18 & 3230 & 16.63 & 14.10 & 12.70 & 11.93 & 3.60 & 1.00 & -4.39 &  9.0 &  0 \\ 
12.640 & -53.861 & 1.73 & 2.40 & 20.9 & 4.48 & 0.31 & 3442 & 14.14 & 11.91 & 10.66 & 9.89 & 2.95 & 1.00 & -3.95 & 17.7 &  3 \\ 
12.928 & -13.886 & 12.98 & 6.03 & 8419.9 & 5.13 & 2.50 & 10593 & 6.03 & 6.14 & 6.15 & 6.21 & 5.55 & 1.00 & -4.16 & 61.4 &  1 \\ 
12.974 & 58.302 & 2.77 & 13.71 & 422.9 & 1.35 & 1.56 & 7603 & 9.02 & 8.77 & 8.57 & 8.48 & 6.45 & 1.06 & -4.50 &  7.9 &  1 \\ 
13.048 & 74.712 & 1.47 & 1.06 & 81.4 & 5.18 & 0.36 & 3490 & 15.02 & 12.84 & 11.60 & 10.82 & 4.25 & 1.30 & -3.61 &  7.9 &  2 \\ 
13.408 & 27.048 & 3.21 & 11.65 & 84.2 & 2.21 & 0.96 & 5689 & 10.53 & 9.81 & 9.30 & 8.93 & 5.50 & 1.01 & -4.84 & 13.2 &  1 \\ 
13.494 & -57.211 & 1.98 & 19.54 & 2.3 & 2.08 & 0.42 & 3606 & 14.03 & 11.97 & 10.80 & 10.01 & 3.90 & 1.00 & -4.53 &  8.0 &  3 \\ 
13.690 & -81.593 & 1.85 & 2.20 & 15.3 & 6.25 & 0.24 & 3324 & 14.95 & 12.57 & 11.24 & 10.48 & 2.85 & 1.15 & -3.84 & 29.3 &  3 \\ 
13.824 & -20.209 & 1.05 & 15.96 & 0.6 & 1.16 & 0.16 & 3228 & 14.83 & 12.30 & 10.91 & 10.15 & 1.60 & 1.00 & -3.96 &  8.3 &  1 \\ 
14.214 & 79.814 & 1.06 & 0.50 & 659.2 & 1.50 & 0.55 & 3996 & 12.87 & 11.21 & 10.21 & 9.40 & 4.20 & 1.05 & -4.10 &  7.9 &  3 \\ 
14.313 & 32.413 & 2.29 & 5.80 & 7.6 & 4.23 & 0.34 & 3470 & 12.33 & 10.15 & 8.92 & 8.15 & 1.40 & 1.00 & -4.54 & 46.1 &  2 \\ 
14.807 & -14.485 & 1.99 & 12.55 & 0.6 & 5.39 & 0.16 & 3027 & 14.98 & 12.76 & 11.10 & 10.32 & 1.45 & 1.00 & -3.53 & 19.7 &  1 \\ 
15.256 & 48.530 & 2.27 & 7.68 & 4.4 & 3.93 & 0.31 & 3435 & 14.47 & 12.23 & 10.98 & 10.21 & 3.25 & 1.21 & -3.70 & 15.1 &  2 \\ 
15.926 & 75.902 & 1.94 & 5.06 & 333.5 & 1.72 & 1.05 & 5888 & 8.68 & 8.02 & 7.56 & 7.23 & 4.05 & 1.00 & -4.64 & 16.7 &  2 \\ 
\enddata

\tablecomments{This catalog is based on one realization of the Monte Carlo
  simulation. The detections are drawn from the $2\times 10^5$ target
  stars that are observed with a 2~min cadence. The larger sample of
  detections from stars that are only observed in full-frame images is
  not provided here. The entirety of this table is available electronically;
  only the first 50 lines are shown here to illustrate its form and content.}

\end{deluxetable*}


\begin{thebibliography}{}

\bibitem[Alonso et al.(2004)]{tres} Alonso, R., Brown, 
T.~M., Torres, G., et al.\ 2004, \apjl, 613, L153 

\bibitem[Andersen(1991)]{ebs} Andersen, J.\ 1991, \aapr, 3, 91 

\bibitem[Auvergne et al.(2009)]{corot} Auvergne, M., Bodin, P.,
  Boisnard, L., et al.\ 2009, \aap, 506, 411

\bibitem[Bahcall \& Soneira(1981)]{bahcall1981} Bahcall, J.~N., \&
  Soneira, R.~M.\ 1981, \apjs, 47, 357

\bibitem[Bakos et al.(2004)]{hat} Bakos, G., Noyes, R.~W., 
Kov{\'a}cs, G., et al.\ 2004, \pasp, 116, 266 

\bibitem[Basri et al.(2013)]{basri} Basri, G., Walkowicz, 
L.~M., \& Reiners, A.\ 2013, \apj, 769, 37 

\bibitem[Beatty \& Gaudi(2008)]{beattygaudi} Beatty, T.~G., \& Gaudi,
  B.~S.\ 2008, \apj, 686, 1302

\bibitem[Bochanski et al.(2010)]{bochanski} Bochanski, J.~J., 
Hawley, S.~L., Covey, K.~R., et al.\ 2010, \aj, 139, 2679 

\bibitem[Borucki et al.(2010)]{kepler} Borucki, W.~J., Koch, 
D., Basri, G., et al.\ 2010, Science, 327, 977 

\bibitem[Bouchy et al.(2005)]{hd189733b} Bouchy, F., Udry, S., Mayor,
  M., et al.\ 2005, \aap, 444, L15

\bibitem[Boyajian et al.(2012)]{chara} Boyajian, T.~S., von 
Braun, K., van Belle, G., et al.\ 2012, \apj, 757, 112

\bibitem[Brown et al.(2011)]{kepmag} Brown, T.~M., Latham, 
D.~W., Everett, M.~E., \& Esquerdo, G.~A.\ 2011, \aj, 142, 112 

\bibitem[Burke et al.(2014)]{burke2014} Burke, C.~J., Bryson, 
S.~T., Mullally, F., et al.\ 2014, \apjs, 210, 19 

\bibitem[Carter et al.(2008)]{carter2008} Carter, J.~A., Yee, 
J.~C., Eastman, J., Gaudi, B.~S., \& Winn, J.~N.\ 2008, \apj, 689, 499 

\bibitem[Cardelli et al.(1989)]{cardelli} Cardelli, J.~A., 
Clayton, G.~C., \& Mathis, J.~S.\ 1989, \apj, 345, 245 

\bibitem[Chabrier(2001)]{chabrier} Chabrier, G.\ 2001, \apj, 
554, 1274 

\bibitem[Charbonneau et al.(2000)]{hd209458b} Charbonneau, D., 
Brown, T.~M., Latham, D.~W., \& Mayor, M.\ 2000, \apjl, 529, L45 

\bibitem[Charbonneau et al.(2009)]{gj1214b} Charbonneau, D., 
Berta, Z.~K., Irwin, J., et al.\ 2009, \nat, 462, 891 

\bibitem[Chabrier et al.(2000)]{dusty} Chabrier, G., Baraffe, 
I., Allard, F., \& Hauschildt, P.\ 2000, \apj, 542, 464 

\bibitem[Claret et al.(2012)]{claret2012} Claret, A., Hauschildt,
  P.~H., \& Witte, S.\ 2012, \aap, 546, AA14

\bibitem[Claret et al.(2013)]{claret2013} Claret, A., Hauschildt,
  P.~H., \& Witte, S.\ 2013, \aap, 552, AA16

\bibitem[Cruz et al.(2007)]{cruz} Cruz, K.~L., Reid, I.~N., 
Kirkpatrick, J.~D., et al.\ 2007, \aj, 133, 439 

\bibitem[Delfosse et al.(2004)]{delfosse2004} Delfosse, X., Beuzit, 
J.-L., Marchal, L., et al.\ 2004, Spectroscopically and Spatially Resolving 
the Components of the Close Binary Stars, 318, 166 

\bibitem[Deming et al.(2009)]{deming2009} Deming, D., Seager, S., 
Winn, J., et al.\ 2009, \pasp, 121, 952 

\bibitem[Demory et al.(2011)]{demory2011} Demory, B.-O., Gillon, M.,
  Deming, D., et al.\ 2011, \aap, 533, AA114
  
\bibitem[Deroo et al.(2012)]{finesse} Deroo, P., Swain, M.~R., 
\& Green, R.~O.\ 2012, \procspie, 8442, 844241 

\bibitem[Dotter et al.(2008)]{dartmouth} Dotter, A., Chaboyer, 
B., Jevremovi{\'c}, D., et al.\ 2008, \apjs, 178, 89 

\bibitem[Dragomir et al.(2013)]{dragomir2013} Dragomir, D., 
Matthews, J.~M., Eastman, J.~D., et al.\ 2013, \apjl, 772, LL2 

\bibitem[Dressing \& Charbonneau(2013)]{dressing2013} Dressing, C.~D.,
  \& Charbonneau, D.\ 2013, \apj, 767, 95
  
\bibitem[Dressing \& Charbonneau(2015)]{dressing2015} Dressing, C.~D., 
	\& Charbonneau, D.\ 2015, arXiv:1501.01623 

\bibitem[Duch{\^e}ne \& Kraus(2013)]{duchene} Duch{\^e}ne, G., \&
  Kraus, A.\ 2013, \araa, 51, 269

\bibitem[Dumusque et al.(2014)]{dumusque2014} Dumusque, X., Bonomo, 
A.~S., Haywood, R.~D., et al.\ 2014, \apj, 789, 154 

\bibitem[Eggleton(2009)]{eggleton} Eggleton, P.~P.\ 2009, 
\mnras, 399, 1471 

\bibitem[Eggleton \& Tokovinin(2008)]{et2008} Eggleton, P.~P., \&
  Tokovinin, A.~A.\ 2008, \mnras, 389, 869

\bibitem[Fabrycky et al.(2014)]{fabrycky2014} Fabrycky, D.~C., 
Lissauer, J.~J., Ragozzine, D., et al.\ 2014, \apj, 790, 146 

\bibitem[Figueira et al.(2012)]{figueira} Figueira, P., Marmier, M.,
  Bou{\'e}, G., et al.\ 2012, \aap, 541, AA139
  
\bibitem[Fortier et al.(2014)]{cheops} Fortier, A., Beck, T., 
Benz, W., et al.\ 2014, \procspie, 9143, 91432J 
   
\bibitem[Fressin et al.(2013)]{fressin2013} Fressin, F., Torres, 
G., Charbonneau, D., et al.\ 2013, \apj, 766, 81 

\bibitem[Fuhrmann(1998)]{fuhrmann} Fuhrmann, K.\ 1998, \aap, 338, 161 

\bibitem[Gaidos et al.(2014)]{gaidos2014} Gaidos, E., Mann, A.~W., 
L{\'e}pine, S., et al.\ 2014, \mnras, 443, 2561 


\bibitem[Gautier et al.(2012)]{gautier2012} Gautier, T.~N., III, 
Charbonneau, D., Rowe, J.~F., et al.\ 2012, \apj, 749, 15 

\bibitem[Gilliland et al.(2011)]{gilliland} Gilliland, R.~L., 
Chaplin, W.~J., Dunham, E.~W., et al.\ 2011, \apjs, 197, 6 

\bibitem[Girardi et al.(2000)]{padova} Girardi, L., Bressan, A.,
  Bertelli, G., \& Chiosi, C.\ 2000, \aaps, 141, 371

\bibitem[Girardi et al.(2005)]{trilegal} Girardi, L., Groenewegen,
  M.~A.~T., Hatziminaoglou, E., \& da Costa, L.\ 2005, \aap, 436, 895

\bibitem[Girardi et al.(2012)]{trilegalbook} Girardi, L., Barbieri, 
M., Groenewegen, M.~A.~T., et al.\ 2012, Red Giants as Probes of the 
Structure and Evolution of the Milky Way, 165 

\bibitem[G{\'o}rski et al.(2005)]{healpix} G{\'o}rski, K.~M., 
Hivon, E., Banday, A.~J., et al.\ 2005, \apj, 622, 759 

\bibitem[Hatzes et al.(2011)]{hatzes2011} Hatzes, A.~P., Fridlund, 
M., Nachmani, G., et al.\ 2011, \apj, 743, 75 

\bibitem[Hayes(1985)]{vega} Hayes, D.~S.\ 1985, Calibration 
of Fundamental Stellar Quantities, 111, 225 

\bibitem[Henry et al.(2006)]{recons} Henry, T.~J., Jao, W.-C., 
Subasavage, J.~P., et al.\ 2006, \aj, 132, 2360 

\bibitem[Howell et al.(2014)]{k2} Howell, S.~B., Sobeck, 
C., Haas, M., et al.\ 2014, \pasp, 126, 398 

\bibitem[Jenkins et al.(1996)]{jenkins1996} Jenkins, J.~M., Doyle, 
	L.~R., \& Cullers, D.~K.\ 1996, {\it Icarus}, 119, 244 

\bibitem[Jenkins et al.(2002)]{jenkins2002} Jenkins, J.~M., Caldwell,
  D.~A., \& Borucki, W.~J.\ 2002, \apj, 564, 495

\bibitem[Jordi et al.(2006)]{jordi} Jordi, K., Grebel, E.~K., \& Ammon, K.\ 2006, \aap, 460, 339 

\bibitem[Kipping(2010)]{kipping2010} Kipping, D.~M.\ 2010,
  \mnras, 408, 1758

\bibitem[Kopal(1979)]{kopal} Kopal, Z.\ 1979, Astrophysics 
and Space Science Library, 77, 21

\bibitem[Kopparapu et al.(2013)]{kopparapu13} Kopparapu, R.~K., 
Ramirez, R., Kasting, J.~F., et al.\ 2013, \apj, 765, 131 

\bibitem[Kouwenhoven et al.(2007)]{kouwenhoven} Kouwenhoven, M.~B.~N.,
  Brown, A.~G.~A., Portegies Zwart, S.~F., \& Kaper, L.\ 2007, \aap,
  474, 77

\bibitem[Kreidberg et al.(2014)]{kreidberg} Kreidberg, L., Bean, 
J.~L., D{\'e}sert, J.-M., et al.\ 2014, \nat, 505, 69 

\bibitem[van Leeuwen(2007)]{hip2} van Leeuwen, F.\ 2007, \aap, 474, 653 

\bibitem[L{\'e}ger et al.(2009)]{corot7b} L{\'e}ger, A., Rouan, D.,
  Schneider, J., et al.\ 2009, \aap, 506, 287
  
\bibitem[L{\'e}pine \& Shara(2005)]{lspm} L{\'e}pine, S., \& Shara, M.~M.\ 2005, \aj, 129, 1483 

\bibitem[Lucy(1967)]{lucy1967} Lucy, L.~B.\ 1967, \zap, 65, 89

\bibitem[Marcy et al.(2014)]{marcy2014} Marcy, G.~W., Isaacson, 
H., Howard, A.~W., et al.\ 2014, \apjs, 210, 20 

\bibitem[Mazeh(2008)]{mazeh} Mazeh, T.\ 2008, EAS 
Publications Series, 29, 1 

\bibitem[McCullough et al.(2005)]{xo} McCullough, P.~R., 
Stys, J.~E., Valenti, J.~A., et al.\ 2005, \pasp, 117, 783 

\bibitem[Morris \& Naftilan(1993)]{morris} Morris, S.~L., \& Naftilan,
  S.~A.\ 1993, \apj, 419, 344

\bibitem[Ofek(2008)]{ofek2008} Ofek, E.~O.\ 2008, \pasp, 120, 1128 

\bibitem[Perryman et al.(1997)]{hip1} Perryman, M.~A.~C., Lindegren,
  L., Kovalevsky, J., et al.\ 1997, \aap, 323, L49

\bibitem[Pecaut \& Mamajek(2013)]{pecaut} Pecaut, M.~J., \& Mamajek,
  E.~E.\ 2013, \apjs, 208, 9

\bibitem[Pepper et al.(2003)]{pepper2003} Pepper, J., Gould, A., \& Depoy, 
D.~L.\ 2003, {\it Acta Astronomica}, 53, 213 

\bibitem[Pepper et al.(2007)]{kelt} Pepper, J., Pogge, 
R.~W., DePoy, D.~L., et al.\ 2007, \pasp, 119, 923 

\bibitem[Price \& Rogers (2014)]{price2014} Price, E.~M., \& Rogers, L.~A.\ 2014, \apj, 794, 92 

\bibitem[Perryman et al.(2001)]{gaia2001} Perryman, M.~A.~C., de Boer, K.~S., Gilmore, G., et al.\ 2001, \aap, 369, 339 

\bibitem[Pickles(1998)]{pickles} Pickles, A.~J.\ 1998, \pasp, 
110, 863 

\bibitem[Pollacco et al.(2006)]{wasp} Pollacco, D.~L., 
Skillen, I., Collier Cameron, A., et al.\ 2006, \pasp, 118, 1407 

\bibitem[Pr{\v s}a et al.(2011)]{prsa} Pr{\v s}a, A., 
Batalha, N., Slawson, R.~W., et al.\ 2011, \aj, 141, 83 

\bibitem[Rauer et al.(2014)]{plato} Rauer, H., Catala, C., 
Aerts, C., et al.\ 2014, Experimental Astronomy, 38, 249 

\bibitem[Reid et al.(2003)]{reid2003} Reid, I.~N., Cruz, K.~L.,
  Laurie, S.~P., et al.\ 2003, \aj, 125, 354

\bibitem[Raghavan et al.(2010)]{raghavan2010} Raghavan, D., 
McAlister, H.~A., Henry, T.~J., et al.\ 2010, \apjs, 190, 1 

\bibitem[Reid et al.(2002)]{reid2002} Reid, I.~N., Gizis,
  J.~E., \& Hawley, S.~L.\ 2002, \aj, 124, 2721

\bibitem[Reid \& Hawley(2005)]{reidhawley} Reid, I.~N., \&
  Hawley, S.~L.\ 2005, New Light on Dark Stars: Red Dwarfs, Low-Mass
  Stars, Brown Dwarfs (Springer-Praxis)
  
\bibitem[Ricker et al.(2015)]{ricker2015} Ricker, G.~R., 
Winn, J.~N., Vanderspek, R., Latham, D.~W., et al.\ 2015, SPIE Journal of
Astronomical Telescopes, Instruments, and Systems, 1, id.014003

\bibitem[Rogers(2014)]{rogers2014} Rogers, L.~A.\ 2014, 
arXiv:1407.4457
  
\bibitem[Skrutskie et al.(2006)]{2mass} Skrutskie, M.~F., 
Cutri, R.~M., Stiening, R., et al.\ 2006, \aj, 131, 1163 
  
\bibitem[Slawson et al.(2011)]{slawson} Slawson, R.~W., Pr{\v 
s}a, A., Welsh, W.~F., et al.\ 2011, \aj, 142, 160 

\bibitem[Stassun et al.(2014)]{tdc} Stassun, K.~G., Pepper, 
J.~A., Paegert, M., De Lee, N., \& Sanchis-Ojeda, R.\ 2014, arXiv:1410.6379 

\bibitem[Groenewegen et al.(2002)]{trilegalold} Groenewegen, M.~A.~T.,
  Girardi, L., Hatziminaoglou, E., et al.\ 2002, \aap, 392, 741
  
\bibitem[Tinetti et al.(2012)]{echo} Tinetti, G., Beaulieu, 
J.~P., Henning, T., et al.\ 2012, Experimental Astronomy, 34, 311 

\bibitem[Tremaine \& Dong(2012)]{tremaine} Tremaine, S., \& Dong, S.\
  2012, \aj, 143, 94
  
\bibitem[Van Grootel et al.(2014)]{hd97658b} Van Grootel, V., 
Gillon, M., Valencia, D., et al.\ 2014, \apj, 786, 2 

\bibitem[Vanhollebeke et al.(2009)]{bulge} Vanhollebeke, E., 
Groenewegen, M.~A.~T., \& Girardi, L.\ 2009, arXiv:0903.0946 

\bibitem[Weiss et al.(2013)]{weiss} Weiss, L.~M., Marcy, 
G.~W., Rowe, J.~F., et al.\ 2013, \apj, 768, 14 
  
\bibitem[Winn et al.(2011b)]{winn16} Winn, J.~N., Albrecht, S., 
Johnson, J.~A., et al.\ 2011, \apjl, 741, L1 

\bibitem[Winn et al.(2011a)]{winn55ce} Winn, J.~N., Matthews, 
J.~M., Dawson, R.~I., et al.\ 2011, \apjl, 737, L18 

\bibitem[Winn(2011)]{winnbook} Winn, J.~N.\ 2011, in Exoplanets, 
edited by S.~Seager.~ Tucson, AZ: University of Arizona Press, 2011, 526 
pp.~ ISBN 978-0-8165-2945-2., p.~55-77

\bibitem[Zheng et al.(2004)]{zheng2004} Zheng, Z., Flynn,
  C., Gould, A., Bahcall, J.~N., \& Salim, S.\ 2004, \apj, 601, 500

\end{thebibliography}
\end{document}